\documentclass[longnamesfirst,numberedappendix,apj]{emulateapj}
\usepackage[it,normalsize]{subfigure}
\usepackage{xspace}
\usepackage{epsfig}
\usepackage{amsmath}
\usepackage{apjfonts}
\renewcommand\arcdeg{\mbox{$^\circ$}\xspace}%
\renewcommand\arcsec{\mbox{$^{\prime\prime}$}\xspace}%
\renewcommand\micron{\mbox{$\mu$m}\xspace}%
\newcommand{\ssim}{\sim\!}

\newcommand{\kms}{${\rm km\;s}^{-1}$\xspace}
\newcommand{\msun}{$M_\sun$\xspace}
\newcommand{\rsun}{$R_\sun$\xspace}
\newcommand{\lsun}{$L_\sun$\xspace}
\newcommand{\msunyr}{$M_\sun$~yr$^{-1}$\xspace}
\newcommand{\secp}[1]{(\S\ref{#1})\xspace}
\newcommand{\sect}[1]{\S\ref{#1}\xspace}
\newcommand{\eqt}[1]{(\ref{#1})\xspace}
\newcommand{\eqp}[1]{(eq.\ [\ref{#1}])\xspace}
\newcommand{\panp}[1]{({\em #1})\xspace}
\newcommand{\pant}[1]{{\em #1}\xspace}
\newcommand{\pa}{P.A.\xspace}
\newcommand{\pas}{P.A.s\xspace}
\newcommand{\eg}{e.g.,\xspace}
\newcommand{\ie}{i.e.,\xspace}

\begin{document}


\shorttitle{The 3-D Map of SN~1987A}
\shortauthors{Sugerman {\em et al.}}

\title{The Three-Dimensional Circumstellar Environment of SN 1987A} 

\author{
 Ben E.\ K.\ Sugerman\altaffilmark{1,2},
 Arlin P.\ S.\ Crotts\altaffilmark{2,3},
 William E.\ Kunkel\altaffilmark{4},
 Stephen R.\ Heathcote\altaffilmark{5},
 and Stephen S.\ Lawrence\altaffilmark{6}
}

\altaffiltext{1}{Space Telescope Science Institute, 3700 San Martin
  Drive, Baltimore, MD 21218; sugerman@stsci.edu}
\altaffiltext{2}{Department of Astronomy, Columbia University,
  New York, NY 10027; arlin@astro.columbia.edu}
\altaffiltext{3}{Guest Observer, Cerro-Tololo Inter-American
  Observatory}
\altaffiltext{4}{Las Campanas Observatory, Carnegie Observatories,
  Casilla 601, La Serena, Chile; kunkel@ociw.edu}
\altaffiltext{5}{Southern Observatory for Astronomical Research, Casilla 603,
  La Serena, Chile; sheathcote@noao.edu}
\altaffiltext{6}{Department of Physics, Hofstra
  University, Hempstead, NY 11549; Stephen.Lawrence@hofstra.edu}

\begin{abstract}
Surrounding SN~1987A is a three-ring nebula attributed to interacting
stellar winds, yet no model has successfully reproduced this system.
Fortunately, the progenitor's mass-loss history can be reconstructed
using light echoes, in which scattered light from the supernova traces
the three-dimensional morphology of its circumstellar dust.  In this
paper, we construct and analyze the most complete map to date of the
progenitor's circumstellar environment, using ground and space-based
imaging from the past 16 years.  PSF-matched difference-imaging
analyses of data from 1988 through 1997 reveal material between 1 and
28 ly from the SN.  Previously-known structures, such as an inner
hourglass, Napoleon's Hat, and a contact discontinuity, are probed in
greater spatial detail than before.  Previously-unknown features are
also discovered, such as a southern counterpart to Napoleon's Hat.
Careful analyses of these echoes allows the reconstruction of the
probable circumstellar environment, revealing a richly-structured
bipolar nebula.  An outer, double-lobed ``Peanut,'' which is believed
to be the contact discontinuity between red supergiant and main
sequence winds, is a prolate shell extending 28 ly along the poles and
11 ly near the equator.  Napoleon's Hat, previously believed to be an
independent structure, is the waist of this Peanut, which is pinched
to a radius of 6 ly.  Interior to this is a cylindrical hourglass, 1
ly in radius and 4 ly long, which connects to the Peanut by a thick
equatorial disk.  The nebulae are inclined 41\degr south and 8\degr
east of the line of sight, slightly elliptical in cross section, and
marginally offset west of the SN.  From the hourglass to the large,
bipolar lobes, echo fluxes suggest that the gas density drops from
1--3 cm$^{-3}$ to $\gtrsim0.03$ cm$^{-3}$, while the maximum
dust-grain size increases from $\sim0.2$\micron to 2\micron, and the
silicate:carbonaceous dust ratio decreases.  The nebulae have a total
mass of $\sim1.7$\msun.  The geometry of the three rings is studied,
suggesting the northern and southern rings are located 1.3 and 1.0 ly
from the SN, while the equatorial ring is elliptical
($b/a\lesssim0.98$), and spatially offset in the same direction as the
hourglass.
\end{abstract}

\keywords{ 
circumstellar matter --- 
dust --- 
scattering ---
stars: mass loss  --- 
supernovae:individual (SN 1987A) --- 
techniques: image processing
}


\section{INTRODUCTION }\label{sec-intro}

Supernova (SN) 1987A in the Large Magellanic Cloud (LMC) is the
nearest SN in 400 years; the only SN for which the progenitor
star has been identified and studied; and the only SN whose evolution
is resolved and observed at all wavebands.  Located approximately 50
kpc away, this system is just marginally resolved by most ground-based
telescopes, even with adaptive optics and high-resolution cameras.
Matters are further complicated by bright, neighboring stars, which
severely crowd the innermost field of interest for ground- (and even
some space-) based imaging and spectroscopy.  Nature has a terrible
sense of irony in placing the first nearby SN since the inventions of
the modern telescope and detector at the far extreme of their
observing limits.

The progenitor has been identified as Sk $-69$\degr202 \citep{San69}, a
B3 I supergiant \citep{Wal89}.  From its spectral type and distance,
one infers a luminosity of $\ssim10^5$ \lsun, a surface temperature
15000--18000~K, and a radius around 43\rsun, which imply that the star
exploded as a blue supergiant \citep[or BSG;][]{Woo87}.  This small
radius is confirmed by the underluminous light curve of the SN
\citep{Arn87} and the extremely-short time delay between
detection of the neutrino burst and optical brightening of the SN
\citep{Shi87,Arn88}.

\begin{figure}\centering
\includegraphics[width=2.in,angle=0]{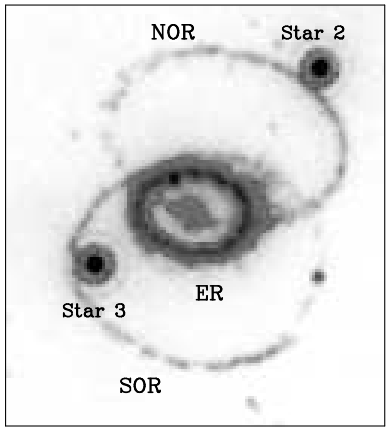}
\caption{{\em HST} Wide Field and Planetary Camera 2 (WFPC2) negative
image of a $4\farcs5\times5\farcs0$ field surrounding SN~1987A, taken
in F656N. North is up, east is left.  The ER (central ring) surrounds
the ejecta from the SN (center), and is flanked by the north and
south ORs (NOR and SOR, respectively).  The companion stars 2 and 3,
using the classification of \citet{WS90}, are positionally coincident
along the line of sight.  To increase the display range, the ER is
displayed with a separate  color stretch.
\label{PCMAP}}
\end{figure}

Sk $-69$\degr202, has a highly-structured circumstellar environment
(CSE), evidenced in large part by the famous three-ring nebula (shown
in Fig.\ \ref{PCMAP}).  This material was flash-ionized by the SN
light pulse, and is observed today through recombination cooling.  The
kinematics of the inner, equatorial ring (ER) show it to be a planar
ring, expanding at $\ssim10.3$ \kms\, rather than a limb-brightened
ellipsoid \citep{CH91}.  Assuming the ER is circular, its observed
geometry implies it is inclined at $\ssim43\arcdeg$
\citep{Pla95,Bur95}, with the northern edge closest to earth. The
morphology of the outer rings (ORs) is poorly understood, in part
because the south OR (SOR) is distorted from the elliptical shapes of
the ER and north OR (NOR).

While SN~1987A exploded as a BSG, the presence of circumstellar
material, as well as significant CNO processing within that gas,
indicates that the star first passed through a RSG phase
\citep{Fra89}.  To form the three rings, many authors have looked to
interacting stellar winds (ISW) between the RSG and BSG outflows.  In
this model \citep{Kwo82,Bal87}, an equatorial overdensity in a
previously-expelled slow, dense RSG wind focuses a fast, tenuous BSG
wind into a polar trajectory, the interaction from which produces a
bipolar peanut-shaped nebula, or wind-blown bubble
\citep{Woo88,WPW88,Arn89,CE89,Fra89,Luo91,WM92}.  Here, the ER is the
overdense waist of the bipolar nebula, and the ORs lie at some point
along its outer edges.  Such a structure of dust has been traced out
by scattered-light echoes as reported in \citet[][hereafter
CKH95]{CKH95}, supporting this general model.  

\citet{BL93} first modeled the ring nebulosity within a full
\mbox{2-D}, time-dependent hydrodynamic realization of the ISW model,
assuming reasonable values for the mass-loss rates and expansion
velocities of the stellar winds in both RSG and BSG stages, as well as
an {\em ad hoc} asymmetry function for the RSG wind. \citet{MA95} also
performed this exercise with a simple cooling law, creating synthetic
images from their model (using an assumed opacity and ionizing UV
flux) which reproduce the ring geometry only if one identifies the ORs
with limb-brightening along peanut-shaped bubbles rather than discrete
rings.  Rather than rely on {\em ad-hoc} asymmetry functions,
\citet{Col99} incorporated the wind-compressed disk theory \citep[or
WCD,][]{BC93}, in which winds leaving a rapidly rotating star converge
at the equator as they orbit the star, to provide the equatorial RSG
overdensity.  They find good agreement with the ER geometry but
require that the star rotate at 30\% of the breakup rotation rate.  To
spin the envelope this quickly, they invoke a binary merger in which a
companion spun up the envelope, then joined the primary core
\citep{LS88}.  None of these models have provided a satisfactory
explanation of the ORs, leading some to consider in more detail the
influence of a binary companion \citep{Pod91,LOK95}.

These are only a few examples of a much larger body of work, and
despite these extensive efforts, many of the critical details of the
final evolution of the progenitor and of the CSE surrounding the SN
remain speculative.  Within the generally-accepted framework of the
ISW formation hypothesis lie a large number of variant models,
invoking magnetohydrodynamics, binary companions, multiple
evolutionary tracks, diverse mass-loss histories and dichotomous final
states.  Unfortunately, no model has accurately reproduced this system
\citep{Sug05}. 

Thus far, formation models have assumed as sufficient
conditions a set of ``reasonable'' (\ie typical of nearby LMC
supergiants) values for the mass, radius, temperature and composition
of Sk $-69$\degr202, $\dot{M}$ and $v_{exp}$ for the RSG ($1-2 \times
10^{-5}$ \msunyr, 5--20 \kms) and BSG ($10^{-6}-10^{-7}$ \msunyr,
300--600 \kms) winds, and have used as the primary constraints the
few physical conditions of the ER of which we are certain: the radius
\citep[$0\farcs81$;][]{Bur95}, and the expansion velocity \citep[$10.3$ km
s$^{-1}$;][]{CH91}.   Results matching only a handful of observed conditions
can only be regarded as suggestive and preliminary.  However, up to
this point, only a handful of constraints exist. Our
understanding of the circumstellar history of SN~1987A is limited
by a lack of observed constraints, not theoretical models.

The SN provides us with two independent, yet equally useful means to
reconstruct the mass-loss history. First, we are now observing a
unique period in the formation of SNR~1987A as the high-velocity SN
debris overtakes the slowly expanding ER \citep{Son98b,Sug02}.  While
this interaction will eventually destroy the circumstellar nebula, it
provides the opportunity to explore the CSE by detailing the material
directly through shock-heating.  Furthermore, the increasing X-ray and
UV flux from this interaction will reionize unseen portions of the
nebular structure, revealing many of them for the first time \citep{Luo94}.

Second, light echoes from the SN have illuminated a significant volume
of the CSE, and can be used to constrain the three-dimensional (3-D)
structure, density, and composition of the stellar outflows.  Three
general structures have been previously reported.

{\em Close circumstellar material:} The inner few light years
surrounding SN~1987A have been studied between days 750--1469 (after
the SN; 1987 Feb.\ 24.23) by CKH95.  Echoes detected within 3\arcsec
of the SN were mostly found to lie on an hourglass-shaped bipolar
nebula, inclined $43\degr\pm 5\degr$ south, and rotated $3\degr\pm
5\degr$ east of the line of sight, while a smaller set were also
identified as roughly coplanar with the ER but at larger radii.

{\em Napoleon's Hat:} Napoleon's Hat (NH) was originally identified by
\citet{Wam90a} as an arc-like feature \citep[also described as an
archer's bow by][]{CK91} looping roughly 6\arcsec north of the SN.  It
has been variously interpreted as a disk coplanar to the ER
\citep{WW92}, a truncated double cone \citep{Pod91}, or a parabolic
bow shock at the terminal edge of a region evacuated by the
main-sequence (MS) progenitor \citep{WDK93}.

{\em Large radius contact discontinuity:} Recognizing that red giants
and supergiants typically have slow, dense stellar winds, \citet{CE89}
proposed that the interface between the RSG wind and surrounding
medium (homogenized and rarified by a prolonged MS wind) would be a
``snowplow'' discontinuity \citep{CMW75}.  As these are dense and
cold, the interface should be both long-lived and have a significant
dust component.  \citet{CE89} predicted that a stellar-wind bubble in
pressure equilibrium with the surrounding medium should occur at a
radius of roughly 16 ly for reasonable RSG wind parameters.  This
corresponds closely to echoes reported by Bond et al.\ (1989,1990),
\citet{CM89}, and Crotts \& Kunkel (1989,1991) roughly 9\arcsec from
the SN, which are interpreted as the contact discontinuity (CD)
between the RSG/MS winds.

Until now, most modeling efforts have focused on the last
$\ssim20,000$ years of evolution (roughly the kinematic timescale of
the rings and evolutionary timescale of the BSG stage) while, in large
part, avoiding the previous 0.3--1 Myr of the RSG phase, during which
the NH and CD features formed.  Observations of the full circumstellar
environment provide a unique opportunity to detail the products of the
progenitor's mass-loss.  A more complete set of boundary conditions is
necessary to properly investigate the SN system, and to distinguish
viable models from unrealistic scenarios.

Using the technique of PSF-matched
difference imaging on 16 years of optical imaging data, we report an
observational effort, at the highest sensitivity and resolution to
date, to understand more fully the entire circumstellar environment
(CSE) of SN~1987A, and to recreate a more thorough history of the
progenitor's mass-loss.  This paper describes the detection and
analysis of light echoes within 30\arcsec of the SN as observed in
ground-based and {\em HST} imaging.  This work is also summarized in
\citet{Sug05}, followed by a detailed analysis of how our results
constrain formation models and the progenitor's evolution.

 The theory and application of light echoes are presented in
\sect{sec-le}, including a dust-scattering model which is used to
constrain the density and composition of echoing material.  The data
are described in \sect{sec-echodata}, followed by the data-reduction
pipeline \secp{sec-reduc}.  The light echoes, their three-dimensional
positions, and the tools necessary to analyze them, are presented in
\sect{sec-lea}.  These data are used to build geometric models of
material within $\ssim30$ ly of the SN, the details of which are
explained in \sect{sec-CD}--\ref{sec-CS}.  A complete picture of the
CSE, including constraints on the density, chemical composition, and
mass of all circumstellar material, is made in \sect{sec-density}.
It is our hope that the results presented here will
serve as a significantly-improved set of constraints for hydrodynamic
models of the CSE formation, for stellar-evolution models of the
progenitor, and for addressing the general question of asymmetric and
bipolar stellar outflows.  


\section{Light Echoes: Theory and Model }\label{sec-le}

\begin{figure}\centering
\includegraphics[angle=0,width=3.25in]{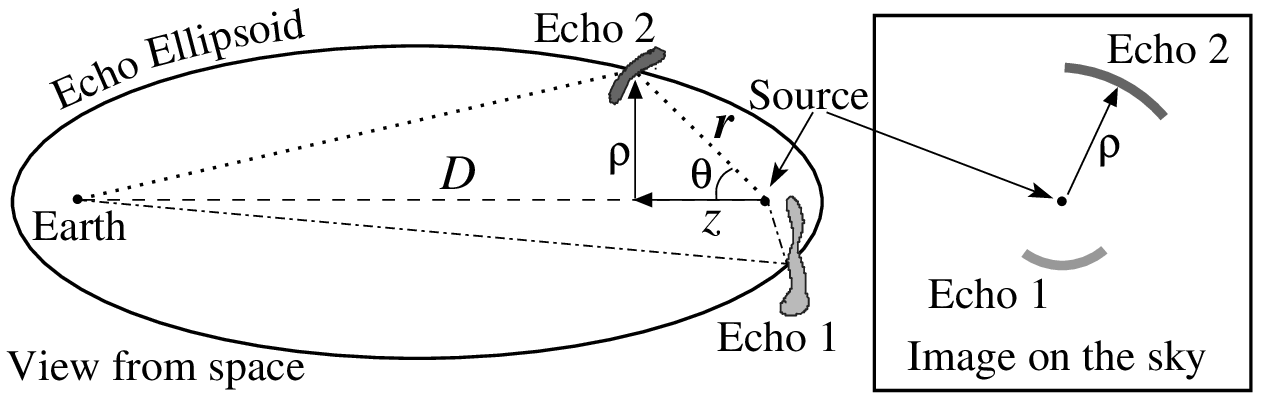} 
\caption{Cartoon schematic of a scattered-light echo.  Echoes lie on
the locus of points equidistant in light travel from the source and
observer--\ie an ellipsoid--and appear as arcs or rings on the plane
of the sky, with a one-to-one mapping between their 2-D image
positions (right) and their 3-D locations in space (left).  Note that
the distances are not to scale:  $D$ is much larger than $z$, such
that the segments from Earth to the scattering dust are nearly
parallel to the line of sight. 
\label{echo_toon}}
\end{figure}

When a light pulse is scattered by dust into the line of sight, an
observable echo is produced, provided the pulse is sufficiently
luminous and the dust sufficiently dense.  An echo observed a given
time after the pulse must lie on the locus of points equidistant in
total light travel from the source and observer, that is, an ellipsoid with
known foci.  This simple geometry, shown in Figure \ref{echo_toon},
directly yields the three-dimensional (3-D) position of an echo, uncertain
only by the assumed distance to the source.  A complete discussion of
light echoes, including single-scattering models and the observability
of echoes around a wide variety of cataclysmic and other variable
stars, can be found in \citet{Sug03}.  For completeness, we briefly
summarize the salient points below.

\subsection{Light Echo Surface Brightness \label{sec-le-sb}}

Consider the geometry shown in Figure \ref{echo_toon}, in which a
source at distance $D$ from the observer emits a pulse of light of
duration $\tau$.  This light scatters off dust located at position
{\boldmath $r$} from the source, and reaches the observer a time $t$
after the arrival of the unscattered light pulse.  Since $D$ is
generally much greater than $r$, the echo depth $z$ along the
line-of-sight can be approximated by the parabola \citep{Cou39}
\begin{equation} \label{sec-le-sb1}
 z=\frac{\rho^2}{2ct}-\frac{ct}{2}
\end{equation}
where $\rho=r\sin{\theta}$ is the distance of the echo from the
line-of-sight in the plane of the sky, and $\theta$ is the scattering
angle (provided $D\gg r$).

Treating the dust as a thin planar sheet of thickness $\Delta z$,
and adopting cylindrical coordinates, the surface brightness of an
echo at a given wavelength $\lambda$ is given by 
\begin{equation} \label{sec-le-sb6}
 B_{\rm SC}(\lambda,t)=F(\lambda) n_{\rm H}(\mbox{\boldmath $r$})
 \frac{c\Delta z}{4\pi r \rho \Delta \rho}
 S(\lambda,\mu).
\end{equation}
To facilitate computations, we adopt units of years and light-years,
making $c=1$.
We define the integrated scattering function 
\begin{equation} \label{sec-le-sb7}
 S(\lambda,\mu) =\int Q_{\rm SC}(\lambda,a)\sigma_g\Phi(\mu,\lambda,a)f(a)da
\end{equation}
where $a$ is the dust-grain radius, $\sigma_g=\pi a^2$ is the physical
cross section of a grain, and $Q_{\rm SC}$ is a grain's scattering
efficiency.  The number density of grains at position {\boldmath $r$}
with radii between $(a,a+da)$ is given by
\begin{equation}
 n_d(\mbox{\boldmath $r$},a)da=n_{\rm H}(\mbox{\boldmath $r$}) f(a)da
 \label{sec-le-sb8}
\end{equation}
where $n_{\rm H}$ is the number density of H nuclei, and $f(a)$ is the
grain-size distribution function.  The total flux from the light pulse
incident on the dust is $F(\lambda)=\int F(\lambda,t') dt'$.  We adopt
the \citet{HG41} phase function $\Phi$.  Since the light pulse and
dust have finite widths, the observed radial size $\Delta\rho$ of the
echo yields an estimate of its line-of-sight depth
\begin{equation}\label{sec-le-width3}
 \Delta z = \frac{\rho}{t}
  \sqrt{\Delta\rho^2-\left(\frac{\rho}{2t}+\frac{t}{2\rho}\right)^2\tau^2 
  }.
\end{equation}
 
Both the observed flux $F({\lambda})$ and area element
$\rho\Delta\rho$ diminish as $D^{-2}$, thus equation \eqt{sec-le-sb6} for
surface brightness is distance independent, as expected.  If the echo
and source are close, they should suffer roughly the same extinction.
This work is restricted to single-scattering events, and the reader is
referred to \citet{Che86} for a discussion of multiple-scatterings,
which are only of significance when the optical depth of the
scattering material $\gtrsim 0.3$.

\subsection{Dust Properties \label{sec-le-dust}}

\begin{figure}\centering
\includegraphics[angle=-90,width=3.25in]{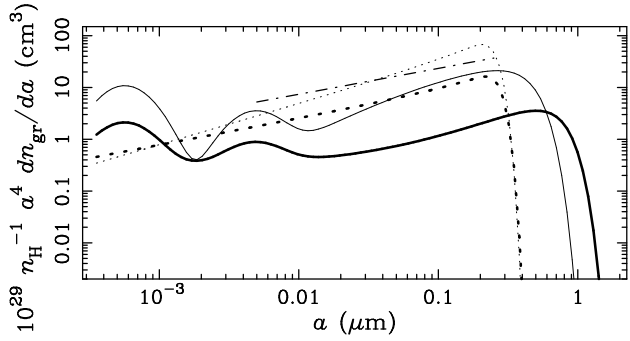}
\caption{Dust grain density distributions for carbonaceous (solid
lines) and silicate (dotted lines) dust from {WD01}.  Thick lines are
for the LMC model with $R_V=2.6,\ b_{\rm C}=1\times 10^{-5}$, and thin
lines are for the Galactic model with $R_V=3.1,\ b_{\rm C}=6\times
10^{-5}$.  For clarity, and to allow the reader to estimate the mass
present in each size range, data are plotted in units of
$a^4\,dn_{gr}/da$.  The dot-dashed line shows the corresponding
density distribution from \citet{MRN77}.
\label{dnda}} 
\end{figure}

\citet{MRN77} found that a mixture of carbonaceous and silicate grains
following a density distribution given by $f(a)\propto a^{-3.5}$
adequately modeled Galactic starlight extinction.  \citet[hereafter
WD01]{WD01} and \citet{Li01} have calculated dust scattering and
absorption properties for graphite and ``astronomical
silicate,''\footnote{Tabulated values of dust properties are available
at
\url{http://www.astro.princeton.edu/$\sim$draine/dust/dust.diel.html}}
from which {WD01} have constructed a new density distribution $f(a)$
to fit observed interstellar extinction.  We refer the reader to their
paper for the actual functional form.  We adopt their LMC dust model
for $R_V=2.6$, and $b_{\rm C}=1\times 10^{-5}$, which is shown by
thick curves in Figure \ref{dnda}.  For comparison, their preferred
Galactic dust model (case A, $R_V=3.1$, $b_{\rm C}=6\times 10^{-5}$)
is also shown.  The reddening ratio is defined as $R_V=A_V/E(B-V)$
\citep{Car89}, and $b_{\rm C}$ is the total C abundance per H nucleus.

\subsection{The Light Pulse from SN~1987A \label{sec-le-flux}}

\begin{figure}\centering
\includegraphics[angle=-90,width=3.25in]{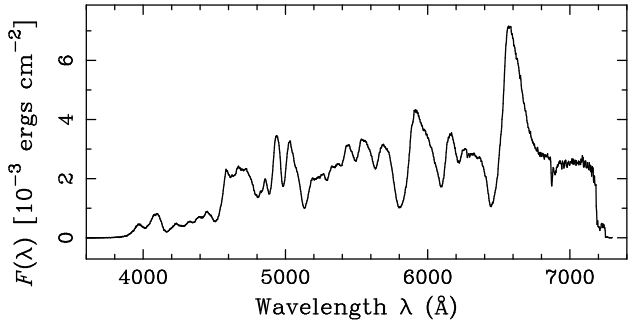}
\caption{Time-integrated spectrum of SN 1987A from 59 days before to
42 days after maximum light.
\label{pl87A}}
\end{figure}

Using the early optical light curves from \citet{Ham88}, SN~1987A
reached maximum light in $V$ and $R$ roughly 87 days after its
discovery on 1987 Feb 24.   The light-curve duration $\tau$ is measured
as the effective width $\tau=\int{F(\lambda,t)dt}/F(\lambda,t_{\rm
max})$.  A homogenous time series of spectra from the first thousand
days are publically available from the SUSPECT supernova
archive\footnote{\url{http://bruford.nhn.ou.edu/$\sim$suspect/index.html}}.
We integrated these spectra over the first 420 days, and found an
effective width of 100 days in the wavelength ranges of interest.
Integrating again over just this width, centered at maximum light, we
generated the time-integrated flux $F(\lambda)$ that will enter
equation \eqt{sec-le-sb6}, shown in Figure
\ref{pl87A}.  Note that at these early times, spectral coverage ended
at 7300\AA, and there is no empirical spectrum available in the
wavelength range of our reddest filter \secp{sec-echodata-filters}.

Referring again to Figure \ref{echo_toon}, since $D\gg r$, the
trajectory from source to dust to earth is very similar to the direct
source-earth path.  Under the assumptions that the ISM is homogenous
near the SN, and that extinction is small within the echoing dust
(implicit in the single-scattering approximation), flux from the SN
and surrounding light echoes should be reddened by the same amount
along the optical path between the LMC to the observer.  This
assumption allows the direct comparison of the observed echo fluxes to
the dust-scattering model, without the additional uncertainty of
extinction corrections \secp{sec-reduc-calib}.  The spectrum in Figure
\ref{pl87A} is therefore not extinction-corrected\footnote{Spectra of
SN~1987A in the SUSPECT archive were dereddened assuming $E(B-V)=0.16$
and $R_V=3.1$, which we have removed to reproduce the observed
values.}.

\subsection{The Light Echo Model \label{sec-le-model}}

To calculate the total echo surface brightness received in a given
bandpass, equation \eqt{sec-le-sb6} must be integrated in wavelength
over the filter response function $T(\lambda)$.  The integrated
scattering function $S(\lambda,\mu)$ must also be integrated over
grain sizes, using equation \eqt{sec-le-sb7} and the dust-density
models from {WD01}.  Removing the geometric factors and gas density
from equation \eqt{sec-le-sb6}, we model echo surface
brightness by computing
\begin{equation}
 I=\iint{T(\lambda)F(\lambda)Q_{\rm  SC}(\lambda,a)
 \sigma_g\Phi(\mu,\lambda,a) f(a)da\;d\lambda}.
 \label{sec-le-model1}
\end{equation}
which is integrated numerically using Simpson's Rule \citep{Press}.
This gives the energy scattered toward the observer in unit
time for unit volume of gas with unit density at unit distance from
the source, and therefore it need only be scaled to the appropriate
physical values to yield the predicted surface brightness.

\begin{figure}\centering
\includegraphics[angle=-90,width=3.25in]{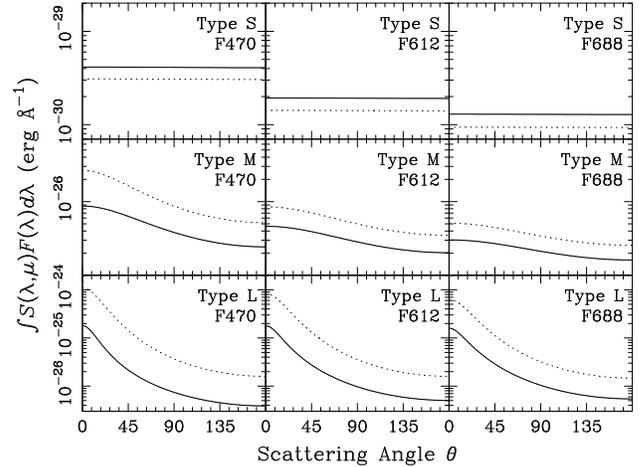}
\caption{The average integrated scattering function $S(\lambda,\mu)$,
integrated over the total spectrum $F(\lambda)$ of SN~1987A (Fig.\
\ref{pl87A}) and filter profiles (Fig.\ \ref{plfilt}).  Line style
designates composition: solid lines show carbonaceous dust, and dotted
lines show silicates.  Columns are calculated for a given filter, and
rows are calcuated for LMC dust with grain-size range types L, M, and
S, as noted at upper right in each panel.  This integral gives the
energy scattered toward the observer in unit time by a unit volume of
gas with unit density located unit distance from the source.
\label{plaisf}}
\end{figure}

Figure \ref{plaisf} shows this integral for the three filters F470,
F612 and F688 (Fig.\ \ref{plfilt}).  Results are given separately for
carbonaceous and silicate dust.  Since the range of dust-grain sizes
in the CSE are unknown, we integrate over three size intervals,
denoted ``S'' (small, $a=3.5\times 10^{-4} - 0.01$\micron), ``M''
(medium, $a=3.5\times 10^{-4} - 0.1$\micron), and ``L'' (large,
$a=3.5\times 10^{-4} - 2.0$\micron).  The type-L dust includes the
full range of grain sizes in $f(a)$ from \citet{WD01}.

Scattering for grains with $a < 0.1\lambda / 2\pi$ occurs within the
Rayleigh limit, for which $S\propto\lambda^{-4}$.  These small grains
have a small scattering efficiency and scatter isotropically, while
distributions including large grains predominantly forward scatter.
It should be noted that while $S\propto\lambda^{-4}$ for Rayleigh
scattering, the largest color shift possible for the integrated SN
spectrum (Fig.\ \ref{pl87A}) in our continuum filters is
$S\propto\lambda^{-3}$, as can be easily calculated from the type-S
dust in Figure \ref{plaisf}.

To apply this scattering model, colors of echoes are compared to the
model values in Figure \ref{plaisf} to constrain the grain-size
distribution, and the relative populations of carbonaceous and
silicate grains.  Dust (or gas) density is calculated from the scaling
between the model and observed surface brightness, using the geometric
factors $\rho$ and $\Delta\rho$, measured from the data, with
$\tau$=100 days, as determined above.

Distinguishing between type S and L dust can be accomplished a few
ways.  First, as noted above, small grains scatter less flux than
larger ones, thereby requiring a higher gas density for a given
observed surface brightness. Since greater density dust has stronger
extinction, practical constraints exist for upper limits to gas
density.  If a single structure is probed at many scattering angles,
and one assumes the density of that structure is constant, the
variation of surface brightness with $\theta$ can also constrain the
grain sizes.  Dust composition can be constrained by comparing colors
of observed and modeled echoes, although carbonaceous and silicate
dust have nearly identical scattering properties at these optical
wavelengths.  A better probe of chemical composition are UV echoes,
since carbon and silicate dust grains have strongly-differing features
at $\lambda<2500$\AA\ \citep{Sug03}.

As a final note, in treating echoes with a planar approximation, this
scattering model assumes that the observed echo is well represented by
a slab of dust of thickness $\Delta z$, all of which is scattering the
integrated flux $F(\lambda)$ toward the observer.  In reality, the
echo is the integral of many volume elements of gas, each of which
scatters a particular time-dependent flux from the SN lightcurve
\citep[see \S2.1 of ][]{Sug03}.  For echo observations in which a
resolution element is much smaller than the outburst duration $c\tau$
\citep[such as those from V838 Mon][]{Bon03}, a much more detailed
scattering model can deconvolve the spatially-resolved dust properties
using the time-dependent lightcurve.  However, such treatment is
unsuited to circumstellar echoes from SN 1987A, since the image
resolution is larger than 100 light days.


\section{Data} \label{sec-echodata}

\subsection{Observations \label{sec-echodata-obs}}

SN~1987A has been observed many times per year since day 375 (after core
collapse) in a ground-based campaign to monitor the appearance and
evolution of its light pulse as it illuminates circumstellar and
interstellar material.  Some data from this campaign have previously been
reported by \citet{CKM89}; \citet{CH91}; \citet{CK91}; and
\citet{CKH95} in studying the circumstellar environment, and in
\citet{Xu94,Xu95} and \citet{Xu99} for interstellar echoes.  

For the present application, we require data with high
signal-to-noise, good seeing (arcsecond or better), and sufficient
resolution to resolve most crowded stars and detail in the narrow surface
brightness features.  Those epochs of imaging that match these
criteria are listed in Table \ref{tbl-echodata-obs}.  Data were
primarily taken on the 2.5-m (100-inch) Dupont telescope at the Las
Campanas Observatory (LCO) at platescales of 0\farcs163 and 0\farcs260
pix$^{-1}$, but one epoch from the Cerro Tololo Interamerican
Observatory (CTIO) 4-m telescope is also used.

Since its launch in 1990, {\em HST} has been used at least yearly to
observe the SN, largely through the efforts of the Supernova Intensive
Survey (SInS) collaboration.  We used archival Wide Field and
Planetary Camera 2 (WFPC2) data taken between 1994 and 2001 for the
current application. Those epochs in which echoes were discovered are
also listed in Table \ref{tbl-echodata-obs}.

\begin{deluxetable*}{l c c c l}
\tablecaption{Light Echo Observations \label{tbl-echodata-obs}}
\tablewidth{5in}
\tablehead{
\colhead{Epoch} & \colhead{Day} & \colhead{Telescope} &
  \colhead{Platescale} & \colhead{Filters}
}
\startdata
1988 Dec 14 & 659  & LCO-100 & 0\farcs163 & B,V,R,I,Z \\
1989 Mar 15 & 750  & LCO-100 & 0\farcs163 & 6023,6067,6967 \\
1989 Dec 18 & 1028 & LCO-100 & 0\farcs163 & F470,F612,F688,F809 \\
1990 Feb 14 & 1086 & LCO-100 & 0\farcs163 & F470,F612,F688,F809 \\
1991 Mar 4  & 1469 & CTIO-4 & 0\farcs124 & F688 \\
1992 Jan 15 & 1787 & LCO-100 & 0\farcs260 & F612,F688,F809  \\
1992 Mar 20 & 1851 & LCO-100 & 0\farcs163 & F612 \\
1992 Aug 14 & 1998 & LCO-100 & 0\farcs260 & F612,F688 \\
1992 Nov 19 & 2095 & LCO-100 & 0\farcs260 & F612,F809 \\
1993 Nov 22 & 2462 & LCO-100 & 0\farcs260 & F470,F612,F688,F809 \\
1993 Dec 27 & 2498 & LCO-100 & 0\farcs260 & F470,F612,F688,F809 \\
1994 Jan  6 & 2508 & LCO-100 & 0\farcs260 & F470,F612,F688,F809 \\
1994 Apr  1 & 2593 & LCO-100 & 0\farcs260 & F612 \\
1994 Sep 24 & 2769 & {\em HST}/WFPC2 & 0\farcs045 & F555W,F675W,F814W \\
1994 Nov 25 & 2830 & LCO-100 & 0\farcs260 & F612,F688,F809 \\
1995 Jan  7 & 2874 & LCO-100 & 0\farcs260 & F612,F688,F809 \\
1995 Mar  5 & 2932 & {\em HST}/WFPC2 & 0\farcs045 & F555W,F675W,F814W \\
1995 Mar 10 & 2935 & LCO-100 & 0\farcs260 & F612,F688,F809 \\
1996 Jan 12 & 3245 & LCO-100 & 0\farcs260 &  F612,F688,F809 \\
1996 Feb 06 & 3270 & {\em HST}/WFPC2 & 0\farcs045 & F555W,F675W,F814W \\
1997 Jul 10 & 3789 & {\em HST}/WFPC2 & 0\farcs045 & F555W,F675W,F814W 
\enddata
\end{deluxetable*}

\subsection{Photometric Bands \label{sec-echodata-filters}}

\begin{figure}\centering
\includegraphics[angle=-90,width=3.25in]{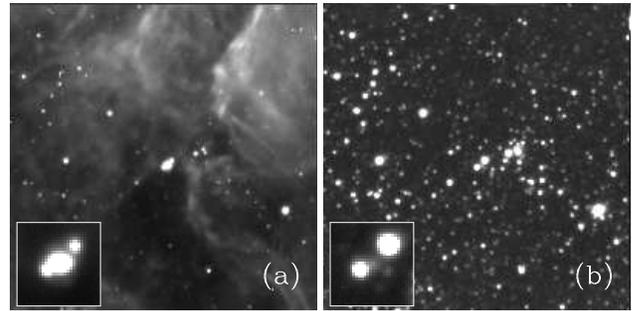}
\caption{Ground-based images of SN~1987A from day 1998.  Each image is
130\arcsec\ on a side, in a standard orientation with north up, east
left.  The inset at bottom left is 10\arcsec on a side.  \panp{a}
Narrow-band H$\alpha$ image.  The inset shows the extended nebula
surrounding the SN, as well as stars 2 and 3. \panp{b} F612
continuum-only image of the same field as \panp{a}.
\label{Ha612}}
\end{figure}

SN~1987A is located in a region of the LMC with considerable nebular
structure, as shown in an $H\alpha$ image from day 1998 in Figure
\ref{Ha612}\pant{a}.  Furthermore, the three-ring nebula surrounding the
SN is the strongest source of line emission in this region.  These
extended surface brightness features create significant sources of
noise when searching for light echoes, especially in the inner few
arcsec around the SN, since the flux from three-ring nebula changes
with time.  To minimize the contribution of nebular-line contamination,
four specially-selected continuum bands were selected for the
monitoring campaign, centered at 4700, 6120, 6880, and 8090\AA.  The
laboratory transmission curves are plotted in Figure \ref{plfilt}.
These bandpasses were chosen to avoid the brighter emission lines of
the diffuse LMC clouds (\eg H$\alpha$, [\ion{N}{2}], [\ion{O}{3}], and
[\ion{S}{2}]), as well as strong emission lines from the SN ejecta.
This has the added benefit of suppressing flux from the central SN
at early times by 1--1.5 mags \citep{CKH95}.  

In keeping with the {\em HST} filter naming convention, we will refer
to each continuum filter by its central wavelength in nm, preceded by
an ``F'', such that the 6120\AA\ filter is F612.  Note however that
these differ from {\em HST} filter names since the latter are appended
with the width of the filter (i.e. F656N).  The effectiveness of these
filters at suppressing nebular-line emission is shown in Figure
\ref{Ha612}\pant{b}, the F612 image from the same epoch as panel
\panp{a}.  There is no detectable signal from the large cirrus-like
clouds or the inner circumstellar nebula.  An echo results from the SN
light pulse scattering off dust, and since its spectrum had a bright
continuum over the duration of its outburst (Fig.\ \ref{pl87A}), these
filters will not suppress echo signal.


\section{Data Reduction }\label{sec-reduc}

Data were reduced in the following stages: (1) CCD processing, (2)
homogenization, (3) pipeline pre-processing, and (4) PSF-matched
difference imaging.  Each stage is designed to be minimally-invasive,
thereby maximally preserving faint structure and flux.  We begin with
a discussion of difference-imaging, followed by the four reduction
stages listed above. 


\subsection{Difference-Imaging Photometry }\label{sec-dip}

\begin{figure}\centering
\includegraphics[angle=-90,width=3.25in]{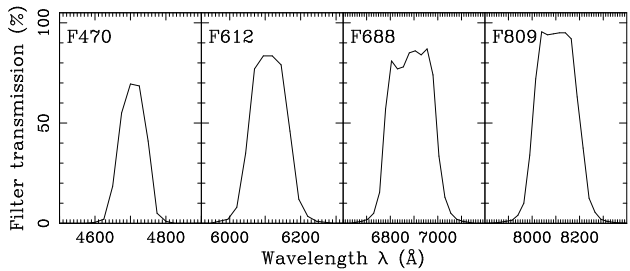}
\caption{Filter transmission curves for the narrow-band continuum
 filters used to observe SN 1987A.
\label{plfilt}}
\end{figure}
\begin{figure}\centering
\includegraphics[height=3.25in,angle=-90]{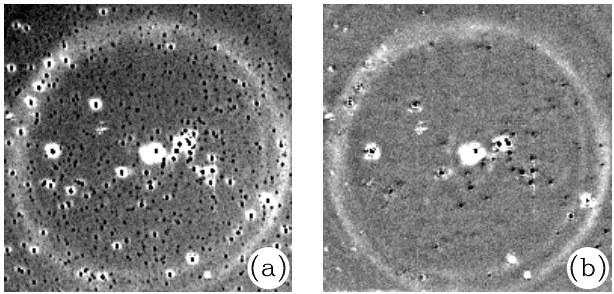}
\caption{Differences of two photometrically scaled images, showing \panp{a}
  the direct subtraction, and \panp{b} the PSF-matched subtraction.
\label{470_14Feb90m_sub}} 
\end{figure}

Consider the problem of subtracting two photometrically-scaled images
to study time variable phenomena.  Figure
\ref{470_14Feb90m_sub}\pant{a} shows a typical direct arithmetic
subtraction.  Even under identical circumstances (same telescope,
detector, airmass, etc.) the point-spread function (PSF) across the
detector can change, caused by e.g.\ flexure of the mirrors, movement
of the filter, or change in atmospheric seeing.  As a result,
residuals remain around every star in the field, leaving holes and
haloes such as those seen here.

\begin{figure}\centering
\includegraphics[height=3.25in,angle=-90]{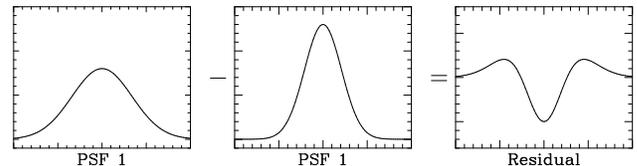}
\caption{Two different PSFs of identical area are subtracted, yielding
the residual in the right-hand panel.
\label{subgauss}}
\end{figure}

This is easily understood by subtracting two Gaussians of differing
peak and FWHM, but identical total area, as shown in Figure
\ref{subgauss}.  The residual has a bright halo in the wings and a
hole in the center, since an uneven amount of flux was subtracted from
each point along the profile.  This can be addressed a few ways.  Each
image can be degraded, by smoothing with \eg gaussian filters. This is an
invasive procedure, smearing out structure and detail that might
otherwise be studied.  Rather, an optimal technique is to remap the
flux in one PSF to that of the other in a uniform and consistent way.

Our technique is to seek a convolution kernel $k$ such that the input
star $i$ can be mapped onto the reference star $r$, or $ i=r \ast k$.
From the Convolution Theorem, the Fourier Transform (FT) of this
mapping becomes
\begin{equation}
 {\rm FT}(i)={\rm FT}(r) \times {\rm FT}(k)
\label{sec-dip2}
\end{equation}
from which we isolate the convolution kernel
\begin{equation}
 k={\rm FT}^{-1}\left[\frac{{\rm FT}(i)}{{\rm FT}(r)}\right].
\label{sec-dip3}
\end{equation}

\citet{TC96} implemented this method of PSF-matching in an external
IRAF\footnote{{\em IRAF} is distributed by the National Optical
Astronomy Observatories, which are operated by the Association of
Universities for Research in Astronomy, Inc., under cooperative
agreement with the National Science Foundation.} package {\em
difimphot} (DIFference IMaging PHOTometry), originally designed for
use with the M31 microlensing project \citep{Cro92}, and subsequently
updated for the MACHO microlensing project \citep[e.g.][]{Alc95}.  In
its simplest implementation, empirical or model PSFs are used with
equation \ref{sec-dip3} to build the kernel, which is then convolved
with the input image, and the reference image is subtracted away.  We
have implemented further modifications to employ the robust
PSF-fitting and building algorithms of {\em daophot} \citep{Ste87}, as
described in Appendix \ref{sec-dip-pipeline}.  See \citet{Ala00} for an
example of alternative methods of PSF-matching.

The same data entering the arithmetic subtraction in Figure
\ref{470_14Feb90m_sub}\pant{a} are shown after PSF-matching in panel
\panp{b}.  Nearly all stars have been cleanly
subtracted.  Since Poisson noise $\sigma$ scales as the square-root of
signal, 
the data are slightly noisier at the positions of subtracted stars than in
the surrounding sky.  Nonetheless, this represents an improvement in
the background noise by a factor of $\sim2$ for faint stars, and
$\gtrsim5$ for those of median brightness.  The brightest stars with
linear profiles are well-subtracted except in the cores, where Poisson
noise is highest.    We can reliably detect extended surface
brightness only $1-2\sigma$ above the background sky, while for noiser
regions (near bright stars or other variable sources) this 
threshold must be evaluated on an individual bases.

PSF-matching can not completely subtract away stars with non-linear or
saturated profiles for many reasons.  (1) These stars do not have a
uniform profile that can be described by a single function.  (2)
The PSF of such a star does not properly represent how flux is
recorded on the chip.  (3) Since these are the brightest objects in
the field, even if the average profile is well-subtracted, the
residuals will be largest.  Thus these represent untreatable sources
of error even in the best of subtractions.

\subsection{CCD Processing \label{sec-reduc-proc}}

Ground-based data were reduced within {\em IRAF} in the standard
manner.  Images were column biased, checked for residual bias and, if
appropriate, a full bias frame was subtraced.  Images were then
flat-fielded.  Bad columns, rows or pixels were {\em not} cosmetically
fixed; however, most images had bad data within 1--4 pixels of each
edge, which were reset to zero value.  Pipeline-calibrated WFPC2
images were used directly from the data archive.

\subsection{Homogenization \label{sec-reduc-reg}}

For a given epoch, all data in a single filter must be combined into a
single ``stack'' image.  To perform difference-imaging, this data must
also be geometrically-registered to a common reference frame.  This
latter step is critical since registration errors lead to large
stellar subtraction residuals, a significant source of noise and
error.  We chose to register all ground-based data to the 0\farcs163
pix$^{-1}$ platescale, for the following reasons.  The primary
interest of this work is to detect and resolve faint structure, thus
data need to be of high-resolution.  Faint structures become brighter
when resampled onto a lower-resolution grid, but the ability to
resolve these from the background (stars, subtraction residuals,
cosmetic defects, statistical pixel noise) is diminished with the
width of such features.  Resampling high resolution data onto a
lower-resolution grid also degrades the FWHM of stellar profiles, and
PSF-matching works best with a well-sampled FWHM$\gg2$ pixels. 

Since the images taken at 0\farcs260 pix$^{-1}$ have high
signal-to-noise and the stellar PSFs are well-sampled, the data can be
interpolated onto a higher-resolution grid without great loss of
information in faint features.  This has been rigorously tested and
discussed by \citet{FH02} within the context of ``drizzling'' data to
higher resolution.  Individual frames were all registered to a common
orientation, then stacked.  Since most frames are slightly offset from
each other over a single epoch, this is effectively the same procedure
as drizzling observations with sub-pixel shifts.

\subsubsection{Geometric Registration} \label{sec-reduc-reg-reg}

One epoch at 0\farcs163 pix$^{-1}$ was chosen, and the F612 image was
geometrically-registered to orient the $(x,y)$ axes with north and
west, using an astrometric solution derived from stars in the UCAC-1
and 2 catalogs \citep{Zac00,Zac04}, which we determine to be accurate
to within 0\farcs03 of the Hipparcos astrometric reference frame in
this region of the sky.  This is the absolute reference orientation.
All other filters within that epoch were registered to this
orientation, then all other epochs were registered to the
corresponding filter in the reference orientation.  For those epochs
with non-standard filters, the nearest equivalent was used.

Registration transformations for ground-based frames were computed for
a third-order polynomial with half-cross terms, and resampling was
accomplished using a bicubic interpolated spline. Since centroids are
the dominant source of error in these transformations, we employ the
techniques of crowded-field PSF-fitting, for which extensive software
exists in the {\em daophot} package \citep{Ste87}.  An empirical PSF
is built using actual stellar profiles from the frame, as outlined in
the {\em daophot} package.  This PSF is then iteratively fit to groups
of stars using a non-linear least-squares minimization, and centroids
are output to a file.  Avoiding stars with saturated or non-linear
profiles, we find PSF-fitting has an average centroid uncertainty of
less than 1\% of the FWHM.  Matched pairs of stars in both input and
reference images were made using {\em starmatch}, a highly efficient
implementation of the triangle-matching algorithm of \citet{Gro86}.
In 98\% of the frames, geometric registration was accomplished with
RMS errors less than 0.1 pixel in both dimensions.

\subsubsection{Image Stacking} \label{sec-reduc-reg-stack}

All registered images in a given filter per epoch are combined into a
single ``stack.''  This begins with determining a list of reference
stars, which are uncrowded, have a clean spatial profiles, high
signal-to-noise, and constant flux.  These lists are used
to build empirical PSFs, and determine the photometric scalings
between images for combining and differencing.  

As a general algorithm, reference stars are identified from a
group of images.  Bright and unsaturated stars are selected by
applying flux cut-offs to the stellar identification algorithm {\em
daofind}, while crowding and clean profiles are addressed by the {\em
daophot} software (e.g.\ {\em pstsel}), and custom software in the {\em
difimphot} package.  Aperture and PSF-fitted photometry is performed
on this star list for all images in the group.  The photometry
provides an initial guess of the photometric scaling between images,
which is applied to the flux of each star per image.  Stars that vary
by more than a user-specified amount are rejected, and the procedure
is repeated.  Stars that vary within 3\% of their median value are
treated as constant, and constitute the reference-star list.

PSF-fitted photometry is performed on each image in the stack using
the reference star list.  This photometry is used to determine an
additive (sky) and multiplicative (scale) scaling to remove the median
sky and photometrically normalize each image to the best-seeing,
highest signal-to-noise image in the stack.  Data with unusually-poor
seeing or large photometric scalings can be rejected, and the
remaining images are combined using a weighted average, where each image
is weighted by
\begin{equation}
 w= \frac{1}{{\rm scale}^2\cdot{\rm sky}\cdot{\rm FWHM}^2},
\end{equation}
renormalized such that $\sum{w_i}=1$.  Bad pixels, from cosmic rays,
warm or dead pixels, cosmetic defects, or large statistical outliers,
can be rejected using sigma-clipping if there are $\gtrsim10$ images
in the stack, otherwise we use the median.  We found through
trial-and-error that when sigma-clipping rejection could not be used,
rejecting the highest and lowest pixel value for a given position in
the data was sufficient to remove most bad pixels.

\subsubsection{The Homogenized Data}\label{sec-reduc-reg-final}

The result of the above is a homogenous set of registered images, one
per filter per epoch, which maximized the signal-to-noise while
minimially-resampling the data.  This last point is important, since
each spatial resampling degrades the image quality, thereby smearing
out faint echo signal, making them more difficult to detect.

Finally, note in Figure \ref{Ha612} that Stars 2 and 3 are crowded
with the SN and the immediately-surrounding nebula.  Star 3 is known
to be variable, and will leave subtraction residuals in many
difference images, depending on its flux at the time of observation.
Since the SN is fading with time, it too will produce confusing
residuals at early times, when it was still bright in the continuum.
We therefore PSF-subtract the SN and Stars 2 \& 3 using {\em daophot}.
This was performed two ways.  First, we allowed the subtraction
algorithm ({\em allstar}) to locate its own centroids for the three
sources.  Second, we generated a geometric transformation between the
ground-based frame and the WFPC2 F656N (H$\alpha$ and [\ion{N}{2}])
field, in which the centroids of these three sources can be measured
to within 5 mas, and fixed the centroids to these transformed values.

WFPC2 data were registered and stacked as described in
\citet{Sug02}.  Tiny Tim model PSFs were used to subtract
Stars 2 and 3, as well as all other bright stars with detectable
diffraction spikes.  Note that in this and the following, ground-based
and and WFPC2 data are never intercompared.

\subsection{Pipeline Preprocessing \label{sec-reduc-pp}}

As explained in \sect{sec-dip-pipeline}, pre-processing requires the
selection of reference stars, building of PSFs, and creation of
reference images.  For both the ground-based and WFPC2 images,
reference-star lists were constructed from those stars that are
present and photometrically constant in each dataset.  High
signal-to-noise and clean PSFs were built using these lists as
outlined in \sect{sec-dip-pipeline-psf}.  In the current application
of difference-imaging, we are searching for extended
surface-brightness features that vary in time.  As such, we require a
reference image \secp{sec-dip-pipeline-ref} that is free of all echo
signal.  Since echoes are present in each epoch in the ground-based
imaging, an iterative procedure must be adopted to generate such a
frame, as described below.

All stacked images in a given filter were median-combined into a
temporary reference stack, using zero-point offsets (for the sky) and
photometric scalings derived from photometry of the reference-star
list.  Each stacked image was then PSF-matched to and differenced
\secp{sec-dip-pipeline-dip} by this temporary reference
stack.  For a given epoch, the difference images from all filters were
co-added into a single echo image.  If a faint structure is truly an
echo, it will appear as positive signal in all continuum filters,
while background noise is a random distribution about a fixed value.
Artifacts, such as flat-fielding errors or ghosts, can either vary or
remain constant between different filters, and therefore can present
themselves as false signals in these images.
 
Light echo candidates in each echo image were identified using a
custom-written implementation of the friends-of-friends algorithm.
The median and standard deviation are computed within a moving box of
$m\times m$ pixels, and any pixel whose value is more than $N\sigma$
above the median is flagged as a ``friend.''  Subsequently, every
pixel touching that friend that is more than $n\sigma$ above the
median is a friend of the friend.
Every friend-of-a-friend is found in the image, with different groups
merging when they touch.  From a practical standpoint, this works best
on block-medianed images, in which each pixel in a $4\times4$ moving
box is replaced by the median value in that box.  This smooths out a
great deal of statistical shot noise, making echoes easier to detect
above the background.  In general, we found that $N=3$ and $n=2$ were
sufficient.

Friend-lists surrounding bright stars or more than 30\arcsec from the
SN were rejected.  The remaining groups were checked against those
from the two adjacent epochs to ensure the signals are real.  It is
highly unlikely that a bright echo will appear in only one epoch, and
not a few months prior or later, since such a structure would have to
be extremely thin on the plane of the sky and along the line of sight,
as well as being fortuitously placed to be illuminated at that
particular epoch.  Additionally, groups were checked to ensure they
did not result from ghosts from nearby bright stars; as there are many
such stars near the SN, many such ghosts were found.  Those lists that
survived all cuts were turned into pixel masks for each image in that
epoch.

In some cases, consistent structures appeared as positive flux in
early difference images and negative flux later, or vice versa,
indicating that an echo was present in the reference image.  In such
cases, stacks were differenced from other individual stacks well
separated in time, and friend lists were generated from these
difference images.

All stacks in a given filter were then again median-combined, this
time with the echo pixels in each image masked out.  The above steps
were repeated with these new reference images, and very few additional
echoes remained.  The third iteration generated reference stacks which
did not produce any detectable echo residuals.  We adopt these as our
reference, echo-free images.  The same technique was used for both the
ground-based and WFPC2 images; however, in the latter case, echoes
disappeared from the PC chip by 1997, after which considerable data
exist. Thus generation of the WFPC2 echo-free images was much more
straightforward.

\subsection{PSF-Matched Difference Imaging \label{sec-reduc-dip}}

PSF-matching and difference imaging proceeded as explained in 
\sect{sec-dip-pipeline-dip}.  For the ground-based data, difference
images were made between each epoch in each continuum
filter and the corresponding reference image, for all three kinds of
stacks: those in which no PSF stars were subtracted, those in which
the SN and Stars 2--3 were subtracted with floating centroids, and
those in which the centroids were fixed by their astrometry.

In many cases, no matter how the PSF was created, it left circular
residuals around star 2, and to a lesser degree around star 3.  The
cause is straightforward: the empirical PSF is scaled to match the
flux of star 2, as are its wings.  Since the wings of any empirical
PSF are noisy (due to the low signal-to-noise of the data that
comprise them), this noise becomes significant when scaled to the
flux of a bright or non-linear star, such as star 2.  One may either
replace the wings {\em a posteriori} by, \eg an elliptical gaussian
(as we do in PSF-matching) or attempt to minimize this effect by
carefully constructing the PSF.  The former option is not built-in to
{\em daophot}, and rather than write a revised version of this
software, we opted for the latter.

In subsequent discussions of structures found within 3\arcsec of the
SN, a ground-based result is not deemed significant unless it appears
in both PSF-subtracted difference images.  Furthermore, a result that
lies within a noisy PSF-subtraction residual is not significant if we
cannot demonstrate that the structure is real and not an artifact of
the noisy wings.  This is easy to test by examining the
non-PSF-subtracted difference image, or by PSF-subtracting other
bright stars and examining their residuals.

For WFPC2 imaging, the {\em Tiny Tim} model PSFs have
no noise in the wings, thus there is no need to consider alternate
PSF-subtractions.  All epochs of data were differenced from the
corresponding reference image in the three filters of interest, F555W,
F675W and F814W.  

\subsection{Photometric Calibration \label{sec-reduc-calib}}

To study the density and composition of the dust producing the
observed light echoes \secp{sec-le-sb}, it is necessary to
photometrically calibrate the continuum filters.
Since fluxes determined from broad-band Johnson filters may be a poor
fit to our narrow continuum bands, we instead determine the
calibration spectroscopically, as follows.

Space Telescope Imaging Spectrograph (STIS) spectra in G430L and G750L
(blue and red) were taken through the 2\arcsec-wide slit in 2000
November as part of a Director's Discretionary {\em HST} allocation to
study hot spot evolution \citep{Sug02}.  This pointing included stars
4, 7, 34 and 35 from \citet{WS90}, respectively 5\farcs5, 9\farcs7,
17\farcs0, and 16\farcs0 from the SN.  Each star is reasonably
un-crowded on our ground-based frames, allowing for accurate
PSF-fitted photometry with {\em daophot}.  All four are close to the
SN on the sky, and assuming they are similarly located within
the LMC, all five sources should should suffer roughly the same
extinction.  

Pipeline-calibrated spectral images were combined and one-dimensional
spectra extracted using optimal techniques packaged within {\em
stsdas}.  Since stars were located at different positions orthogonal
to the dispersion axis, their spectra suffer constant,
positional-dependent offsets, which we corrected empirically by
matching absorption features.

The red and blue spectra were integrated over the filter response
functions $T(\lambda)$, as shown in Figure \ref{plfilt}, to determine
each star's observed fluence.  The average of these fluences, compared
to the observed counts (per second) of the stars in the reference
images, give the calibrated flux of one count per second.  The
difference between the four stars per filter are consistent to within
$\sim5$\%.

Note that this calibration is for {\em observed} fluxes, and has not
been dereddened.  \citet{WS90} find that the field surrounding
SN~1987A is consistent with a reddening of $E(B-V)=0.17$, while
\citet{Scu96} determine $E(B-V)=0.19$ from {\em HST} Faint Object
Spectrograph spectra of star 2.  These two papers assume $R_V=3.1$ for
the LMC, while \citet{Mis99} give an average value of $R_V=2.6$ using
ten sightlines.  However, if one assumes the SN, surrounding dust, and
these nearby stars all suffer the same extinction, then the
dust-scattering model \secp{sec-le-sb} can be applied without any
reddening corrections.


\section{Echo Measurement and Visualization }\label{sec-lea}

In this section, we present the PSF-matched difference images,
describe how light echoes were identified and measured, and discuss
analysis and visualization tools that will be used in this work.

\subsection{Light Echo Data \label{sec-lea-data}}

All epochs of imaging listed in Table \ref{tbl-echodata-obs} were
pipeline-reduced, PSF-matched and differenced as explained in
\sect{sec-reduc}.  A subset of the resulting difference images are
shown in Figure \ref{dimages}.  Recall that the WFPC2 PC has 3.6 times
higher resolution per pixel and therefore a smaller field-of-view.
Since {\em HST} observations were rarely taken at the same
orientation, the images are clipped with respect to ground-based data,
and the clipped regions change with epoch.

For all observations, images are shown in a standard north-up,
east-left orientation, with major tick marks denoting angular
distances of 10\arcsec each.  The innermost echoes were particularly
bright up to day 1469, and are washed out at the brightness-stretch
used to display the background and outer echoes.  In such cases, a
12\arcsec-square inset centered on the SN is shown at the top right of
each panel, scaled in brightness to better-resolve the innermost echo
signal.  All images are shown with a linear greyscale scaling, the
limits of which differ between images to better show the echoes in each
particular epoch.  

\begin{figure*}\centering
\includegraphics[width=5in]{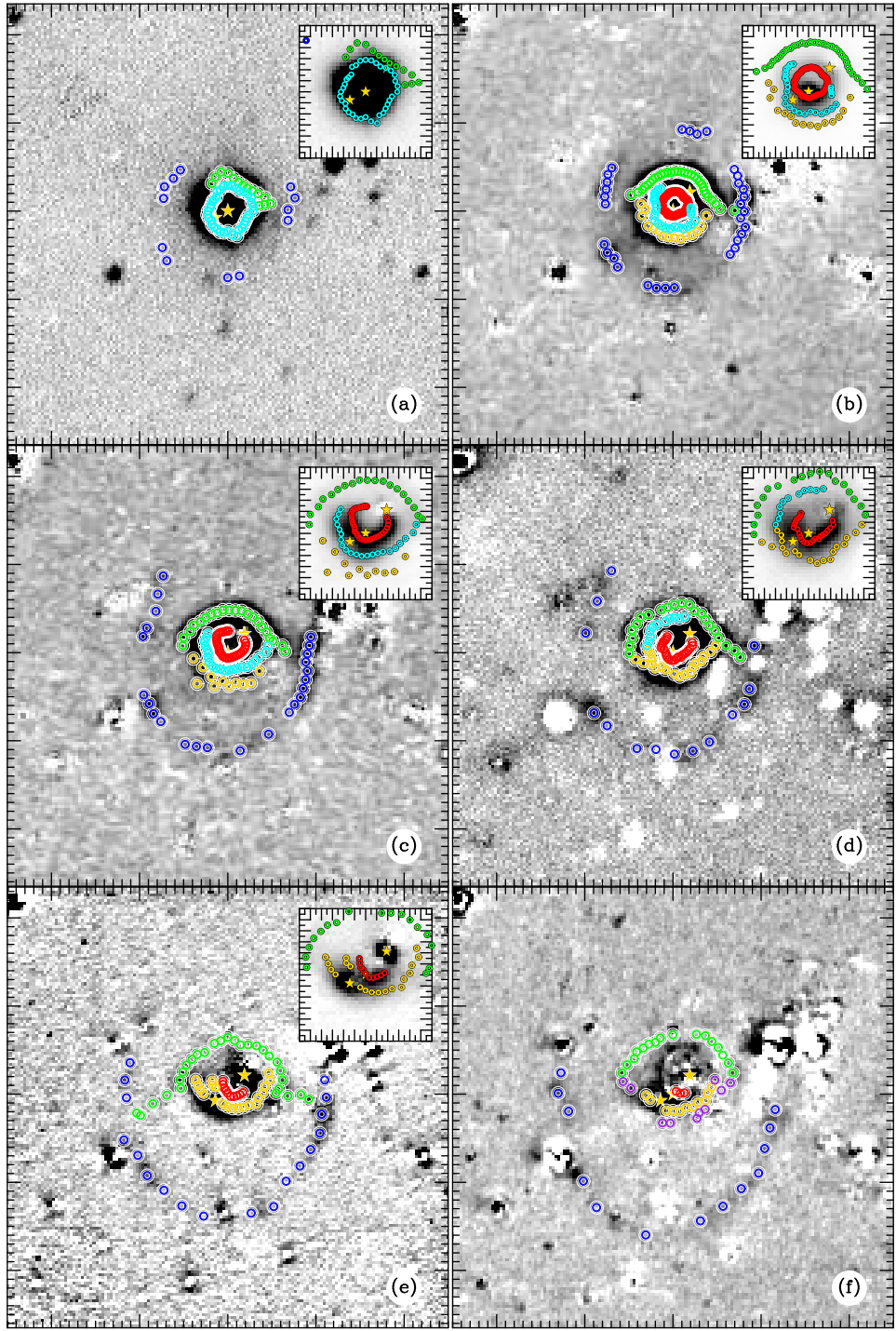}
\caption{Selected $50\arcsec\times50\arcsec$ difference images.  North
is up, east is left, and major ticks mark 10\arcsec.  The position of
the SN, Stars 2 and 3 are marked with yellow stars.  Echoes are marked
by colored circles.  The inset shows the central 12\arcsec at a
different color stretch to resolve the innermost echoes.  Colors are
explained in \sect{sec-lea-data-disc}. 
\panp{a} $V$ image from day 659. 
\panp{b} 6067\AA\ image from day 750.
\panp{c} F612 image from day 1028.
\panp{d} F612 image from day 1086.
\panp{e} F612 image from day 1469.
\panp{f} F612 image from day 1787.
\label{dimages}}
\end{figure*}

\addtocounter{figure}{-1}
\begin{figure*}\centering
\includegraphics[width=5.in]{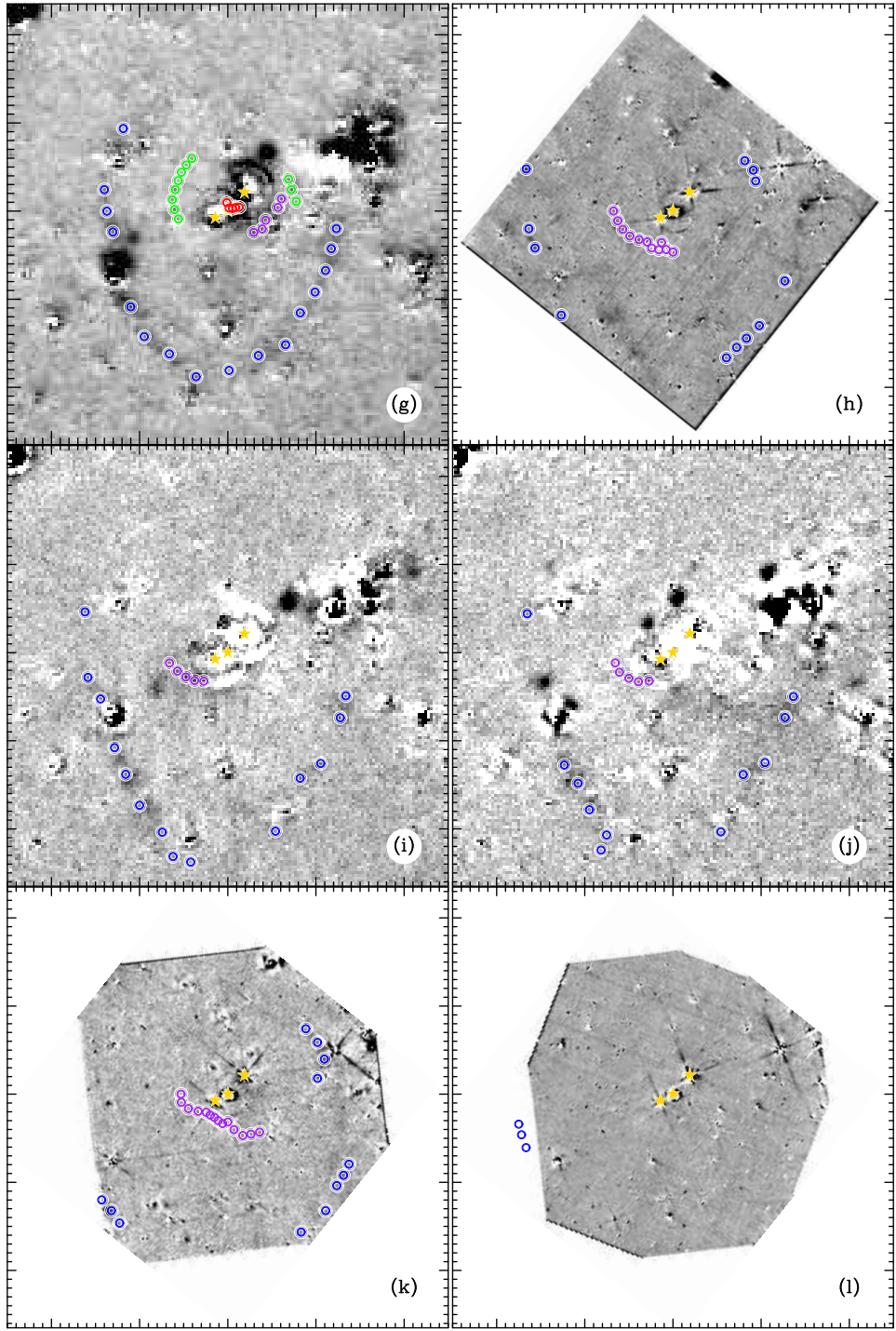}
\caption{continued.
\panp{g} F612 image from day 2095.
\panp{h} WFPC2 image from day 2769.
\panp{i} F612 image from day 2830.
\panp{j} F612 image from day 2874.
\panp{k} WFPC2 image from day 3270.
\panp{l} WFPC2 image from day 3789.
}
\end{figure*}

Panels \panp{a} and \panp{b} were taken through the
$V$-band and a narrow continuum filter centered at 6067\AA,
respectively.  In the first case, this image (day 659) was differenced
from a sum of the F470 and F612 reference images, weighted to
approximately reproduce the throughput of the Johnson filter.  This
scheme worked well for removing most faint background sources.  We
compared this image to a difference made against another $V$-band
observation from 1990 Nov.\ 5 (day 1171) and found the background was
sufficiently flattened using the combination of narrow filters.  The
non-standard narrow continuum image from day 750 was differenced
against the F612 image, the result of which is of equivalent
quality to other F612 difference images.  In both cases, it was
preferable to use the echo-free reference images rather than the same
filter at a later epoch, since this latter option introduces additional
confusion from echoes present in the later data.  

The data in panel \panp{d} were taken in a dithered pattern by
imaging the SN on each of the four extreme corners of the detector.
The resulting mosaic has different noise characteristics depending on
the amount of overlapping data.

A number of the images, \eg panels \panp{f}, \panp{g}, and \panp{i},
show the limitations of PSF-subtracting Stars 2 and 3 prior to
differencing, also described in \sect{sec-reduc-dip}.  Star 2 is
especially bright and often has significant flux in its wings at a
radius larger than can be adequately modeled by an empirical PSF.\ \
This can be well seen in panel \panp{g}.  Additionally, the wings
of the model PSF are intrinsically noisy at large radii, and the
bright flux from star 2 made even small-scale fluctuations large,
producing subtraction residuals within the model-PSF radius.  This is
most obvious in panel \panp{i}.

There are many noteworthy artifacts present in these data.  As noted
in \sect{sec-dip}, the Poisson noise is highest at the
position of a bright star, and will therefore leave a subtraction
residual at its core.  Such residuals can be seen throughout the data.
Non-linear and saturated stars leave much larger residuals, both as
noisy cores and diffuse wings, such as can be seen in the upper left
corner or in the bright-star cluster $\sim15\arcsec$ WNW of the SN in
ground-based images.  Ghosts of bright stars, caused by internal
reflections within the detector optics, are particularly noticeable in
panels \panp{g} and \panp{i}, but can be easily
identified since they generally occur with the same positional offset
from all bright stars.  

WFPC2 data suffer from a host of image artifacts that complicate echo
detection.  Foremost are the extended diffraction spikes from bright
stars, which can extend over 6\arcsec (128 pixels) from the source.
These have been included in the {\em Tiny Tim} model PSF, but in
practice can never be completely subtracted away.  Linear banding from
charge transfer of bright stars is also omnipresent.  Stray light from
a bright star positioned within a well-defined avoidance region
outside the PC field of view will cause broad arcs and thick bands of
flux (fully discussed in the WFPC2 ISR 95-06), which may be seen as
dark regions along the left side of panel \panp{h}.  A smaller,
roughly-horizontal dark arc on the right side, roughly 6\arcsec south
of the SN, is also a bright-star artifact.  There may be much more
echo signal in this image than we could reliably identify (next
section), due to these considerable sources of noise.

\subsection{Light Echo Detection \label{sec-lea-data-echoes}}

Echoes are present at a variety of positions and over a wide range of
surface brightnesses.  To measure its position, 
an echo may either be specified by the total locus of bright pixels,
or as the central position of the surface-brightness profile.
However, quantifying the structure of extended, diffuse echoes,
particularly in the wings of the brightest ones, is difficult, for a
variety of reasons.

Identifying all pixels associated with an echo is comprehensive in
giving a better representation of the complete spatial region
producing the echo.  However, it is important that the echo positions
can be visualized in a meaningful way.  We experimented with
describing echoes three-dimensionally as the complete locus of all
pixels (identified with our friends-of-friends algorithm,
\sect{sec-reduc-pp}) within the surface-brightness profile, and found
that when many epochs were rendered in 3-D, the result was very
difficult to analyze, since background data were completely obscured
by foreground material.  Since visualization of the data is critical to
interpreting the positional information contained in the echoes, we
instead adopted a discretized measurement by which the signal is
described as a series of convolved profiles.

Consider a bright, resolved echo with wings that extend to roughly
three-times the FWHM of a PSF.  This could result from a large, single
distibution of dust, or from the convolution of many discrete dust
distributions, each with their own density.  In the first case, if we
identify the echo only at its center, we reject information contained
in the wings, namely, that there is dust there.  Rather, if we treat
the extended flux as the convolution of many discrete echoes, even if
there is only one echo present, we will better represent the spatial
dust distribution.

The PSFs in most of our images are better described by moffats than
Gaussians.  The functional form for a moffat is given in Appendix
\ref{app-model-LM-gauss}; it is essentially a modified Gaussian that
can have a more rapid or shallow profile in the wings.  Echoes were
identified by fitting multiple moffats to radial surface-brightness
profiles, generally measured by summing over wedges of arclength
10\degr.

Moffats were fit using a highly ``hands-on'' implementation of
Levenberg-Marquardt (LM) minimization (Appendix
\ref{app-model-LM-gauss}).  Once the surface-brightness profile is
measured, the user determines the radial range over which moffats will
be fit, and can reject data that are known to be bad \ie points
coincident with subtraction residuals.  The user specifies the number
of moffats to be fit, and an initial guess for the peak, center,
width, and exponent $\beta$.  Parameters are relaxed into the
``best-fit'' through iterative calls to the LM algorithm, with
parameters replaced or reset when they deviate from reasonable values.
``Reasonable'' is fairly subjective but not without practical limits.
For example, the width of a moffat was not allowed to be smaller than
that of a stellar PSF, but was also not allowed to grow larger than
$\sim2$FWHM.

\begin{figure}\centering
\includegraphics[height=3.25in,angle=-90]{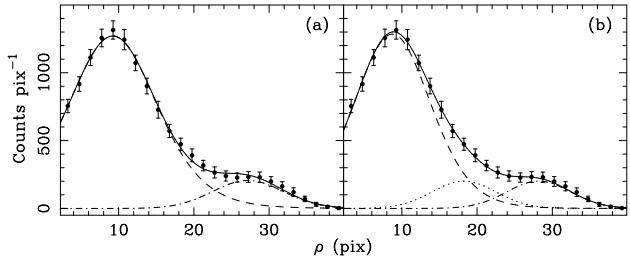}
\caption{Sample radial profile from PA$\sim50\degr$ on day 1028 (black
points), fit by two and three moffats in panels \panp{a} and \panp{b},
respectively.  Each individual function is shown with a different line
style, and the sum of all functions is drawn in black.
\label{plmoffat}}
\end{figure}

Each profile is first fit with a single function, and additional
moffats are added only as needed to reproduce the flux distribution.
It is rare that more than three moffats were needed for a given echo.
An example is shown in Figure \ref{plmoffat}.  The profile (points and
error bars) is poorly described by a single Gaussian or moffat, as it
has an extended plateau around $\rho=25$ pix, and will require at
least two functions.  The best-fit solution for two moffats is shown
in panel \panp{a} in black, with the individual functions plotted in
different line styles.  Note that the fit is only marginal between
$\rho=18-26$ pixels.  Subtracting these from the profile reveals a
third peak centered near $\rho=18$ pixels.  Indeed, including a third
function (panel \pant{b}) is a much better fit, improving the $\chi^2$
residual threefold.

Once a successful fit is made, parameters are varied to ensure the fit
is robust.  If so, the parameters are recorded, with the width of the
echo $\Delta\rho$ taken to be the moffat width $\sigma$.  Radial
profiles were examined at all position angles in each image, and out
to a radius of at least 16\arcsec at early times, and 32\arcsec at
later times, to ensure that faint, distant echoes were not overlooked.
The results have been marked in the difference images, and color coded
to compare echoes believed to originate in physical associations
between epochs.

As can be seen from the difference images in Figure \ref{dimages},
many echoes are faint, lie near bright stars, or in the case of the
innermost echoes, near the positions of the SN and Stars 2--3.  As
with generating the echo-free images \secp{sec-reduc-pp}, faint echo
signal was better resolved from the background by coadding the
difference images from all filters in a given epoch.  Echoes beyond
$\sim5\arcsec$ of the SN were generally measured in these summed
images.  Identification of echoes within 5\arcsec is complicated by
the unknown quality of the PSF-subtraction of the SN and Stars 2--3.
Profiles were fit from difference images made using both methods of
PSF-subtraction explained in \sect{sec-reduc-reg}, and carefully
compared for consistent results.  When inconsistent,
or we felt the subtraction of these sources could influence the
result, the measurement was thrown out.  In some cases echoes
may be present but not reported.

The difference image from day 659 (Fig.\ \ref{dimages}\panp{a})
deserves special note.  At this early date, the SN was sufficiently
bright to saturate the detector long before sufficient surrounding
signal could be gathered.  The PSF of the SN is too bright and
extended to perform a PSF-subtraction, however we did find that
subtracting a single moffat from the radial profiles left consistent
echo signal in the PSF wings, as marked.

\subsubsection{Photometry \label{sec-lea-data-photo}}

Once identified as described above, echo brightness was measured as
follows.  All stars bright enough to leave a detectable subtraction
residual were masked from each image.  Echo fluxes were measured from
these images with annular apertures of the same angular width as used
during detection, and with radial widths given by the FWHM of each
echo at each position.  Since many echoes are extremely faint, we
measured each twice, first as the surface brightness through the
aperture allowing all data, and again counting only positive pixel
values.  The corresponding median sky value was subtracted from each
surface brightness, and the two values were rejected if less than the
propagated noise, or averaged if greater.  Counts were converted to
fluence using the scaled exposure time from difference imaging and the
photometric calibrations described in \sect{sec-reduc-calib}.

\subsubsection{Discussion \label{sec-lea-data-disc}}

During their detection and measurement, echoes were categorized as
belonging to one of three expected structures, listed here in order of
increasing radius on the plane of the sky: a circumstellar
hourglass-shaped nebula reported in \citet{CKH95}, Napoleon's Hat
\citep{Wam90a}, or circular echoes at large radii
\citep{Bon89,CM89,CK89} interpreted as the contact discontinuity
between the progenitor's RSG and MS winds \citep{CE89}.  These have
been color coded in Figure \ref{dimages} with the primary colors red,
green, and blue, respectively.  However in fitting moffat profiles, we
found that additional identifications could be made between the
hourglass and Napoleon's Hat, generally colored cyan for echoes
looping to the north of the SN, or yellow for positions south.  Many
of these southern echoes, once examined in 3-D, appeared at a larger
distance, consistent with a southern counterpart to Napoleon's Hat,
and have been marked in purple.

That diffuse structure exists between the inner hourglass echoes (red
points) and Napoleon's hat (green points) has been previously
reported.  \citet{CK91} detected diffuse emission from
$3\arcsec<\rho<12\arcsec$ from the SN, measured in semi-circular
apertures.  \citet{CKH95} reported the detection of a ``jet-like''
feature, several arcsec in length, projecting northeast from star 3
between days 1028 and 1727.  These are consistent with the cyan and
yellow echoes identified in the same region of the current data, which 
now map out a much more extensive volume of the innermost
circumstellar environment.

Contact-discontinuity echoes have been previously reported by
\citet{CKM89}, \citet{Bon90}, and \citet{CK91} in observations only up
to day 1028, while we are able to trace this structure through day
3270.  Napoleon's Hat was imaged as early as days 850 and up to day
1650 by \citet{WW92}, while we detect its signal from day 659 to 2095.
We believe this is the first discussion of the southern-counterpart
(purple) to Napoleon's Hat.

\subsection{Visualization and Analysis Tools \label{sec-lea-analy}}

The 3-D analysis of these echoes is divided into the three broad
categories defined above, and will be presented in
\sect{sec-CD}--\ref{sec-CS}, making use of visualization and analysis
tools presented here.

\subsubsection{Coordinate Definitions \label{sec-lea-analy-coo}}

In the following discussion of echo positions, we will make use of a
few coordinate systems.  On the plane of the sky, we use
standard two-dimensional cartesian coordinates.  The origin is always
defined to be the SN, west is to the right and north is up.  Polar
coordinates $\rho$ denote the radial distance from the origin, and
position angle (\pa) is measured as the counter-clockwise (CCW) angle
from north.  

\begin{figure}\centering
\includegraphics[height=2.7in,angle=0]{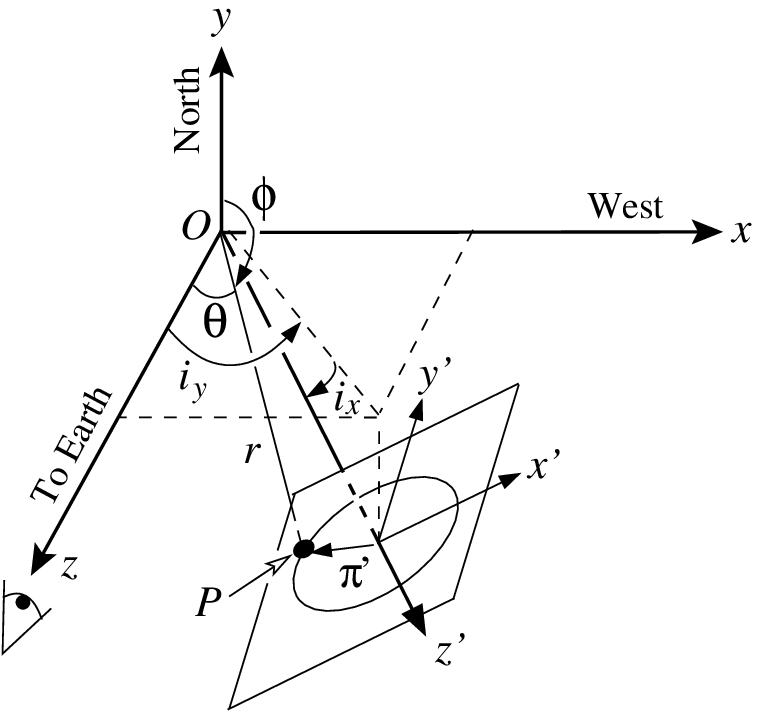}
\caption{3-D coordinate definitions for discussing light echoes. 
\label{axis}}
\end{figure}

Three-dimensional coordinate systems are shown in Figure
\ref{axis}.  Let us define the standard orientation to be the
orthonormal cartesian axes with $z$ toward the observer, orthogonal to
the 2-D image plane and centered at the SN, with $x$ increasing west,
and $y$ increasing north.  Thus, the line-of-sight distance, or depth,
is measured by $z$.  A point $P$ in space can be described by its
cartesian, spherical, or cylindrical coordinates.  The
spherical distance $r$ is measured from the origin $O$, with
polar angle $\phi$ measured from north ($y$), and azimuthal angles are
measured in the $x-z$ plane CCW from the line-of-sight ($z$).  We
reserve an unscripted $\theta$ for the scattering angle at point $P$,
which is the angle between vector $\overrightarrow{OP}$ and $\hat{z}$
(under the reasonable assumption that $d\gg r$).  cylindrical
coordinates will be used for structures with rotational symmetry about
an inclined axis $z'$.  The point $P$ has axial radius $\pi'$ measured
orthogonally from $z'$.  Transformed cartesian axes $x'$ and $y'$ may
also be defined in the plane orthogonal to $z'$ containing $P$.

Rotational transformations are performed using a permutation of the
pitch-yaw-roll (PYR) convention, where rotations are performed about
each of the three cartesian axes through the triplet
$(\psi_z,\theta_x,\phi_y)$, where we first rotate (roll) about $z$ by
$\psi_z$, then about the rotated $x$ (pitch) axis by $\theta_x$, and
finally about the rotated $y$ axis by $\phi_y$ (yaw).  Positive
rotations are made about an axis following the right-hand rule.  When
referring to inclinations, the rotation is named for the axis about
which the rotation is made, \ie a rotation north or south is made
about the $x$ axis and named $i_x$.  The cylindrical axis $z'$ is then
inclined south/north by $i_x$ from the $x-z$ plane, and rotated $i_y$
west/east from the $y-z$ plane.

\subsubsection{Three-Dimensional Rendering \label{sec-lea-analy-3de}}

\begin{figure}
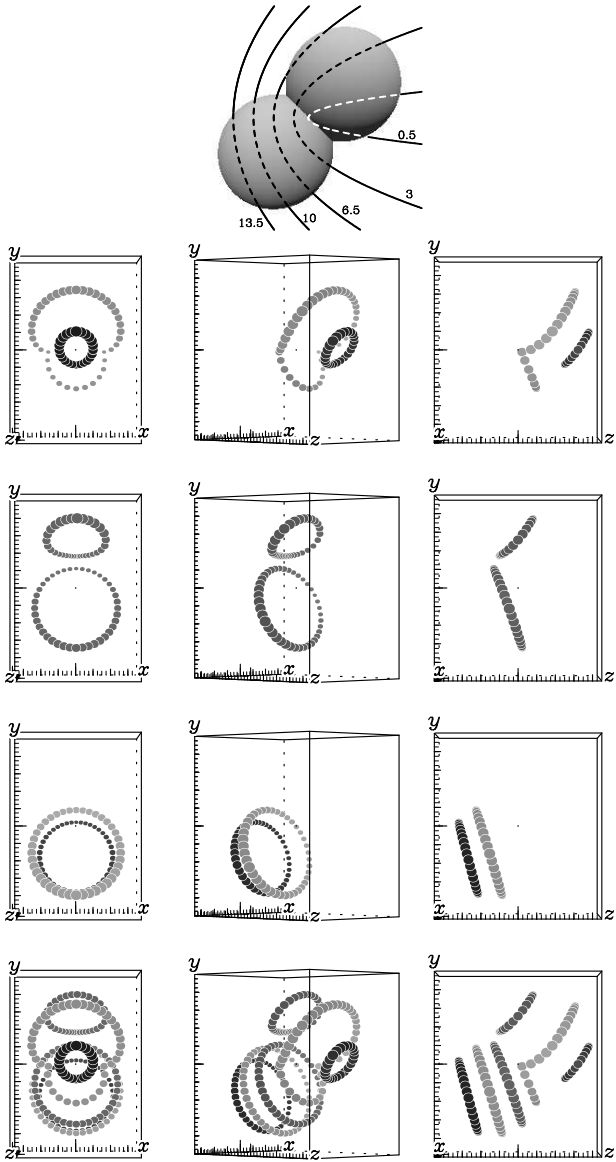
\centering
 \includegraphics[width=1.2in,angle=-90]{f13a} \\
 \includegraphics[width=3.25in,angle=0]{f13b}
 \caption{ Illustrative examples of 3-D renderings of light echoes in
 a simple bipolar nebula.  This non-trivial example is intended to
 guide the reader in visualizing light echoes, and to annotate the
 rendering technique.  The nebula, depicted at top, is inclined
 45\degr to the line of sight, which is to the right; each lobe is 5
 ly in radius, with the centers separated by 8 ly.  Five echo
 parabolae are drawn, each occuring at the time (in years) indicated
 next to each curve.  Below, renderings show the light echoes for the
 first two parabolae (first row), third parabola (second row), last
 two parabolae (third row), and all parabolae (last row).  Axes are
 defined as in \sect{sec-lea-analy-coo}; axis labels indicate the
 positive direction.  The origin is set at the SN, and is always
 marked by a black dot.  Major ticks mark 2 ly, and the origin is
 indicated by the longest tick along each axis.  The coordinates and
 axes have been given a slight perspective transformation (Appendix
 \ref{app-3de}).  Points are shaded using simple ray tracing, and
 larger points are closer to the observer.  {\em Left Column:} Face-on
 view (the plane of the sky). {\em Middle Column:} Oblique view
 rotated 45\degr. {\em Right Column:} Side view from far to the east.
 Note that the geometry of the nebula, as revealed in light echoes, is
 only clear when many echoes well-separated in time are considered at
 once.
 \label{3de_parab}}
\end{figure}

To study echo positions in 3-D, we have custom-written graphics
software to perform simple renderings, in which the positions of the
echoes are transformed to allow viewing from any angle.  An
illustrative example is shown in Figure \ref{3de_parab}.  

To realize a 2-D visualization of 3-D data, the following steps are
performed.   First, the user specifies the shift, rotation, scaling,
and if desired, translated point of view of the observer and
perspective.  Using the matrix transformations explained in Appendix
\ref{app-3de}, the data are transformed into a 2-D matrix of positions
on the page, with the third dimension saved as a ``$z$-buffer'', which
gives the depth of the point in or out of the page.  Data are sorted
by the $z$-buffer position so that points furthest from the observer
are plotted first; in this way, distant points behind nearer ones will
be covered.  To facilitate an interpretation of depth, point size
scales with distance from the viewer, with smallest points furthest
away.   

In viewing 3-D structures, we find variation of point size is not
sufficient to give the viewer an immediate sense of depth.  We have
added a simple ray-traced shading, assuming a single light source
positioned behind the observer.  In general, ray-tracing on an
arbitrary shape is very difficult and involves complicated algorithms
such as tesselation.  However, all our echo data lie on parabolae of
known functional form, thus the normal, or gradient, is always known.
Points are shaded by the angle between the normal and line of sight,
such that 0\degr\ (modulo 180\degr) is shaded light, and 90\degr\
(modulo 180\degr) is dark.  In practical terms, that part of the
paraboloid that is orthogonal to the viewer's line of sight will
appear whiter, and limbs will appear darker.  For completeness, the
line of sight in this context is orthogonal to the page, intersecting
the origin (black dot), and the observer is out of the page looking
down.

\subsubsection{The Volume of Space Probed by Echoes \label{sec-lea-analy-parabs}}

\begin{figure}\centering
\includegraphics[height=3.25in,angle=-90]{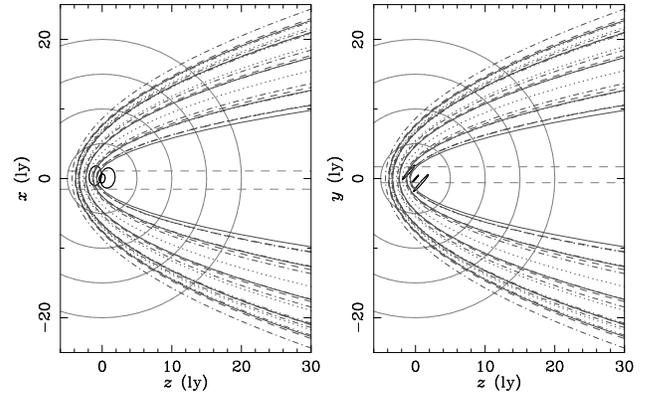}
\caption{Regions of space surrounding SN~1987A probed by light echoes.
The left and right panels show the view from far to the north and
east, respectively.  The probable positions of the three rings
\secp{sec-CS-rings} are centered at the origin.  Circles of constant
radii, in multiples of 5 ly, are drawn in light grey.  The positions
on the plane of the sky of Stars 2 and 3 are indicated by horizontal
grey dashed lines.  Echo parabolae corresponding to the epochs of
observation used in this work (and listed in Table
\ref{tbl-echodata-obs}) are indicated in dark grey, with line style
cycling from solid -- dashed -- dot-dashed -- dotted --
dot-dot-dot-dashed in order of increasing epoch.  \label{parabs}}
\end{figure}

Interpreting light echoes is often confusing, since they lie along
paraboloids, a surface one rarely encounters when considering
geometric intersections.  A plane, perpendicular to the line of sight,
will appear as a concentric ring, growing in radius with time, until
the paraboloid is larger than the material boundary.  A thin,
spherical shell of radius $r$, centered on the echo source, will
appear first as a ring of increasing radius as the paraboloids expand
with time, until the paraboloid sweeps into the rear half of the
sphere, after which the ring will collapse and eventually disappear at
$t=2r$.  

Other, more complicated geometries can appear in
particularly-unrecognizable forms until the echoes have illuminated a
substantial part of the structure.  For example, the inclined, bipolar
nebula in Figure \ref{3de_parab}, can appear as (1) centered or offset
rings, both circular, elliptical, or highly distorted; (2) a distorted
figure-eight; (3) multiple distorted rings; or (4) small curved
segments, depending on the time and orientation at which echoes are
observed.  

Figure \ref{parabs} is more relevant to echoes from SN~1987A, showing
the positions of echo parabolae corresponding to all epochs of data in
Table \ref{tbl-echodata-obs}.  Each panel shows a planar cross section
of space passing through the origin: the $x-z$ plane viewed from above
on the left, and the $y-z$ plane viewed from the side on the right.
Concentric circles at radius intervals of 5 ly are plotted to show how
these echoes will illuminate structures of different sizes.  Also
shown is the probable geometry of the three rings, as determined in
\sect{sec-CS-rings}, and the line-of-sight positions of Stars 2 and 3
(as horizontal dashed lines), whose relative positions with respect to
the supernova are unknown.

\subsection{The Three-Dimensional Echo Positions}\label{sec-lea-3D}

\begin{figure*}\centering
\includegraphics[angle=0,width=6in]{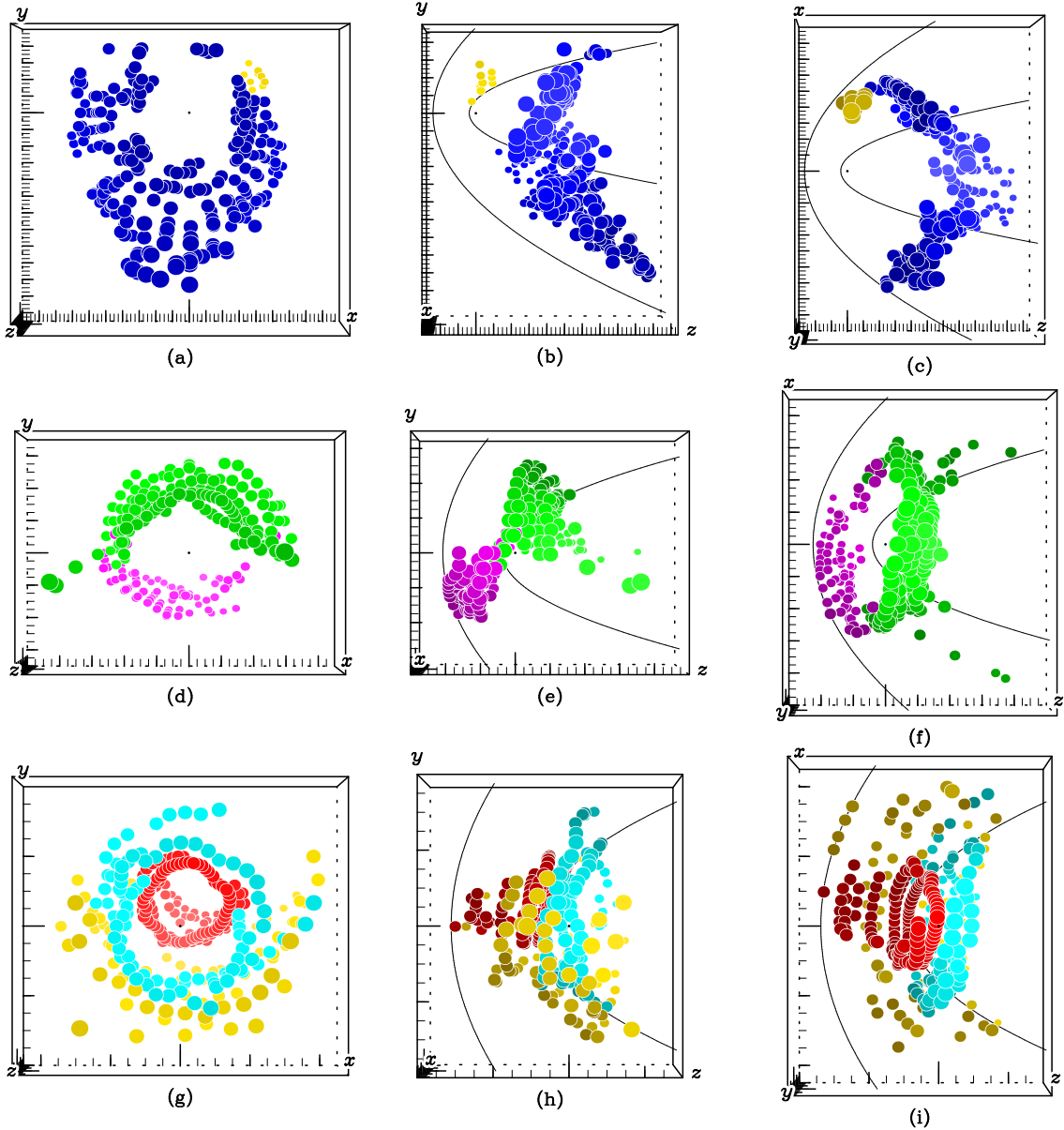}
\caption{Rendered views of all light echoes identified in Fig.\
 \ref{dimages}.  Points have been rendered using the method described
 in \sect{sec-lea-analy-3de}; also see Fig.\ \ref{3de_parab}.  The
 left column shows observed views in the plane of the sky, the middle
 column shows views from the side (far to the east) and the right
 column shows views from the top (far to the north).  Parabolae from
 the earliest and latest epochs at which echoes were observed are
 indicated in the side and top views.  Point colors correspond to
 those in Fig.\ \ref{dimages} except for the gold points in the top
 row, which denote echoes found in WFPC2 images which could not be
 resolved from the residuals of nearby bright stars in ground-based
 images.  Note that the field-of-view changes between rows, but major
 tick marks always denote 2 ly.  
 {\em Top row:} Contact discontinuity echoes.  
 {\em Middle row:} Napoleon's Hat and southern counterpart.
 {\em Bottom row:} Inner circumstellar hourglass echoes.  
 \label{3dle}}
\end{figure*}

Echo positions in 3-D are computed from the 2-D positions on the plane
of the sky using equation \eqt{sec-le-sb1}.  We have implicitly
assumed that the echo center is produced by the light pulse maximum
(which we take to be 87 days after discovery, \sect{sec-le-flux}), and
correct the observation date in Table \ref{tbl-echodata-obs} by this
amount when it enters equation \eqt{sec-le-sb1}.  The angular size
$\rho$ is converted to physical units assuming the SN of 50 kpc away.
This value is a good average between many of the derived distances to
the LMC, which vary between 45--54 kpc \citep{Gou98,Fea99,Rom00}.

The 3-D positions of all echoes identified in Figure \ref{dimages} are
rendered in Figure \ref{3dle}, using the methods described in
\sect{sec-lea-analy-3de}.  The top, middle, and bottom rows show the
data for the contact discontinuity (CD), Napoleon's Hat (NH), and the
inner circumstellar (CS) material, respectively.  Except for the
gold-colored points in the top row (see below) colors correspond to
those indicated in Figure \ref{dimages}.  A rigorous analysis of these
data is presented in \sect{sec-CD}--\ref{sec-CS}, however we give a
short qualitative discussion of the echoes below.

\subsubsection{The Contact Discontinuity}\label{sec-lea-3D-CD}

In the plane of the sky (panel \pant{a}), there is no data within
$\rho=5$ ly of the SN.  This results from the lack of echo imaging
prior to 1988 December; this is seen in panels \panp{b}--\panp{c}, for
which there can be no data interior to the inner parabolae.  There are
two noticeable holes in the eastern blue echoes at $(x,y)$ coordinates
$(-7,-1)$ and $(-10,-6)$ ly, since a saturated star lies at each
position.  The bulk of the echoes are located between 20 ly south and
4 ly north of the SN, with very few echoes positioned further north.
Since light echoes symmetrically probe all regions of space around the
line of sight, this suggests that there is little structure
in front and to the north of the SN, with the structure being
illuminated inclined predominantly south of the line of sight.

The distribution of blue echoes east and west of the line of sight is
fairly symmetric, however echoes extend to only $\rho=11$ ly west,
while they are present to the east up to $\rho=14$ ly.  Comparing the
two halves, there are eastern counterparts for all western echoes, yet
there are no western counterparts bewteeen \pa 250\degr--300\degr\ to
the eastern echoes at $\rho>9$ ly from \pa 60\degr--110\degr.  Since
there is a cluster of bright and saturated stars in this western
region, which makes reliable detection impossible, the absence of
westerly echoes may not be meaningful.  To the south, the echoes extend
beyond $\rho=14$ ly.  These southern points do not appear symmetric
about the $y$ axis, but rather about an axis rotated 5\degr--10\degr
eastward.

When viewed from the side (panel \pant{b}), the eastern/western
points lie 4--14 ly in front of the SN, in a mainly
vertical distribution extending 8 ly north and 12 ly south of the line
of sight.  The southern echoes are primarily oriented linearly along an
inclined axis that points toward the SN.  These echoes extend 10 to 20 ly
in front of the SN, corresponding to radial distances of 14 to 28 ly.
From this point of view, the southern echoes have a thin, sheet-like
quality to them.

Panel \panp{c} shows the top view, from which we see that the
eastern/western material follows a thin, curved surface that may be
wrapping around the SN.  These points lie at a fairly constant radius
(in the $x-z$ plane) of 12--14 ly from the SN.  There is greater
scatter in the eastern echoes than to the west, however both sets
appear thin in the radial direction.  This suggests the echoes may be
the walls of a shell-like structure, such as a cylinder or spheroid.
In contrast, the southern points again appear thin and sheet-like.

The geometry of the complete structure is unclear from this subset of
points.  It may be a conical or prolate shell, although from only
these data, it appears the structure may not wrap completely around
the SN.  Note that, aside from the small number of gold-colored points
(see below) echoes do not appear north of the SN.  Instead, the
material seems to stop abruptly along the northern edge of the early
echo parabola (panel \pant{b}).  

A small subset of the echoes have been colored gold.  These points
were identified in WFPC2 images very close to the bright northwest
star cluster (\pa 300\degr, $\rho=15\arcsec$).  As mentioned above,
large residuals from these closely-spaced, bright stars made echo
identification difficult-to-impossible in ground-based images.  As
they were observed at late times, these points lie 6--8 ly behind
other echoes similarly located on the sky.  No counterparts to these
echoes are seen to the east, or at any other position angle at this
radius and depth.  These points do not immediately appear associated
with the rest of the CD echoes, however we include them for
completeness.

\subsubsection{Napoleon's Hat}\label{sec-lea-3D-NH}

Figures \ref{3dle}\pant{d}--\pant{f} show the 3-D
positions of the NH echoes.  As the northern echoes are already a
known feature, we begin by describing them.  These ``NH-north echoes''
are shown in green in Figure \ref{3dle}.

The NH-north echoes are reasonably symmetric between east and west on
the plane of the sky (panel \pant{d}).  Viewed from the side (panel
\pant{e}), the bulk of the material lies along a thin shell that loops
around the northern half of the three rings.  At their closest to the
observer, these echoes appear aligned with the equatorial plane
containing the ER, and extend outwards roughly 1.5 ly beyond the NOR.
This shell looks to be aligned with the inclination of the rings, both
south (as viewed from the east) and east (viewed from the north).
From the side, the echoes could lie on a cylinder or spheroid aligned
with the ring axis.  Viewed face-on, the northern points are closer to
the SN than to the east or west, suggesting that the structure has an
elliptical cross section.

Two nearly-horizontal features protrude from this shell toward the
observeer, appearing horn-like when viewed from above (panel \pant{f}).
In the following discussion, we refer to these as the ``horns,'' to
distinguish them from the rest of the NH material.  These are the
echoes seen in early imaging that bridged between NH-north and the
larger-radii CD echoes (Fig.\ \ref{dimages}).  

Between 1992--1996, arc-like echoes appear 1.5--4 ly south of the SN.
These are disguished from echoes arising in the same region at earlier
times for two reasons.  First, most of these echoes are isolated
features on the sky, \ie in radial profile they have distinct flux
minima on either side of the peak, as opposed to the earlier southern
echoes, which appeared as a convolved complex of nearby features.
Second, the 3-D positions of these points are consistent
with a thin shell-like counterpart to the NH-north echoes, while earlier
echoes are widely dispersed around positions interior to this shell.

We discuss these interior echoes within the context of circumstellar
material below, while the shell-like points are identified as
NH-south.  These are shown in Figure \ref{3dle} in purple.  Considered
independently of the NH-north points, these echoes appear elliptical
(panel \pant{d}) and lie along a thin shell looping around the
southern half of the three-rings.  It therefore appears that both
northern and southern shells lie on the same structure, which (despite
the temptation to call these ``Napoleon's Collar'') we will simply
refer to as NH.

\subsubsection{Inner Circumstellar Material}\label{sec-lea-3D-CS}

The 3-D positions of the CS echoes are rendered in
Figures \ref{3dle}\pant{g}--\pant{i}.  Echoes passing within 1\arcsec
of the SN were identified as part of the hourglass nebula
\citep{CKH95}, and have been colored red.  These form (nearly)
complete loops at early times (up to day 1028), collapsing with time
into increasingly-shorter arcs until only a single point was
distinctly-identifiable on day 2462.  Outside of these innermost
loops, those echoes that form (nearly) complete loops or pass north of
the SN have been color coded cyan, and those that remain primarily
south of the SN are shaded gold.  This particular choice of
categorization is arbitrary, and has been made only for clarity in
viewing the echo positions.

Viewed from the side (panel \pant{h}), the red echoes do indeed appear
to trace out a hollow, conical or tube-like surface.  Visually, these
points appear inclined roughly 45\degr\ to the line of sight, with the
opening oriented north and away from the observer, while viewed from
the top (panel \pant{i}), they appear to lie along an axis rotated
10--15\degr from $z$.  That there are no loop-like counterparts in
front and slightly north of the SN is a selection effect of the
positions of the echo paraboloids, as shown in panel \panp{h} and
Figure \ref{parabs}.  This latter figure does suggest that the
earliest echoes may have skirted the southernmost edge of the
hourglass.  Panel \panp{h} does show cyan and gold echoes positioned
roughly at these expected positions.  There is, however, no {\em a
priori} means to establish which of such echoes intersect the
hourglass, versus which echoes are from material external to that
structure.  The gold echoes, when viewed from the side, are suggestive
of a larger double-lobed structure surrounding the inner hourglass.
Again, physical association of these points is difficult to establish,
and we defer further discussions to our quantitative analyses in
\sect{sec-CS}.

\section{Contact Discontinuity Echoes }\label{sec-CD}

\begin{figure}\centering
\includegraphics[height=1.75in,angle=0]{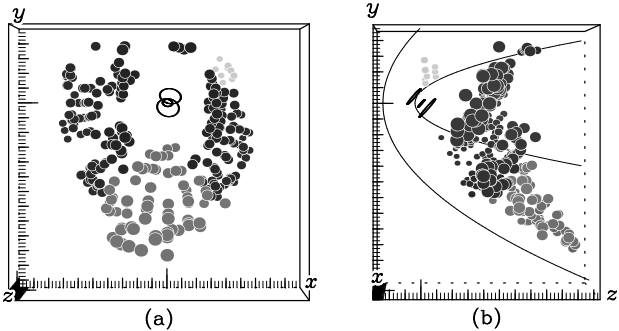} 
\caption{As Fig.\ \ref{3dle}\pant{a--b}, showing only the CD echoes,
  where now 
  ``shell'' echoes are dark grey, ``radial'' echoes medium grey, and
  light-grey points denote the WFPC2 echoes that are difficult to
  interpret \secp{sec-CD-geom-sph}.  The probable positions of the
  three rings \secp{sec-CS-rings} are indicated in each panel.
\label{3de_CD}}
\end{figure}

As explained briefly in \sect{sec-intro}, a strong body of evidence
suggests that the progenitor star passed through a RSG phase before
terminating as a BSG.  Those echoes identified as part of the contact
discontinuity (CD) between the RSG and MS winds were color coded blue
in Figure \ref{dimages}, and blue or gold in Figure \ref{3dle}\pant{a--c}.
Figure \ref{3de_CD} shows the same 3-D echo positions, but in
greyscale.  In \sect{sec-lea-3D-CD}, we noted that the geometry of the
eastern/western points appeared different from those to the south (\pa
150\degr--210\degr).  Here, the former points have been shaded are
darker tone than the latter, and the WFPC2 data for which there are no
ground-based analogs are shaded the lightest.

\subsection{Geometry of the CD \label{sec-CD-geom}}

\begin{figure}\centering
\includegraphics[width=3.25in]{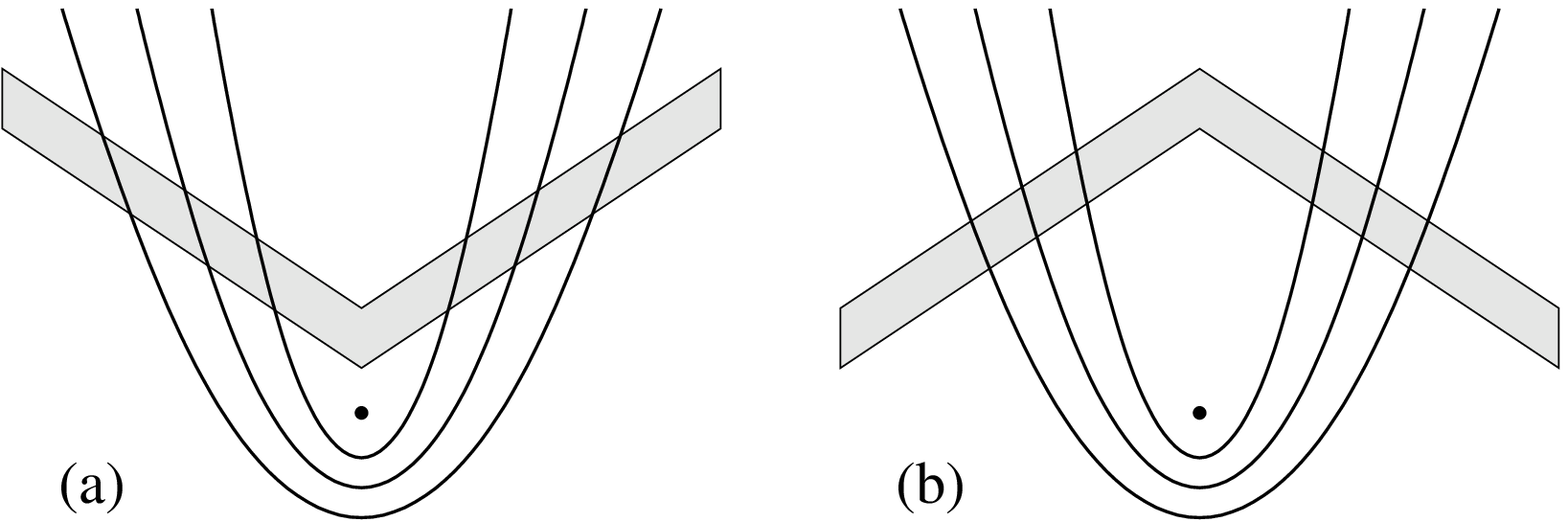} 
\caption{Schematic intersection of \panp{a} outward-facing and
\panp{b} inward-facing cones with light echoes.  The source is
marked with a solid dot.
 \label{cone-echo}}
\end{figure}

The reader may wish to skip to \sect{sec-CD-shape} for a summary. From
the side (Fig.\ \ref{3de_CD}\pant{b}), the CD echoes appear to lie
along a conical surface.  A cone, opening away from the SN, and
inclined roughly 10\degr south, will contain many of the echo points.
However, the fit is only good for the southern material (medium-grey
points).  Consider the intersections of echo parabolae with cones, as
shown in Figure \ref{cone-echo}.  Recall that the parabolae become
larger with time.  Light echoes through an outward-facing cone (panel
\pant{a}) intersect it at increasingly-large distance from the source
with time, consistent with the southern points (medium grey points).
However, eastern/western material (dark-grey points) remain at
constant or decreasing distance with time, more consistent with a
spheroidal shell, or an inward-oriented cone (panel \pant{b}).  This
implied ``saddle'' geometry is unlikely, since no counterparts to the
southern echoes are observed to the north.  Qualitatively, the
structure of which these echoes are a subset is unclear, and we
proceed to a more detailed analysis.

\subsubsection{Geometry in Spherical Coordinates 
 \label{sec-CD-geom-sph}}

\begin{figure}\centering
\includegraphics[height=3.25in,angle=-90]{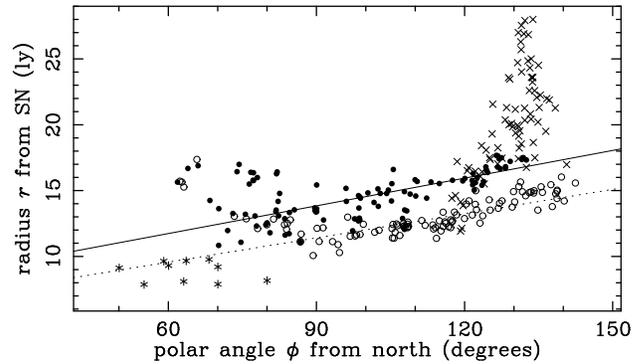} 
\caption{CD light echo positions in spherical coordinates $\phi$
versus radius $r$.  Shell (dark grey) echoes are plotted with circles,
filled for those east of the line of sight, and open for those to the
west.  Radial (medium grey) echoes are plotted as $\times$, and the
small subset of WFPC2 echoes to the northwest (light grey) by
asterisks.  The best-fit lines through the east and west shell echoes
are indicated by solid and dotted lines, respectively.
\label{plCD1}}
\end{figure}

As the eastern/western echoes appear to form a curved surface
surrounding the SN, let us first look at these data in the
spherical coordinates $\phi$ and $r$, shown in Figure
\ref{plCD1}.  Medium-grey points in Figures \ref{3de_CD} are
plotted as crosses, light-grey with asterisks, and dark-grey
as circles, filled or open if they lie east or west of the
line of sight.  Roughly speaking, the circles lie in a horizontal band
between 10 and 18 ly from the SN, while crosses are distributed
vertically at constant $\phi$.  A radial line will appear vertical in
such a plot, while a sphere will appear horizontal.  In the broadest
terms, the circles seem to form a shell-like feature at roughly
constant radius from the SN, while the crosses form a radial feature
at roughly constant inclination.  In the following discussions, we
will distinguish these sets of points by referring to the crosses as a
``radial'' feature and the circles as a ``shell.''

The radial echoes extend 14--28 ly from the SN.  Their distribution is
nearly vertical about $\phi=130\degr$, consistent with a radial line
inclined 40\degr south of the line of sight.  A few points are located
at $r\lesssim17$ ly and $\phi=120\degr$.  Whether these points are
part of the shell, or yet another structure inclined 30\degr to the
line of sight, will be addressed in the next subsection.

To the west (open circles), the shell is radially thin at all angles.
At any given $\phi$, these points show a scatter of roughly 2 ly,
which combined with the intrinsic echo width \citep[$\Delta z\sim 1-3$
ly; see][]{Sug03} suggests a shell thickness of $\sim4$ ly.  To the
east (filled circles), the shell thickness increases from a scatter of
1 ly at large $\phi$ to up to 6 ly at small $\phi$.  Both eastern and
western sets of points are visually suggestive of curved structures,
nearest to the SN around $\phi=90\degr$ (\ie the $x-z$ plane), and
flaring outward in radius above and below that plane.  To test this,
we first examine how well each set is described by a linear fit.
These are shown as solid and dotted lines for the eastern and western
points, with an RMS scatter about each line of 1.1 and 0.8 ly,
respectively.  Both lines are generally a good fit to the data,
however at $\phi<90\degr$ both sets of points deviate to larger $r$.
Points at $\phi\gtrsim120\degr$ may marginally deviate from these
lines to larger radii.

The western half of the shell is closer to the SN by $2.7\pm1.4$ ly,
using the best-fit lines described above.  Although only a 2-$\sigma$
difference, this could indicate an asymmetry in this
structure.  If the CD, whatever its actual shape, has symmetry, then
its axis is spatially offset from the SN.  If we require this axis to
pass through the SN, then the asymmetry may lie in the hydrodynamic
interaction that formed this structure, including the outflows from
the progenitor, and the medium into which they propagated (e.g.\
interaction with the remnant of a previous, nearby SN).

A cartesian line that does not intersect the origin will appear as a
convex curve in spherical coordinates, symmetric in slope about
the minimum in $r$.  That the shell points appear to lie on such
curves suggests they are part of a nearly-linear surface of
revolution, such as a cylinder or a cone.  Comparing Figure
\ref{plCD1} to the expected distributions for various geometric
surfaces, shell points could be consistent with an inclined ellipsoid,
cylinder, or inward-opening cone.  Heuristically, the inward-facing
cone may be the best fit since its apex translates into a highly
inclined, nearly vertical distribution, as seen for the radial points.

The northwestern WFPC2 echoes, marked as stars in Figure \ref{plCD1},
are nearly colinear with the western shell material.  This could be
consistent with the conical distribution described by the rest of the
CD echoes if the cone becomes more cylindrical as it passes the SN.
However, it is difficult to reconcile this structure with the shell
material that flares to larger radius at $\phi<90\degr$, indicative of
a cone growing to larger radii as it nears the SN.  The association of
these points with the rest of the CD echoes is unclear, and they may
instead arise from isolated inhomogeneities in the surrounding ISM, or
an unusual, extended image artifact, rather than from stellar outflows.

\subsubsection{Finding the Orientation \label{sec-CD-geom-axis}}

If the shell and radial material are part of a spheroidal or conical
structure, it should have a well-defined axis of symmetry.  We assume
this axis passes through the SN, and attempt to find it by $\chi^2$
minimization.  Note that only a portion of the physical CD structure
has been illuminated.  A numeric minimization of some merit function
will always be more successful at recovering the axis of an object
with a complete cross section (\eg a full cylinder), than with a
partial one (\eg a cylindrical segment).  Thus, exploring this
material's geometry is inherently uncertain.


First, we directly fit the axis by finding the line that minimizes the
geometric distance to all echo points (Appendix
\ref{app-model-LM-line}).  Constraining the fit to have zero offsets,
we find the best-fit line is inclined $37\degr$ south and rotated
$8\degr$ east of the line of sight.  The formal error of this fit is
$0\fdg5$ per axis of rotation, and while it is robust to perterbations
in the parameters, we expect the actual uncertainty is higher.

\begin{figure}\centering
\includegraphics[height=3.25in,angle=-90]{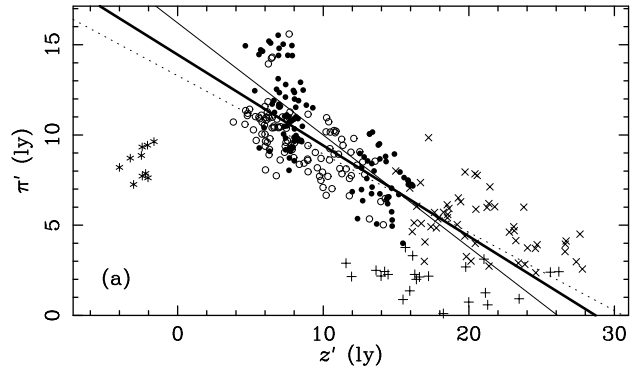} 
\caption{CD echoes, transformed into cylindrical coordinates with $z'$
along the axis of symmetry given by $i_x=37\degr$ and $i_y=8\degr$.
Coordinates are shown in Fig.\ \ref{axis}, with $z'$ along the axis,
and $\pi'$ measured radially from it.  Symbols and line-styles are the
same as Fig.\ \ref{plCD1}, with the heavy solid line is the best fit
through all data.  A subset of the radial echoes have been marked by
$+$, indicating ``cylindrical jet'' material.
\label{plCD2}}
\end{figure}

The echo points are transformed onto a cylindrical coordinate system
about this line, designated by the primed coordinates in Figure
\ref{axis}.  Figure \ref{plCD2} shows the ``radial profile'' of axial
position versus radius for the transformed data. As noted in
\sect{sec-lea-data-echoes}, we only measure the center of echoes,
therefore points have 1--4 ly incertainty in their spatial extent.
Symbols are the same as in Figure \ref{plCD1}, except for the ``+''
marks, which we discuss shortly.  In these coordinates, a cylinder
will appear as a horizontal feature, a regular cone as an inclined
line, and a biaxial ellipsoid (with major axis along $z'$) as a curved
feature with $d\pi'/dz'<0$ and $d^2\pi'/dz'^2<0$.  We have again
plotted the best-fit lines through the shell material, which show that
the bulk of it lies on a linear incline, and thus delineates a conical
structure.


The shell material is best fit by the line
$\pi'=(14.4\pm0.4)-(0.5\pm0.04)z'$, giving the conical structure an
opening angle of $\sim53\degr$.  Considered independently, the eastern
and western halves are fit with zeropoints of $16.2\pm0.5$ and
$13.3\pm0.5$ ly, and slopes of $-0.62\pm0.05$ and $-0.43\pm0.05$,
respectively.  Although these values differ, the eastern points appear
only marginally offset to higher radii than those to the west.  Close
to the SN, the shell flares outward to a larger radius of 15--16 ly,
as first suggested in Figure \ref{plCD1}.  The cone continues with
roughly constant slope and thickness until shell data ends at $z'=16$
ly.  Much of the radial material is colinear with the shell echoes
between $z'=15$ and 28 ly, but with slope $-0.25$, for an opening
angle of $\sim30\degr$.  Treating the shell and radial points as part
of a single ``conical-shell'' structure, we find that it does have a
curvature to its profile.

\begin{figure*}\centering
\includegraphics[height=1.75in,angle=0]{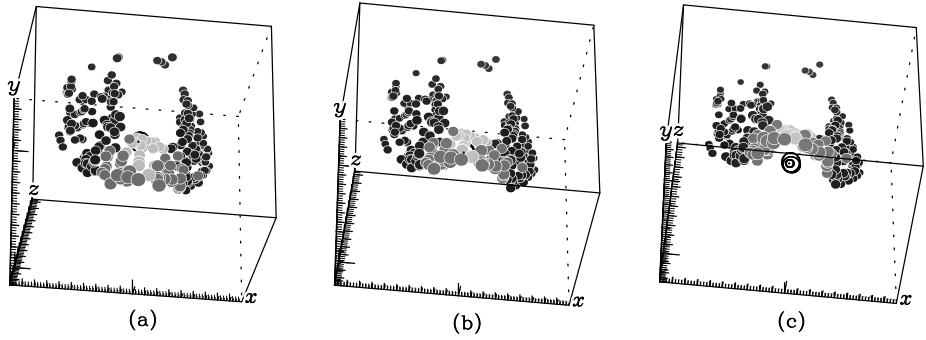}
\caption{Renderings of CD echoes (Similar to Fig.\ \ref{3de_CD}),
 showing possible symmetry axes at \panp{a} 30\degr, \panp{b} 40\degr,
 and \panp{c} 50\degr.  The northwest WFPC2 data are not
 plotted \secp{sec-CD-geom-sph}, and the ``jet'' material
 \secp{sec-CD-geom-axis} is now shaded lightest grey.
 \label{3de_CD1}}
\end{figure*}

A small subset of radial material between $z'=11-26$ ly lies at $\pi'$
just below the shell material.  After considering many possible
inclinations \secp{sec-CD-geom-alt}, we have identified the points
marked by ``+'' symbols as behaving differently from the bulk of the
radial data.  If we consider these points to be part of the conical
shell, then the entire ensemble of radial points ($\times$ and +) have
the same slope as the shell-only echoes (circles).  In such a case,
the conical shell thickens with distance from the SN, and approaches
an apex or termination at a distance of $r=28$ ly.  The subset of
radial points in Figure \ref{plCD1} that are consistent with a 30\degr
inclination \secp{sec-CD-geom-sph} are all found within this new set
of points.  If we treat them as an independent feature, they are well
fit by a horizontal line.  These echoes lie within $\pi'<3$ ly, and
extend inward toward the SN, suggestive of a
quasi-cylindrical region.  If a narrow jet pierced through the CD,
its interaction with the RSG wind might have formed such a narrow
conical or cylindrical high-density feature.  With this in mind, we
will treat these points as a separate feature from the conical shell,
and will refer to them as the ``jet.''

Finally, the WFPC2 echoes northwest of the SN are located at a radius
inconsistent with the slope of the other CD echoes.  Since there are
no echoes located between $z'=0$ and 4 ly, one can only speculate how
the profile of the CD structure changes as it passes the equatorial
plane.  If, as noted above, the shell flares out to large radius at
small $z'$, these WFPC2 echoes do not lie on the boundary of a
structure, but instead may simply be part of a local overdensity of
gas and dust, as noted previously.  If the profile does curve in, and
it contains these echoes, then one must ask why no material joining
the shell to these points was illuminated.  These echoes appear
isolated, and have no counterparts at other P.A.s, thus we give them
little weight when considering the CD structure.

Since the CD echoes appear to lie along a conical shape, its
inclination should be better studied by performing the $\chi^2$
minimization of a cone rather than a single line.  Unfortunately, the
minimization of this merit function (Appendix \ref{app-model-LM-cone})
provided few useful results.  Because the echoes trace out only a
limited portion of the structure on which they are located, the
best-fit cones have inclinations larger than $i_x=60\degr$.  Such
highly-inclined cones are poor fits to the total dataset, but minimize
$\chi^2$ by passing very close to a large subset of points at large
distance from the SN.  As a result, the remaining discrepant points
effect the merit function very little, even though the overall fit is
poor.  While these fits give little information on the inclination of
the conical structure, inclinations in the range of $i_x=30-50\degr$
consistently require rotations of $i_y=5-10\degr$.  In the following
subsections, we describe alternate techniques for establishing the
most probable orientation of this conical structure.

\subsubsection{Alternate Orientations \label{sec-CD-geom-alt}}

As it cannot be established whether the proposed inclination is
correct, we adopt a different approach by studying how the radial
profiles vary over a range of inclination angles.  We consider
inclinations $i_x$ between 30\degr and 56\degr from the line of sight,
tested every 2\degr, with the corresponding rotation $i_y$ given by
the best-fit 3-D line to the data at each inclination.  For clarity,
the CD echoes are rendered for $i_x=30\degr$, 40\degr, and 50\degr in
Figure \ref{3de_CD1}.  The jet material has been shaded light grey,
to distinguish it from the radial echoes.  Qualitatively, the
echoes appear to lie on a conical surface of revolution only for the
40\degr and 50\degr inclinations.  An inclination near 40\degr has the
most overlap (in projection) of the jet echoes and the SN, which
strengthens its intepretation as a narrow, cylindrical jet.  At
50\degr, the radial and jet material lie neatly on a surface of
circular cross-section, suggestive of demarcating the edge of an
evacuated, cylindrical region, perhaps connecting to the ER and SOR.
Although this inclination is not suggested by the previous analysis,
we question whether it could be a viable fit to the full set of CD
echoes.

\begin{figure}\centering
\includegraphics[height=2.75in,angle=-90]{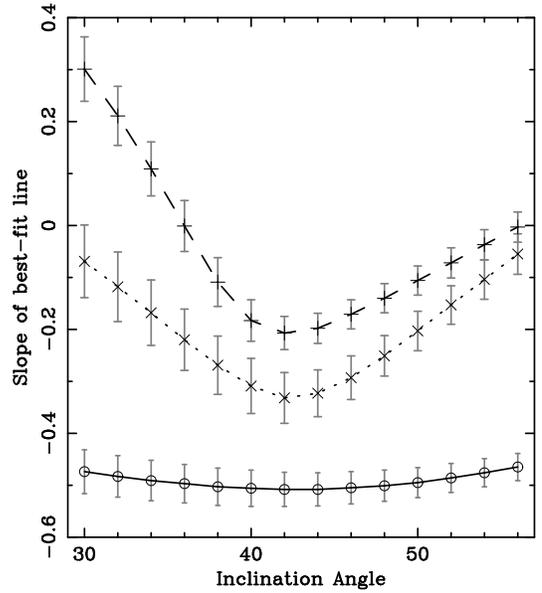}
\caption{Slopes of the best-fit lines $\pi'=a+bz'$ through the CD
echoes, inclined to the south as given along the abscissa.  Symbols
denote the same material as Fig.\ \ref{plCD2}, except all shell
material is represented by open circles.  Grey error bars show the
formal uncertainty from the least-squares fit, assuming equal errors
for all data.\label{incCD}}
\end{figure}

As in Figure \ref{plCD2}, we transform the CD material into
cylindrical coordinates about each inclination axis, and generate
radial profiles through which we measure the best fit lines.  The
slopes of those lines are summarized in Figure \ref{incCD}.  The slope
$(d\pi'/dz')$ of the shell material is fairly uniform between $-0.45$
and $-0.51$, with this steepest slope occuring at
$i_x=42\degr-44\degr$.  Although the variations appear smaller than
the quoted error bars, we note that uniform positional errors of unity
were assumed in the least-squares fits.  The radial echoes are nearly
horizontal at the smallest and largest inclinations, with the steepest
slope also occuring near $i_x=42\degr$.  At no point are the slopes of
the radial and shell echoes equal.  This reinforces the interpretation
from \sect{sec-CD-geom-axis} that radial points, if axisymmetric, lie
on a structure with much shallower opening angle than that of the
shell material.  The slope of the jet material is very sensitive to
inclination, with the implied axisymmetric structure appearing as a
wide cone opening away from the SN at $i_x=30\degr$, a narrow cylinder
at $i_x=36\degr$, or a narrow cone opening toward the SN at
$i_x>38\degr$.  Are these results consistent with viewing the
hypothesized conical shell at different inclinations?

\begin{figure}\centering
\includegraphics[width=2.in,angle=0]{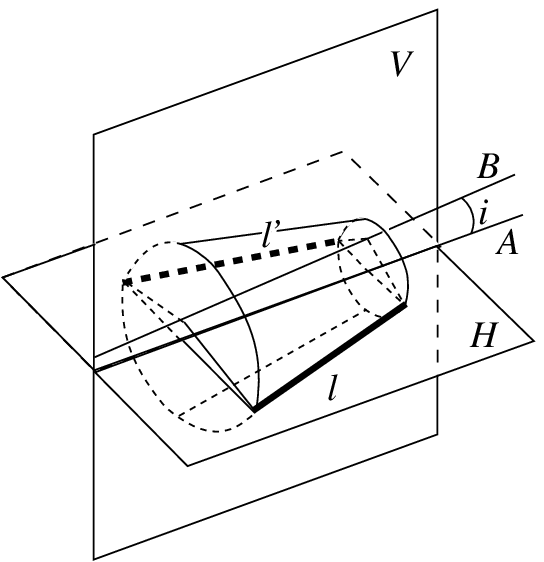}
\caption{Cartoon sketch of a frustum, and its intersection $(l,l')$
with the horizontal plane $H$.  The axis of symmetry is line $A$,
common to planes $V$ and $H$, where $V$ and $H$ are orthogonal.  We
consider inclinations $i$ of the axis relative to line $B$, also in
plane $V$.
\label{inc_cone}}
\end{figure}

\begin{figure}\centering
\includegraphics[width=2.5in,angle=0]{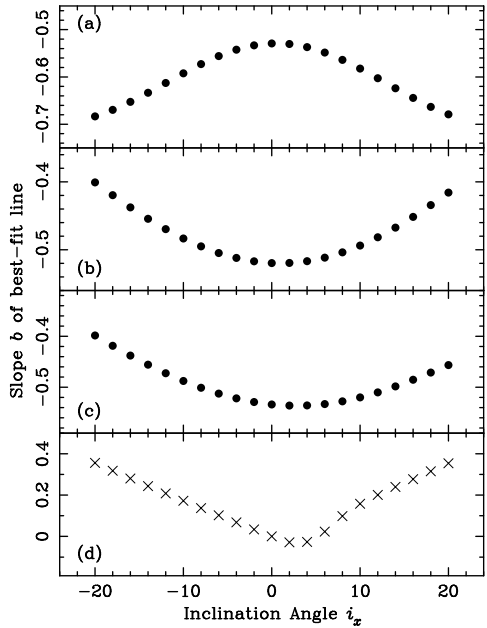}
\caption{Data similar to Fig.\ \ref{incCD} but for a simulated
frustum, oriented as in Fig.\ \ref{inc_cone}, with its axis along line
$A$, and inclinations $i_x$ made in the plane $V$.  Panels
\panp{a}-\panp{c} use a frustum with large opening angle, simulating
the shell echoes.  Best-fit slopes are calculated as follows: \panp{a}
Points sampling the complete surface of revolution. \panp{b} Points
within 30\degr of plane $H$. \panp{c} Points from 30\degr below to
60\degr above plane $H$, meant to simulate the positions of actual
shell echoes.  \panp{d} Points along the upper-half of a
frustum with small opening angle, simulating the jet material.
\label{rotcone}}
\end{figure}

Consider viewing a complete frustum at different inclinations, as
sketched in Figure \ref{inc_cone}.  The radial profile of $\pi'$
versus $z'$ is linear only close to the true axis, marked as line $A$.
Away from $A$ (\eg line $B$) the plots become highly non-linear, and
the slope of the best-fit line becomes more vertical, with increasing
inclination away from the true axis.  This is shown in Figure
\ref{rotcone}\pant{a}, which is generated by viewing a simulated
frustum at different inclinations.  Here, the slopes are steepest at
the highest inclinations, contrary to the data in Figure \ref{incCD}.
Examination of the renderings in Figure \ref{3de_CD1} shows that for
the shell and radial echoes, material is primarily located left and
right of the symmetry axis, not above and below.  Since the
inclinations move this axis through the region with no echoes, there is
no material present to cause the effect just described.

Rather, the situation is akin to rotations in the plane of reflection
between two curved sheets.  Reconsider Figure \ref{inc_cone}, in which
two lines $l$ and $l'$, with reflectional symmetry about plane $V$,
lie in plane $H$.  The slope of line $l$ as measured from line $B$
(inclined in $V$) will be similar to, but shallower than, the slope
measured from line $A$, for moderate inclinations.  If we instead
consider thin, curved sheets located on the frustum (similar to the
shell and radial material illuminated by echoes), but only extending
some angle above and below $H$, the same effect will occur.  This is
shown in Figure \ref{rotcone}\pant{b}.  Under these circumstances, the
slopes are the shallowest at the highest inclinations, consistent with
the conical-shell data.

If the shell and radial material were unformly distributed above and
below the axis of symmetry, then Figure \ref{incCD} would imply the
actual inclination is near 43\degr.  However, as seen in Figure
\ref{3de_CD1}, more material is above the axis than below it.  Figure
\ref{rotcone}\pant{c} shows the expected variation for a frustrum
wedge chosen to approximate the actual distribution of data.  Now the
steepest slope occurs at an inclination a few degrees larger than that
of the true axis, leading us to favor an inclination
$\sim$2\degr--4\degr less than 43\degr, \ie near $i_x\simeq40\degr$,
for the conical-shell material.

If we interpret the jet material as a segment of narrow, semi-cylindrical
sheet, its change in slope should appear as Figure \ref{rotcone}\pant{d}.
Similar in appearance to the jet data in Figure \ref{incCD}, these
simulated points also suggest the extremal slope is
offset by 2\degr--4\degr from the true axis.  If we continue to
interpret the jet as an independent feature, its inclination may also
be in the range 38\degr--40\degr.

\subsubsection{The Conical-Shell Cross Section
 \label{sec-CD-geom-ell2}} 

\begin{figure*}\centering
\includegraphics[height=\linewidth,angle=-90]{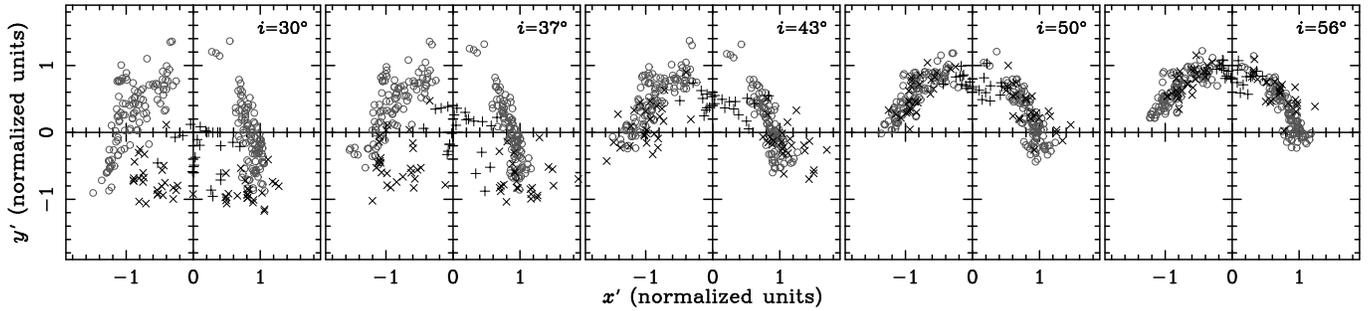}
\caption{Cross sections of CD material, viewed along a hypothetical
symmetry axis with inclination as noted in each panel.  All data are
normalized by their average distance $\langle\pi'\rangle$ measured in
bins along $z'$.  Labels $x'$ and $y'$ are the cartesian axes defined
in Fig.\ \ref{axis}.  Symbols are the same as Fig.\ \ref{plCD2},
except all shell material is designated by open circles, which are
shaded grey to facilitate the viewing of other symbols.  Comparison of
this figure with Fig.\ \ref{rotcone2} shows the most probable CD
inclination is near $i_x=40\degr$ (see text, \sect{sec-CD-geom-ell2})
\label{rotCD}}
\end{figure*}

From the previous arguments, the favored inclination of the conical
shell is around $i_x=40\degr$, although the alternate values of
37\degr and 50\degr have yet to be ruled out.  If the echo points do
lie on a surface of revolution, the cross section as viewed along its
axis should appear circular.  An elliptical cross section would either
indicating a viewpoint inclined to the symmetry axis, or that the
surface is inherently elliptical.  We investigate this by fitting
ellipses to cross-sectional profiles through $\chi^2$ minimization, as
explained in Appendix \ref{app-model-LM-ell}.  Since the axial radius
$\pi'$ varies with $z'$ for the conical-shell, points must be
normalized to a common radius.  The radial profile at a given
inclination is divided into bins of width 1 ly in $z$, then axial
radii are normalized by dividing out the average value in that bin.

For each inclination tested in the previous subsection, the
cross-section of the echoes is calculated (viewed along the axis) as
described in \sect{sec-CD-geom-axis}.  A subset of these are shown in
Figure \ref{rotCD} for the inclinations noted in each panel.  The
shell echoes lie closely scattered along two arcs of a circle or
ellipse.  With increasing inclination, less material appears under the
abscissa.  At small $i_x$, the radial points appear to form the bottom
edge of the ellipse, and become intermixed with the shell echoes at
higher inclination.  The jet material form a smaller circular shape at
moderate inclination, and eventually merge with the other material
into a single ellipse at $i_x=56\degr$, supporting the suggestion from
Figure \ref{3de_CD1}\panp{c} that the inclination axis is large.

\begin{figure*}\centering
\includegraphics[height=5in,angle=-90]{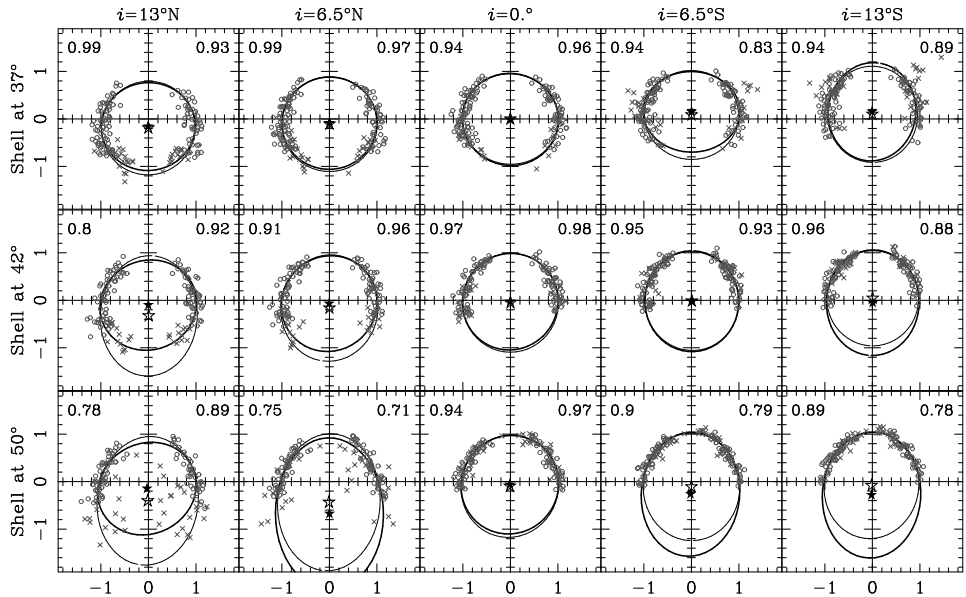}
\caption{Cross-sectional views similar to Figure \ref{rotCD}, but for
simulated conical shells.  Each shell reproduces the CD data viewed at
the incinations listed at the left of each row, and is viewed at the
inclinations listed at the top of each column.  Open circles represent
simulated shell material, and simulated radial material is marked by
$\times$.  Also plotted are the best-fit, 5-parameter ellipses and
centers through the full dataset (thick lines, solid star) and only
through the shell points (thin lines, open star).  The ellipticity
$b/a$ for each fit is given at top right and left of each panel,
respectively.
 \label{rotcone2}}
\end{figure*}

Assuming one of these to be the true inclination, we can build the
conical shell implied by its radial profile, then test whether the
other cross sections are consistent with inclined views of that data.
This is shown in Figure \ref{rotcone2}.  Linear fits to the radial
profiles at inclinations of 37\degr, 42\degr, and 50\degr define the
function $\pi'(z')$ for the shell and radial data.  For each
inclination, the function is rotated about the $z'$ axis through
position angles chosen to approximate the actual cross section.
Points along these surfaces are chosen at random, and given random
scatter, to form a simulated conical shell.  Each shell is shown in a
single row, as labeled on the left side, and inclined north or south
of $z'$ by the angle given at the top of each column.

Suppose the CD is actually inclined at $i_x=37\degr$.  That panel in
Figure \ref{rotCD} should be compared with the 0\degr inclination of
the top row in Figure \ref{rotcone2}; it is a reasonable fit since it
was so constructed.  When inclined north or south, the data should
transform as the simulation does.  If inclined to the north, the data
in Figure \ref{rotCD} at $i_x=30\degr$ is reasonably similar to the
simulation at $6\fdg5$ in the top row of Figure \ref{rotcone2}.  When
inclined to the south, however, the simulated radial material in
Figure \ref{rotcone2} appears above the abscissa with
considerable scatter, a poor fit to the data in Figure \ref{rotCD} at
$i_x\ge43\degr$.  Thus, an inclination of 37\degr seems unlikely.  The
same exercise for $i_x=42\degr$ results in a much better overall
agreement between the data and simulation, with the simulated radial
points matching the rough positions and scatter of the real ones. At
$i_x=50\degr$, the simulated views to the north (corresponding to the
data viewed close to 43\degr and 37\degr) again fail to correctly
place the radial material.  Specifically, the simulated radial
material viewed from the north at 6.5\degr looks more like echo data
at $i=37\degr$, whereas it should match up with the view at
$i=43\degr$.  Within its limited predictive capacity, these results 
favor a true inclination in the vicinity of $i_x=42\degr$.

The model can also test whether fitting ellipses to these particular
data can provide meaningful results.  For each panel in Figure
\ref{rotcone2}, we fit ellipses through all points, and through only
the simulated shell points.  Those fits, as well as their
ellipticities ($b/a$), are indicated in each box.  The actual
ellipticities, from left to right, should be 0.96, 0.99, 1.0, 0.99,
and 0.96, with the centroid's $y$-offset changing roughly as shown in
the top row.

In very few cases does either fit give the correct ellipticity, and
only in one-third of the cases are the offsets correct. Most fits
favor an orientation with the major axis along $y'$.  Considering only
the face-on view (middle column) the fits have an error of 3--5\%
including all data, and 3--14\% using only the shell points.  In the
top two rows, the fits are, on average, good to 5\%, with considerable
room for additional error.  The bottom panel shows highly-eccentric
fits with large centroid offsets $y_0$. These are ``false positives,''
since $\chi^2$ has been minimized, but the ellipse does not reproduce
the actual cross section.  Note however that in all panels, the
centers rarely deviate east or west, thus a measured offset in $x_0$
may be genuine.

Since the uncertainty in a fit is much larger than the change in
ellipticity with inclination, it is difficult to perform an analysis
similar to that in \sect{sec-CD-geom-alt} using ellipse fitting.
Nonetheless, we fit a variety of ellipses to the cross-sectional
profiles of the CD echoes for $i_x=40\degr-42\degr$.  We consistently
find the cross section's center is shifted east by $\sim1.5$ ly.  This
is consistent with the offset in shell material first noted in
\sect{sec-CD-geom-sph}.  At these inclinations, cross sections also
have an ellipticity of $b/a=0.95\pm0.05$, with the major axis aligned
toward north.  However, as noted above, this is not a robust
measurement for this particular dataset.

\subsection{A Complete Picture of the CD \label{sec-CD-shape}}

\begin{figure}\centering
\includegraphics[height=3.25in,angle=-90]{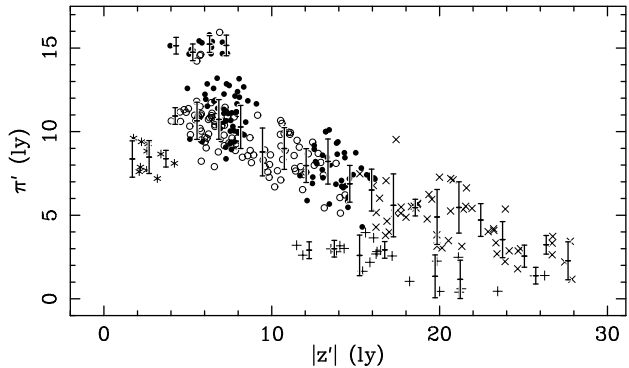}
\caption{Similar to Fig.\ \ref{plCD2}, but for the favored orientation
of $i_x=40\degr$ and $i_y=8\degr$, and showing $|z'|$ along the
abscissa.  Overplotted is the average radial profile, measured as the
average position (and standard deviation) of points binned along $|z'|$.
\label{plCD4}}
\end{figure}

The analyses of the previous subsections suggest the CD echoes lie on
a surface of rotation about a single axis, with a favored orientation
of $i_x=40\degr$ and $i_y=8\degr$.  The radial profile for this
orientation is shown in Figure \ref{plCD4}.  All features are
quantitatively very similar to those in Figure \ref{plCD2}.  Shell
points are described by nearly the same linear fits as for
$i_x=37\degr$ \secp{sec-CD-geom-axis}, while the radial and jet points
are fit by slightly-steeper lines.  Both the radial profiles and
ellipse-fitting suggest that this structure's axis may be offset to
the east of the SN by $\lesssim1$ ly.  This is not a robust result,
and since such a small offset will have a negligible effect on the
shape of the CD structure, we ignore it for now.

Assuming these data also trace a structure that is symmetric about
$z'=0$, an average radial profile of $\langle\pi'\rangle$ versus
$|z'|$ can be constructed by binning the points along the axis.  The
average radial profile (and standard deviations in each bin) is shown
over the data in Figure \ref{plCD4}.

The probable CD structure is visualized by revolving this function
about the symmetry axis, reflecting it about the equator, and
reinclining it to the favored orientation, as shown in Figure
\ref{3dpl}\pant{a--b}.  Panels \panp{c}--\panp{d} show this complete
structure plotted in monotone grey, overlaid with the actual echo
points from Figure \ref{3dle}.  For these panels, shell points are
colored blue, while radial points are red and the jets are green.

\begin{figure}\centering
\includegraphics[width=2.5in,angle=0]{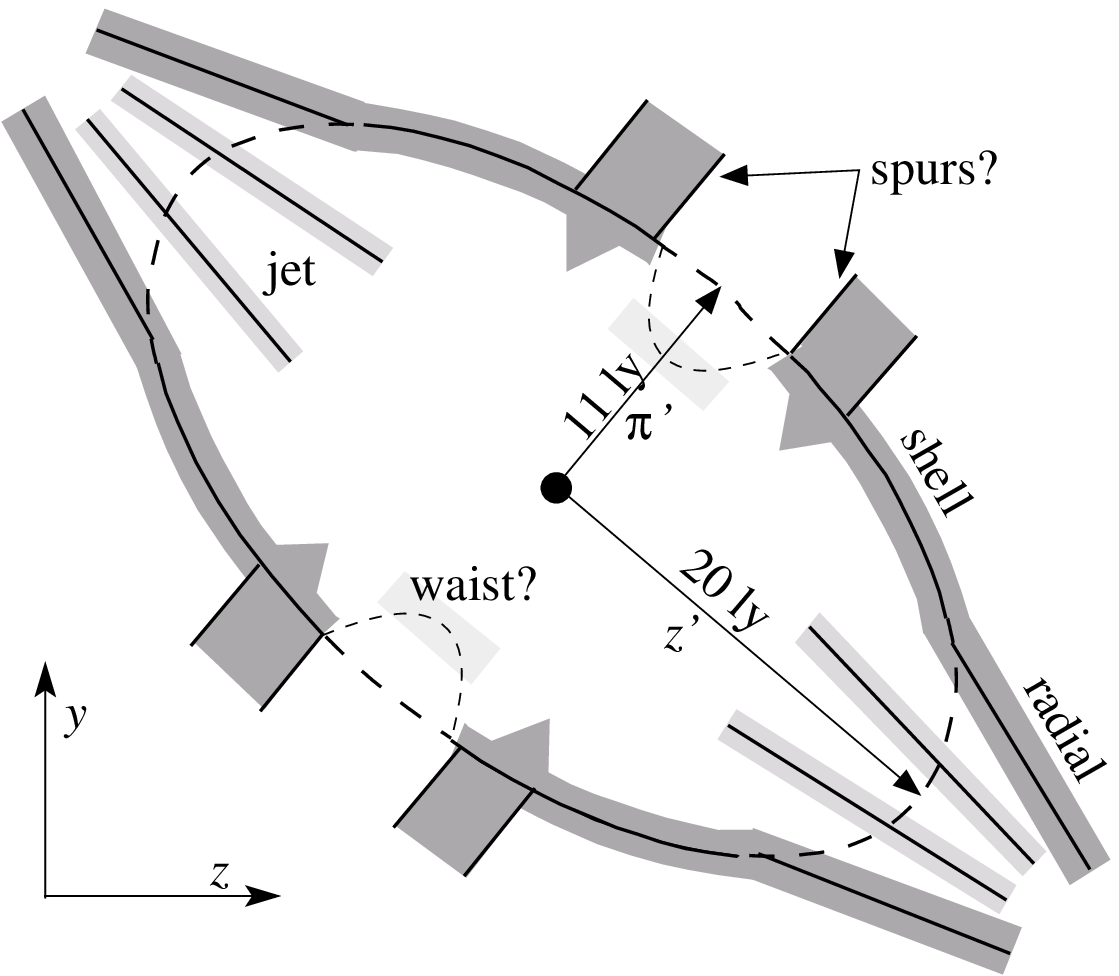}
\caption{Cartoon sketch of the salient structures traced out by the CD
model from Fig.\ \ref{plCD4}.  This figure is to scale, with the
length of the two orientation arrows equal to 10 ly.  Solid lines
trace the average radial profiles, and the scatter above those
profiles are indicated by the solid greyscale regions surrounding each
line. \label{CD_toon}}
\end{figure}

This structure is fairly complicated, and can not be described by a
simple geometric function.  Figure \ref{CD_toon} shows a scaled
cartoon of the salient features.  The ``shell'' echoes appear to lie
along a prolate spheroid with a polar axis of 20 ly and equatorial
axes of 11 ly.  However, the ends of this spheroid have been extended
into tapering cones with opening angles of about 35\degr (the
``radial'' echoes), extending from 16 to 28 ly from the SN.  Embedded
within this prolate structure are narrower, tapering cones (the ``jet
echoes'') extending from 10--26 ly from the SN, with an opening angle
of about 20\degr, and a maximum radius of 3 ly.
Some shell points in Figure \ref{plCD4} are located at $\pi'\sim 15$
ly, and appear disconnected from the rest of the material.  These
points, which we call the ``spurs,'' lie along a cylindrical
annulus that smoothly encircles the CD.

It is unclear whether the prolate shell is continuous in $z'$ along
its equator (dashed equatorial line in Figure \ref{CD_toon}),
since no echoes were observed from that region.  Whether this is due
to shadowing from material closer to the SN is addressed in
\sect{sec-density}.  The northwestern-WFPC2 echoes are positioned at
$z'\sim4$ ly along the axis, and at roughly 8 ly in radius.  This
could suggest the prolate shell is pinched at its waist to a smaller
radius of 8 ly.  However given the very limited spatial sampling of
these inner echoes, there is little evidence that they lie on a
uniform structure.

At this point in the analysis, evidence that the structure containing
these echoes has rotational and mirror symmetries is indirect, and
since construction of the probable CD nebula shown in Figure
\ref{3dpl} relies on these two assumptions, they merit a brief
discussion.  First, we recall that Figures \ref{plCD1} and \ref{plCD2}
are consistent with surfaces of rotational symmetry.  However,
rotational symmetry need not imply mirror symmetry, as in the cases of
a paraboloid or cone.  As we will see in the next two sections, light
echoes from material interior to the CD nebula lie on axisymmetric
surfaces which do exhibit mirror symmetry.  We therefore make the
(reasonable) assumption that stellar winds producing symmetric
structures at small distances would also produce such structures at
larger ones.  

\begin{figure}\centering
\includegraphics[width=3.25in,angle=0]{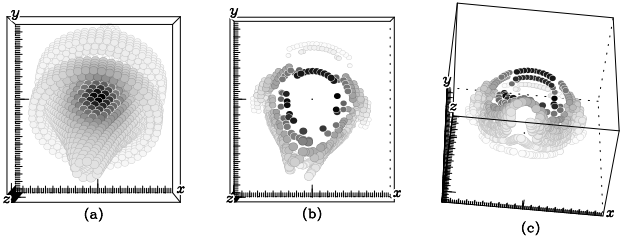}
\caption{Simulated CD showing relative brightness of echoes, where
 greyscale indicates relative brightness, with darker grey
 corresponding to brighter echoes.  To distinguish faint points, they
 have been circumscribed with darker grey circles; the reader is
 reminded that they shade of grey inside these circles indicates
 brightness.  \panp{a} Full CD, viewed face-on.  \panp{b} Random
 points illuminated by the echo parabolae from all epochs.  \panp{c}
 Panel \panp{b} inclined along the CD axis of symmetry.
 \label{CDint}}
\end{figure}

In concluding this section, we first comment on the suggestions in
\sect{sec-CD-geom-alt} and \sect{sec-CD-geom-ell2} that the CD has an
inclination of $i_x=50\degr$.  This is a result of the particular
portion of the CD that was imaged in echoes, rather than an intrinsic
property of the CD itself.  The probable CD is rendered in Figure
\ref{CDint}, now shaded by the relative brightness of dust scattering
\citep{Sug03}, wich assuming uniform density and composition is a
function of distance and scattering angle.  Panel \panp{a}
shows what the entire CD would look like though a complete set of
observations well-sampled in time.  Panel \panp{b} shows only that
portion of the complete CD that could have been observed in our epochs
of observation.  When viewed from the CD symmetry axis (panel
\pant{c}), we see that brighter echoes form a sharp, circular ridge
that could easily be mistaken for delineating the sharp edge of a
narrow cylinder that was seen from $i_x=50\degr$ in Figure
\ref{3de_CD1}\panp{c}.

Finally, if the CD is indeed mirror symmetric, why were echoes not
seen from the far half of the nebula (material at $-z'$)?  In
particular, there is a small portion of material north of the SN
interior to the late-time paraboloids (northern points just interior
to the large parabola in Figure \ref{3dle}\panp{b}).  Figure
\ref{CDint}\panp{b} shows that echoes of this small northern portion
were very faint due to the large scattering angle, and hence were
unlikely to be detected.  On the other hand, a segment of the spurs
along the closer half of the CD should have created very bright echoes
at a radius of $\rho=7$ ly, roughly from PA 315\degr--30\degr.  Very
little signal was detected in this region, which may suggest that the
spurs are isolated clumps, rather than a uniform feature. 

\section{Napoleon's Hat }\label{sec-NH}

In contrast to the previous case of the CD echoes, there is much less
 ambiguity over the possible structures of which the NH
echoes are a part.  As shown in Figure \ref{3dle}\pant{d--f}, NH
appears to be some surface of revolution with a well-defined axis,
such as a prolate ellipsoid, cylinder or frustum.  However, it is not
immediately clear whether the two horns are part of such a structure.
If not, where do they lie, and if so, why has such a limited portion
been illuminated?  The geometry of the NH echoes will be quantified in
the following subsections.

\subsection{Geometry of NH Echoes \label{sec-NH-geom}}

\subsubsection{ Spherical Coordinates
 \label{sec-NH-geom-sph}}

\begin{figure}\centering
\includegraphics[height=3.25in,angle=-90]{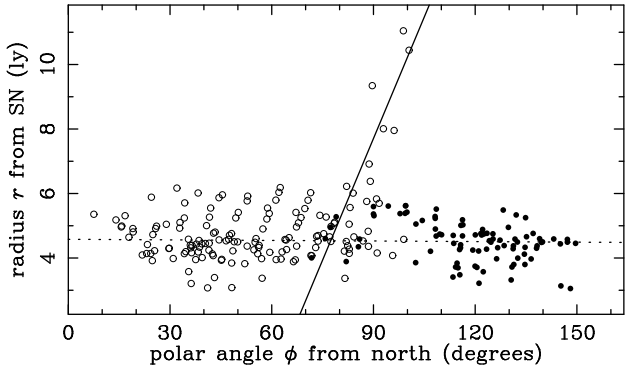} 
\caption{Spherical positions of NH echoes (see Fig.\ \ref{plCD1}).
NH-north are plotted with open circles, and NH-south with filled
circles.  The best-fit lines through the horns (solid line) and the
other data (dotted line) are indicated.
 \label{plNH1}} 
\end{figure}

As with the CD, we examine the NH echoes in spherical
coordinates, shown in Figure \ref{plNH1}.  The bulk of the echoes fill
a horizontal region in $\phi$, with points distributed between
$\pi'=3-6$ ly, but concentrated around $r=4-5$ ly.  In these
coordinates, a thick spherical shell with radius $r\sim4.5$ ly and
width $\Delta r=1$ ly would populate the same region.  However, such a
structure is unlikely: if the data lay along a spherical shell,
echoes in front and south (or behind and north) of the SN should have
been observed.

The horns are also inconsistent with a spherical structure, as they
are instead located in a highly inclined, almost radial region
extending from 6--11 ly from the SN.  Rather, the total ensemble of NH
echoes can be understood to lie along a cylinder or hourglass.  The
distributions of points on such structures, viewed in spherical
coordinates, populate a horizontal region in $\phi$ near the equator,
while points far from the equator will be highly
inclined.  

\subsubsection{Cylindrical Coordinates
 \label{sec-NH-geom-cyl}}

As with the CD echoes, studying points in cylindrical coordinates
requires knowing the axis $z'$.  Since the candidate structures for
the NH echoes are an hourglass or cylinder (a specialized case of an
hourglass), we start by fiting a biconical frustum to all NH echo
points, as explained in Appendix \ref{app-model-LM-cone}.
We first restrict the figure's cross section to be circular, and the
SN center to lie on the symmetry axis.  With these constraints, the
best-fit structure (RMS scatter of 0.73 ly) is an hourglass with waist
radius of 3.7 ly and opening angle 36--39\degr, inclined
$i_y\sim40\degr$ south and $i_x=2-4\degr$ east of the line of sight.
Allowing for an elliptical cross section improves the $\chi^2$ by
nearly a factor of two (RMS=0.55 ly), yielding semi-major and minor
axes of 4.6 and 3.8 ly, with opening angles at the axes of 28 and
40\degr, respectively.  The major axis is rotated $96\degr\pm4\degr$
east of north, and the axis of the hourglass is oriented with
$i_y=39\fdg5\pm 0\fdg5$ south and $i_x=6\fdg5\pm 0\fdg5$ east of the
line of sight.  This fit is robust to perturbations in the fitted
parameters, including spatial offsets of the waist from the SN.  

\begin{figure}\centering
\includegraphics[height=3.25in,angle=-90]{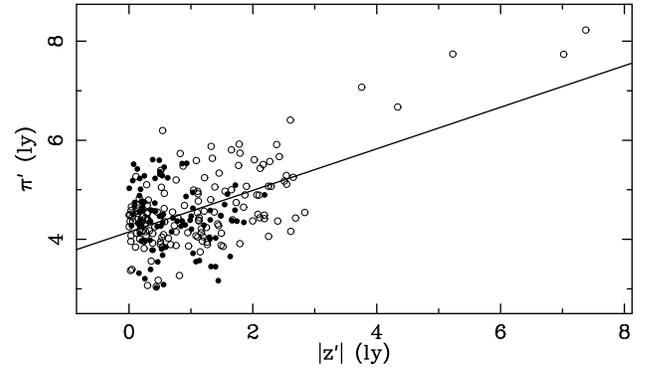} 
\caption{Cylindrical positions of NH echoes (see Fig.\
\ref{plCD2}) for the best-fit inclination of 40\degr south and 6\fdg5
east.  Symbols are the same as Fig.\ \ref{plNH1}.  We assume NH is
symmetric about the equator, and plot $|z'|$.  The solid
line is the best-fit through all points.
\label{plNH2}}
\end{figure}

Figure \ref{plNH2} shows the radial profile of these points as a
function of absolute position along the axis (this assumes the data
have reflectional symmetry about $z'=0$.  This figure is consistent with an
hourglass shape.  The best fit through all data is given by
$\pi'=(4.13\pm0.06)+(0.42\pm0.04)|z'|$, which corresponds to a waist
radius of 4.1 ly and an opening angle of 46\degr.  While there is a
large scatter (RMS=0.5 ly) about this line, note that there are
similarly-sloped loci of points $\lesssim 1$ ly above and below it.
This is further suggestive of a structure with elliptical
cross-section.

The horns extend along the axis from 1--7 ly in front of the SN, and
radially 4--8 ly from the axis.  They are best-fit by the line $\pi'=
(4.4\pm0.2)+(0.52\pm0.07)z'$, which corresponds to a frustum with
inner radius 4.4 ly and an opening angle of 55\degr.  The horns do not
appear to be truly linear, but curve to shallower slope at larger
$z'$, as if they lie on a surface of parabolic, rather than linear,
profile.

\begin{figure}\centering
\includegraphics[height=1.75in,angle=-90]{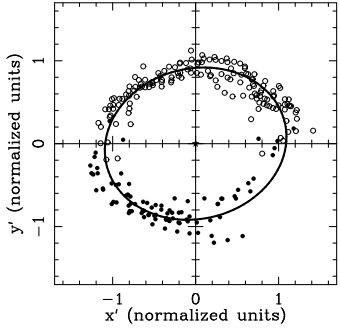}
\caption{Cross-sectional view of NH echoes looking down the best-fit
inclination axis.  Symbols are the same as Fig.\ \ref{plNH1}.  The
best-fit ellipse is indicated by the solid line.
 \label{plNH3}} 
\end{figure}

We directly examine the ellipticity of the NH cross section by
obbserving the distribution of points looking down the best-fit axis
of inclination (40\degr south, 6\fd5 east).  Similar to
\sect{sec-CD-geom-ell2} with the CD, echo positions $(x',y')$ in the
plane orthogonal to $z'$ (see Figure \ref{axis}) are normalized by the
average radius $\pi'$ as a function of $z'$ given by the best-fit line
in Figure \ref{plNH2}.  The resulting distribution, shown in Figure
\ref{plNH3}, is fit with a 5-parameter ellipse, as described in
Appendix \ref{app-model-LM-ell}.  The best-fit ellipse
has negligible offset from the origin, and $b/a=0.82\pm0.02$ with the
major axis rotated $103\degr+/-3$ east of north.  This is consistent
with the hourglass fit above, which yielded $b/a=0.83$ and major-axis
rotations of 92--101\degr.
 
\begin{figure}\centering
\includegraphics[height=3.25in,angle=-90]{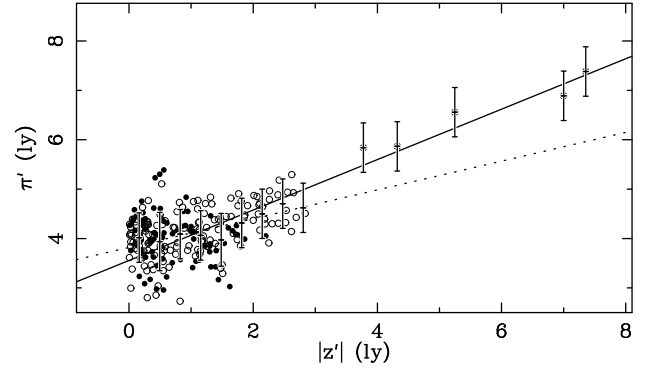}
\caption{Data from Fig.\ \ref{plNH2}, with the ellipticity from
Fig.\ \ref{plNH3} removed.  Overplotted is the average radial profile,
measured in bins along $|z'|$.  Single points have fixed errors.  The
best-fit line through the complete profile is indicated by a solid
line, and the best-fit through points excluding the horns ($|z'|<3$
ly) by the dotted line.
 \label{plNH4}}
\end{figure}

We use these values to remove the ellipticity from the radial profile,
yielding the distribution shown in Figure \ref{plNH4}.  As expected,
the horns have shifted to smaller radii, and the scatter between the
rest of the NH points has diminished.  Also shown is the average
radial profile of the NH echoes, generated by binning the points along
the inclination axis and computing the average radius about that
value.  The implied structure appears cylindrical with a semi-minor
axis of 4 ly up to $|z'|=1.5$ ly, after which it has a fairly constant
slope.  The best-fit line through the entire profile (solid line) is
given by $\pi'=(3.53\pm0.23)+(0.52\pm0.06)z'$, corresponding to an
hourglass with semi-major and minor waist axes of 4.4 and 3.5 ly, and
an opening angle near 54\degr.  Excluding the horns, the fit is
considerably shallower (dotted line), suggesting an opening angle of
36\degr, and axes of 4.8 and 3.8 ly.  Despite the ambiguity in
functional form, these fits are all consistent with the values found
by the hourglass fitting.

\subsubsection{What are the horns? 
 \label{sec-NH-geom-horns}}

\begin{figure}\centering
\includegraphics[width=3.25in,angle=0]{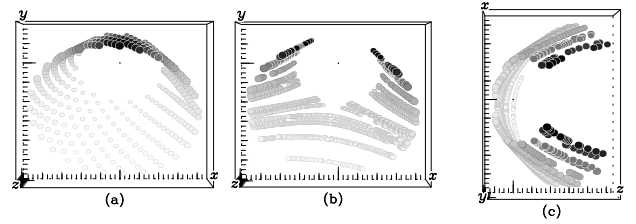}
\caption{As Fig.\ \ref{CDint} but showing the frustum containing the
 horns.  Darker shading indicates increased surface brightness of an
 echo.  \panp{a} The complete frustum as revealed through a complete,
 regularly sampled set of echoes.  Random points along the
 intersection of the frustum with echo parabolae from our observations
 are shown in panels \panp{b} (face-on) and \panp{c} (from the top).
 Greyscale is normalized to the brightest and faintest points plotted,
 thus shadings differ between panels \panp{a} and \panp{b}--\panp{c}.
\label{NHint}}
\end{figure}

If the horns are part of the same hourglass-like structure containing
the rest of the NH echoes, why was such a limited section of the much
larger surface illuminated?  Assume the horns lie along the solid line
in Figure \ref{plNH4}.  This line can be revolved about the $z'$ axis,
then reinclined from the line of sight by the best-fit NH orientation
to generate a complete frustum, as shown in Figure
\ref{NHint}\pant{a}.  As in Figure \ref{CDint}, greyscale indicates
relative brightness of echoes based on their scattering angle and
distance from the SN. Points along the northern limb are the
brightest, since the scattering angle is lowest.

Random points on this frustum along echo paraboloids observed in our
data are shown in panels \panp{b}--\panp{c}.  The northern limb was
never illuminated by echoes as it lies inside the earliest paraboloid.
The earliest intersection occured just south of this limb, and over
time, increasingly-southern slices were illuminated, becoming fainter
with increasing epoch.  The brightest echoes came from the first few
epochs, appearing as linear, radial features at roughly constant
north-south position.  Furthermore, at the earliest epochs, the
illuminated material is oriented close to the line of sight, thus
these echoes appeared even brighter due to a limb-brightening effect.
Thus the limited portion of this structure that was illuminated as
``the horns'' is consistent with these points lying on the same
hourglass as the rest of the NH material.

\subsection{A Complete Picture of the NH Structure \label{sec-NH-shape}}

The probable structure containing the NH echoes is an hourglass, the
waist of which has an ellipticity of $b/a=0.82$, with a semi-major
axis of $\sim4.9$ ly rotated about 102\degr east of north.  Bootstrap
analyses of these fits yield errors of about 3\% for each parameter.
If the horns lie on a surface with similar eccentricity, the hourglass
flares to a semi-minor axis of $\sim 7$ ly about 7 ly from the SN
along the axis of symmetry.  The entire structure is oriented at
$i_y=40\degr$ south and $i_x=7\degr$ east of the line of sight.  To
visualize this, the average radial profile from Figure \ref{plNH4} is
reflected about the equator and revolved around the axis, then
reinclined from the line of sight by the inclinations just given.  The
resulting distribution is rendered in Figure \ref{3dpl}\pant{e--h}.

A much larger portion of this nebula was sampled by echoes than was
actually observed.  This was discussed in some detail above within the
context of the horns, from which we concluded that the majority of the
signal from the near half of the hourglass was very faint due to the
larger distance and scattering angle of that material compared to the
dust at the waist.  The same argument applies to the far half of the
hourglass, explaining why no discrete signal was observed from the
northern parts of the structure at later times.  

\begin{figure}\centering
\includegraphics[height=2.5in,angle=0]{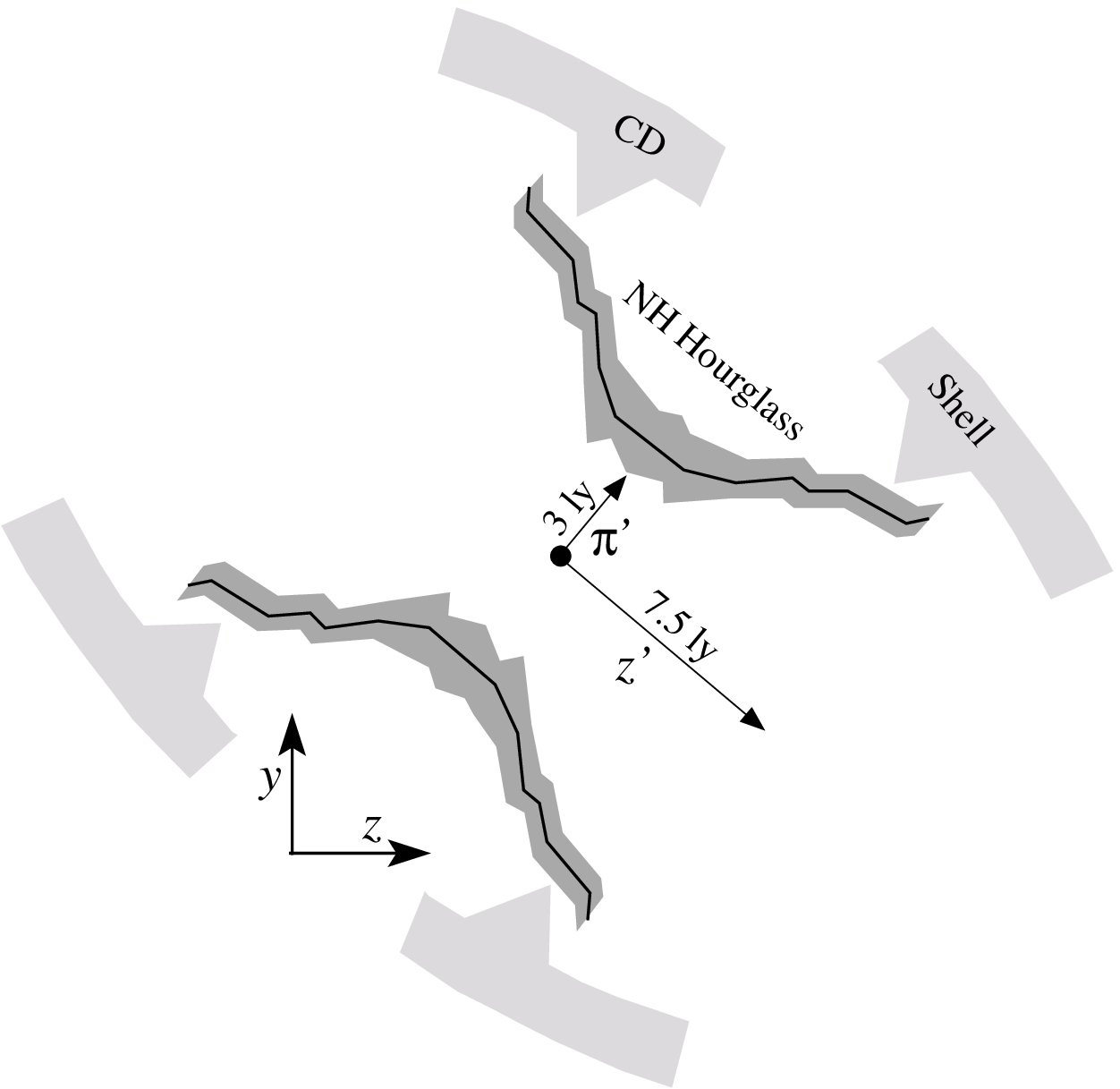}
\caption{As Fig.\ \ref{CD_toon}, but for the NH echoes, with the
 orientation arrows measuring 4 ly.  The solid line traces the radial
 profile from Figure \ref{plNH4}, and the width of the greyscale
 region traces the scatter of points about that profile.  
\label{NH_toon}}
\end{figure}

A number of small-scale features not represented in this figure are
better discussed in the cartoon sketch of the NH echoes shown in
Figure \ref{NH_toon}, where grey-shading represents the scatter in
points about the average profile.  In the first two bins ($|z'|<0.5$
ly) in Figure \ref{plNH4}, material extends beyond the average scatter
to smaller and larger radii of $\pi'\lesssim 3$ and $\pi'\gtrsim 4.5$
ly.  Furthermore, over one-third of the NH points lie within these
first bins, which implies these data trace out an equatorial
enhancement in both density and radial position.  This is shown in
Figure \ref{NH_toon} as small spurs pointing toward and away from the
SN in the equatorial plane.  A similar, but less significant,
inward-pointing feature is seen roughly 1.5 ly along the axis.
CKH95 reported a small set of echoes within 3 ly of the SN
that are coincident with the ER plane, which they interpreted as
evidence for an extended circumstellar equatorial disk.  The
NH equatorial echoes appear to extend this disk to a
radius of 5 ly, with a thickness of $\gtrsim 0.75$ ly.

\subsection{Comparison with Previous Work \label{sec-NH-comp}}

As NH was the subject of interest in the early 1990's, we compare the
newly revealed material with previous interpretations.  \citet{Wam90a}
first reported the discovery of the ``Napoleon's Hat'' echo in
narrow-band [\ion{O}{3}] and [\ion{N}{2}] imaging as diffuse
nebulosity, which blended smoothly into  background filamentary
structure that defines the eastern and western rims of a ``V''-shaped
``dark bay'' south of the SN (see Fig.\ \ref{Ha612}).
\citet{Wam90b} further noted that the larger, arcmin-radius echoes
from interstellar dust \citep[\eg ][]{Xu95} fade when crossing over
this bay.  A natural interpretation was that this bay is an evactuated
region, and no light echoes were seen there because of its
particularly low density.  As the background filaments loop
around the SN, they proposed that the progenitor's MS winds evacuated
the bay, and its V-shape can be attributed to the proper motion of the
progenitor through the ISM.

\citet{WW92} reanalyzed their ground-based data and were able to trace
the NH echoes as early as day 890, but the echoes had faded by day
1650.  An interpretion that these echoes lie along a disk coplanar to
the inner ER was disfavored, since the flux had the wrong distribution
and shape for a solid, dusty disk, and since the echoes moved very
little in time.  Rather, \citet{WW92} and \citet{WDK93} considered
these structures consistent with a parabolic bow-shock to the north of
the SN, which could be further interpreted as the terminal edge of the
dark bay evactuated by the MS progenitor.

Given the limited data at hand, and considerable confusion from the
background filaments in the LMC, the bow-shock model was an attractive
scenario.  However, the current analysis has revealed a much more
extensive volume of material illuminated by the SN, the 3-D positions
of which are inconsistent with a bow shock.  Rather, these data
imply that the disfavored interpretation of \citet{WW92}--that NH is
in the equatorial plane--was correct.

Suppose the star does have a large proper motion, and that the dark
bay is the region evacuated by the MS wind.  Whatever the star's
proper motion through the ISM, it must have moved at least 25 ly north
during its lifetime to create this bay and end at its observed
position.  The CD echoes from this region have illuminated a prolate
structure that ends at roughly the same position on the sky as the
center of this bay.  This is also inconsistent with the bow-shock
model. Most likely, the filamentary nebula is not associated with the
SN, and we propose that the correlation between the position of the NH
echoes and the ``V''-shaped nebula is a positional coincidence.


\section{Circumstellar Echoes }\label{sec-CS}

As noted in \sect{sec-intro}, the inner 3\arcsec surrounding SN~1987A
have been previously studied by CKH95 using a subset of the data
presented here.  Employing PSF-subtraction and rudimentary
PSF-matching, the authors detected echoes along a bipolar nebula
with an hourglass shape, in the sense that a narrowed waist separates
two open-ended lobes.  A smaller set of echoes coplanar with the ER
were identified as evidence of an extended circumstellar disk.

The current analysis differs from CKH95 in a number of ways.  In the
previous work, the authors payed particular attention to the inner few
arcsec around the SN, and their reduction may have less noise from
PSF-subtractions of the SN and stars 2 and 3 than from our automated
reduction pipeline \secp{sec-reduc-reg-final}.  The general
PSF-matching algorithm has been greatly improved since 1995, and we
have implemented a number of custom modifications to minimize
difference-residuals and to increase the detection threshold of echo
signal.  Since we were interested in echoes from a much larger range
of radii, we also have increased the spatial extent over which
circumstellar (CS) echoes have been sought.  Finally, we used a very
different technique for identifying the positions of echoes.
Sufficient differences exist between the two methods that the current
data merit an independent analysis.

\subsection{Geometry of CS Echoes \label{sec-CS-geom}}

An exploration of the geometry of these echoes (summarized in
\sect{sec-CS-shape}) is facilitated by the work of CKH95, which
established the cylindrical symmetry axis of the double-lobed CS
hourglass.  We cannot immediately explore such a feature in our data
since, as noted in \sect{sec-lea-3D-CS}, it is unclear which
extended-flux echoes are associated with this hourglass.  However,
CKH95 reported a small number of echoes along this southern lobe, and
we can use their data to help disentangle our points.

\subsubsection{Geometry in Cylindrical Coordinates \label{sec-CS-geom-cyl}}

\begin{figure}\centering
\includegraphics[height=3.25in,angle=-90]{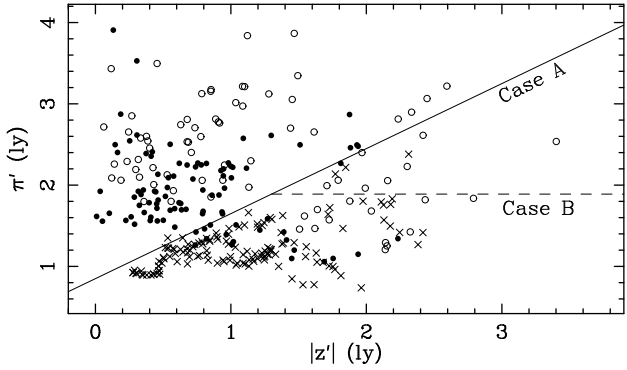}
\caption{Cylindrical profile of $\pi'$ versus $|z'|$ for the CS
echoes inclined by $i_x=45\degr$ and $i_y=8\degr$.  Symbols correspond
to points in Fig.\ \ref{3dle}\pant{g--i}, with closed and open circles
denoting cyan and yellow extended flux points, and $\times$ marking the
red hourglass.  Points below the solid line are potential
members of the CS hourglass, and comprise distribution ``A,'' while
those also under the dashed line are distribution ``B''
\secp{sec-CS-geom-cyl}.
\label{plCS2}}
\end{figure}

We begin by fitting a biconical hourglass to the CKH95 data, using the
$\chi^2$ minimization described in Appendix \ref{app-model-LM-cone}.
Assuming the frusta have circular cross-secitons, the data are
best-fit by a structure inclined by $i_x=45\degr$ south and
$i_y=8\degr$ east of the line of sight.
This compares favorably to the orientation reported in CKH95 using a
minimization-of-scatter estimator, but our rotation $i_y$ is larger
than theirs ($3\degr\pm5\degr$).  We adopt this as the preliminary
orientation axis, and transform all echo points into cylindrical
coordinates (see Figure \ref{axis}) about this axis.  Assuming the
structures are symmetric about the equator ($z'=0$), we plot in Figure
\ref{plCS2} the radial profile of $\pi'$ as a function of absolute
distance $|z'|$ along the axis.  The region containing
northern-hourglass echoes (marked by $\times$) includes some
extended-flux material, which may correspond to additional points
along the northern and southern lobes.

The northern-hourglass echoes in Figure \ref{plCS2} can be considered
as lying along one of two regions.  Case ``A'' is roughly-delimited as
points below the solid line, which follows the outer edge of the
echoes at any given point along the axis.  When binned, the
northern-hourglass points follow a somewhat curved trajectory that
approaches constant radius with increasing $|z'|$.  Such a region is
approximately bounded by the dashed horizontal line, which we denote
case ``B.''  While a few extended-flux points straddle the solid
boundary, the majority are located above it, suggesting that points
above the solid line are part of a different structure, and only
points below it can be part of the hourglass.  We proceed by testing
which distribution of points is most-likely on such a structure.

\begin{figure}\centering
\includegraphics[width=3.25in,angle=0]{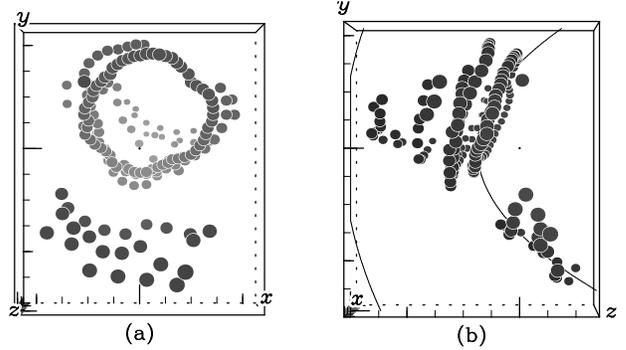}
\caption{Same as Fig.\ \ref{3dpl}\pant{g--h}, but showing only those
CS points believed to lie along the hourglass (the case B distribution
from Fig.\ \ref{plCS2}).
 \label{3de_CS}}
\end{figure}

Biconical hourglasses are fit to the case A and B distributions.  Case
A points are consistent with an hourglass inclined 41\degr--42\degr
south and 10\degr--13\degr east of the line of sight.  The fit to
distribution B is similar, with an inclination of 40\degr--41\degr
south, 12\degr--15\degr east.  This fit has a $\chi^2$ nearly three
times smaller than for case A, while there are only 16 fewer points,
suggesting that the points above the dashed line in Figure \ref{plCS2}
are not directly part of the hourglass structure.  Accordingly, we
choose the B distribution as the probable set of hourglass points.
These points are rendered in Figure \ref{3de_CS}.  As expexted from
the earliest echo parabolae, only a small arc along the southernmost
limb of of the southern lobe is present.

The best-fit hourglass to these data is inclined at $(i_x,i_y)$ of
$(-40.7\degr\pm0.2\degr, 14.7\degr\pm0.7\degr)$, with a westward shift
of $0.04\pm0.01$ ly.  The waist has an ellipticity $b/a=0.94\pm0.03$ and
semi-major axis $1.01\pm0.2$ ly, rotated $10\degr\pm1\degr$ north of
east.  As noted in Appendix \ref{app-model}, these errors
are the scatter in parameters fit to random subsets of the data.

\subsubsection{Constraints from the Equatorial Ring \label{sec-CS-geom-ER}}

If we assume the three-ring nebula lies symmetrically about the
hourglass, then the positions of the rings on the sky should
intersect, or be symmetrically bounded by, the hourglass.  These rings
are commonly assumed to be planar circles, appearing elliptical due to
their inclination.  As such, their observed positions should be
reproduced by simple geometric figures viewed in the hourglass frame
of reference.  Unlike the ORs, for which only limited velocity data
exists, the velocity shear across the ER indicates it is a planar
structure \citep{CH91,CH00}, and its projected geometry is very-well
constrained (see below).  The ER is therefore the best candidate for
this exercise.

\begin{deluxetable}{lcc}
\tablecaption{Observed Geometric Parameters of the ER \label{tbl-geo}}
\tablewidth{0pt}
\tablehead{\colhead{Parameter} & \colhead{Unit} & \colhead{Value}
}
\startdata
 Centroid Offset & (mas W)   & $18.7 \pm 0.4$ \\
 Centroid Offset & (mas S)   & $2.0 \pm 0.7$  \\
 Inclination Angle & (\degr) & $43.0 \pm 0.1$  \\
 Semi-Major Axis & (\arcsec) & $0.820 \pm 0.005$ \\
 Semi-Minor Axis & (\arcsec) & $0.600 \pm 0.005$ \\
 P.A.\ of Major Axis & (\degr E of N) & $81.1\pm0.8$  
\enddata
\end{deluxetable}

\begin{figure}\centering
\includegraphics[height=2.75in,angle=-90]{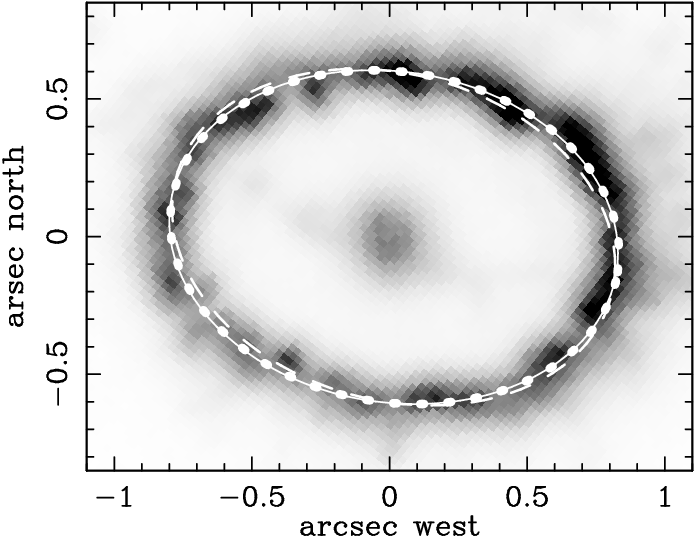}
\caption{Drizzle-reconstructed $(0\farcs23$~pix$^{-1})$ WFPC2 F656N
image of the ER from \citet{Sug02}. The best fit ellipse to the
observed ER (Table \ref{tbl-geo}) is indicated with the thick dotted
line.  The projected position of an ellipse oriented by the PYR
triplet (-16\fdg5,-14\degr,41\fdg5) is shown with a thin dashed line,
corresponding to the best fit to the hourglass from
\sect{sec-CS-geom-cyl}.  The projected view of the ellipse that
best-matches the ER from \sect{sec-CS-geom-ER} oriented by
(-9\degr,-8\degr,41\degr) is indicated by the thin solid line, which
nearly exactly matches the thick dotted line.
 \label{plrings_ER}} 
\end{figure}

\citet{Sug02} measured the ER geometry by fitting ellipses to a
high-resolution WFPC2 image.  The reader is referred to that paper for
a thorough description of the method.  Briefly, the ER was divided
into bins of equal arclength, and the flux in each bin was normalized
to the second brightest pixel value.  Ellipses were then fit to the
brightest N\% of the pixels in each bin, where $N$ varied from 50\% to
95\%, and the resulting parameters averaged to give final values.
However, their choice of fitting algorithm was not as robust as that
described in Appendix \ref{app-model-LM-ell}.  We
reproduced their fit by substiting our improved minimization
algorithm, the results of which are given in Table \ref{tbl-geo} and
drawn with heavy dots in Figure \ref{plrings_ER}.  

If the ER is a planar ellipse, then viewed at the hourglass
orientation, it should overlap closely with our best-fit.  For all the
orientations listed in the previous section, no combination of ellipse
parameters (offsets, axes and major-axis rotation) can fully reproduce
the observed geometry.  The best fit achieved using those orientations
is shown with a thin dashed line, and has an average deviation of 18
mas.  This error is more than double the uncertainty of the ER axes,
and is as large as its centroid offset, thus the orientation provided
by the hourglass fit is not ideal.

Allowing the orientation of the ellipse plane to vary as well, the
absolute best fit to the observed ER deviates by only 3 mas.  Shown by
the thin, solid line, this ellipse has an orientation of $i_x=41\degr$
south and $i_y=8\degr$ east, and with the major axis rotated 9\degr
north of east.  The ellipse has a major axis 0\farcs82, $b/a=0.98$,
and the centroid is shifted 19 mas west of the SN.  A circle at this
same orientation fits the ER with three-times larger $\chi^2$.
Provided the ER is connected to the CS hourglass, this is the first
direct measurement of the ER's orientation and deprojected geometry.

\subsubsection{Geometry in Cylindrical Coordinates II \label{sec-CS-geom-cyl2}}

Why do the hourglass and ER fits yield values of $i_y$ discrepant by
over 6\degr?  The hourglass is well-sampled only along the southern
edge of the northern lobe, otherwise, the data are sparsely sampled.
This distribution is sufficient to constrain the north-south
inclination $i_x$, but allows great freedom in east-west rotations.  We
test the newly-proposed orientation by repeating the hourglass fit,
this time holding $i_y$ fixed at 8\degr.

The result is an hourglass inclined $i_x=41\fdg2\pm0\fdg2$ and shifted
$0.1\pm0.02$ ly west; a waist with semi-major axis $1.04\pm0.02$ ly
rotated $9\degr\pm1\degr$ north of east; and $b/a=0.94\pm0.01$.  The
same caveat for the errors from \sect{sec-CS-geom-cyl} applies.  The
frusta have a half-opening angle of $12\degr\pm1$, which is close to
that found by CKH95.  This is very similar to the orientation
suggested by the ER fit, and therefore we adopt it for the CS
material.   Note that fitting an hourglass
confirms the findings in \sect{sec-CS-geom-ER} and by \citet{Sug02}
that this nebula is offset slightly west of the SN.

\begin{figure}\centering
\includegraphics[height=3.25in,angle=-90]{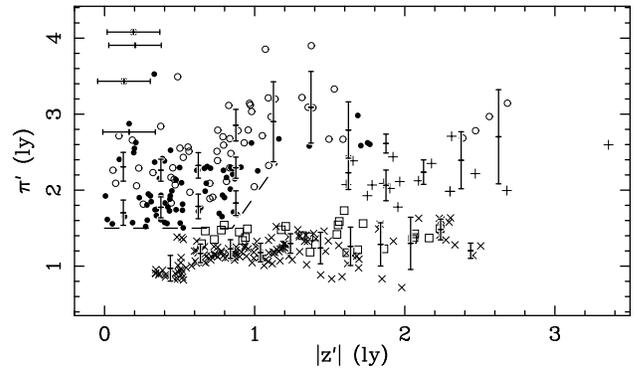}
\caption{Same as Fig.\ \ref{plCS2}, but with data inclined by the
favored CS orientation of $i_x=41\degr$ and $i_y=8\degr$ and an
ellipticity of 0.94 removed from the radii.  Circles that are now
considered part of the hourglass are marked by open squares, and the
case-A only points (Fig.\ \ref{plCS2}) are marked by pluses.  The
average radial profiles to each structure are overplotted.  The dashed
line shows the inner edge of the belt, used in \sect{sec-CS-rings}.
Note that there are many binned averages in this plot, which are
discussed in the \sect{sec-CS-shape}.
\label{plCS3}}
\end{figure}

The cylindrical-radial profile for the data oriented about this new
axis is shown in Figure \ref{plCS3}.  An ellipticity of $b/a=0.94$ has
also been removed from each radius, based on its position angle along
the elliptical hourglass fit given above. Extended flux points that
are now associated with the hourglass are plotted as squares, and the
points belonging only to the case A distribution are shown with plus
marks.  A large subset of the extended-flux echoes (circles) are
concentrated within 1 ly of the equatorial plane $(z'=0)$, between
1.5--4 ly from the SN, and appear to form a thick equatorial waist or
belt circumscribing the hourglass.  As in the previous two sections,
we measure the average radial profile of these points in bins along
$|z'|$.  There is much structure to these data, which requires the
fitting of six profiles.  These, and their interpretation, are
discussed in the next section.

\subsection{A complete picture of the CS nebula \label{sec-CS-shape}}

Surrounding SN~1987A is a complex CS nebula, inclined at $i_x=41\degr$
south and $i_y=8\degr$ east of the line of sight, translated
$\lesssim0.1$ ly to the west of the SN, with an elliptical
cross-section ($b/a\simeq 0.94$) whose semi-major axis is rotated
9\degr clockwise from east.  We visualize this nebula by reflecting
each radial profile about the equator, revolving it around the
cylindrical axis, then reinclining it by the best-fit orientation.
The ellipticity removed in making Figure \ref{plCS3} was reintroduced
when revolving the profiles about the cylindrical axis.  The result,
rendered in Figure \ref{3dpl}\pant{i--j}, shows the probable structure
of the CS material.  Colors are discussed below.  The next two panels
\panp{k--l} compare the complete structure in grey to the observed
echoes.  Here, colors are similar to Figure \ref{3dle}\pant{g--i},
only now all hourglass points are colored red, and the Case-A only
points (Fig.\ \ref{plCS2}) are colored green.  For completeness, we
show only the CS hourglass in panels \panp{m}--\panp{p}.

\begin{figure}\centering
\includegraphics[height=2.in,angle=0]{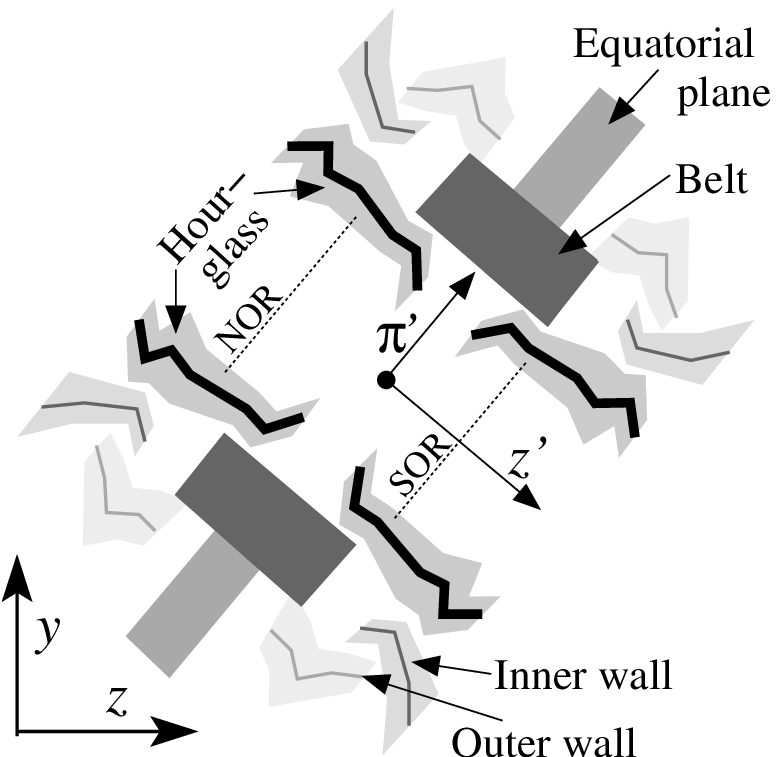}
\caption{As Fig.\ \ref{CD_toon}, but for the CS echoes, with the
 orientation arrows measuring 2 ly.  The expected positions of the ORs
 are indicated by thin dotted lines \secp{sec-CS-rings}.
\label{CS_toon}}
\end{figure}

Figure \ref{CS_toon} shows a scaled cartoon of the relevant CS
features.  In the following subsections, we discuss each feature based
on its radial profile and three-dimensional shape.

\subsubsection{The CS Hourglass \label{sec-CS-shape-hourglass}}

The innermost radial structure is the hourglass, shaded red in Figure
\ref{3dpl}\pant{i--j} and shown by itself in panels
\panp{m}--\panp{n}.  In Figure \ref{plCS3}, these are the crosses
($\times$) and squares within about $\pi'\le1.5 ly$.  This is fairly
cylindrical, with a semi-minor axis of 1.2 ly, tapering close to the
equator and opening just slightly at large $z'$.  The profile is
well-described by a line of slope $\Delta\pi'/\Delta z'=0.14$ and
zeropoint $\pi'=1.02$, suggesting an opening angle of the wall of only
8\degr.  A very small number of points suggest that it also tapers
around $z'=1.5-2$ ly, but this feature is marginal at best.

A number of small-scale features are worth noting in the hourglass.
The northern-lobe points extend to a relative maximum in radius
($\pi'\sim1.5$ ly) at $|z'|=1.4$ ly, tapering slightly in radius on
either side; a similar feature is also seen around $|z'|=0.9$ ly.  In
Figure \ref{plCS3}, many of the extended-flux points lie just outside
the average position of the hourglass.  Since most of these points are
part of the southern lobe, this could suggest the southern lobe is
wider than its northern counterpart.  A more likely interpretation is
that many of the southern-lobe points represent dust just outside the
densest part of the hourglass, which was illuminated prior to our
earliest epoch (see Fig.\ \ref{3dpl}\pant{p}).

Consistent with CKH95, we do not see evidence for a ``capping
surface'' to this structure, as would be expected from a bipolar
peanut-like nebula, the result of many interacting-winds formation
models.  To distinguish this hourglass from the larger one
containing the NH echoes, we refer to each by their association--\ie
the latter is the ``NH hourglass.''

As in previous sections, we address why only a limited amount of this
complete hourglass structure was illuminated by echoes.  Referring to
Figure \ref{3dpl}\panp{p}, only the southernmost portion of the
near (or south) half of the hourglass was probed by the earliest echo.
While the greater majority of the far (or nothern) half of the CS
hourglass lies inside the earliest and latest echo parabolae, the
particular sampling of our observations probed the northern edge only
at the earliest epochs.  A comparison of this panel with Figure
\ref{3de_CS}\panp{b} shows that we have detected roughly all the
material that could have been observed.

\subsubsection{The Equatorial Plane and Belt 
  \label{sec-CS-shape-belt}}

Immediately surrounding the CS hourglass is the thick waist of
extended high-surface brightness, which we call the ``belt,'' and have
colored green in Figure \ref{3dpl}\pant{i--j}.  It extends $\pm1$ ly
along the $z'$ axis, and from 1.5 to 2.5 ly in radius.  To better
represent the extent of this material, two profiles have been fit in
Figure \ref{plCS3}, binned both in radius and along the axis.  Beyond
the outer radius of this belt, echoes lie fairly well constrained to
the equator, tracing a thinner (0.5 ly thickness in $z'$) but extended
equatorial plane to an outer
radius of $\pi'=4$ ly.  This feature has been colored blue.
The belt does not appear to taper smoothly
into the thinner equatorial plane, but rather has a fairly sharp
transition at its outer boundary.

\subsubsection{The Inner and Outer Walls  \label{sec-CS-shape-walls}}

The rest of the CS material lies along one of two ``walls.''  In
Figure \ref{plCS3}, the innermost of these is marked mostly by plusses
at $|z'|>1.5$ ly, while the outer wall is marked with circles from
$0.9\le |z'|\le 1.9$ ly.  The inner wall extends 1.5 to 2.6 ly along
the axis, and may be considered to join with the hourglass around
$|z'|=1.5$ ly.  The profile is positively sloped, which translates
into a wide cone with half-opening angle of 40\degr.  This may simply
trace a decreasing dust-density profile extending out from the sharp
peak of the hourglass.  A single datum is located at $|z'|=3.4$ ly,
and it is unclear whether this can be trusted as a real feature.

The outer wall bridges the belt and the inner wall at radii between
2.6 and 3.4--4.0 ly. These have been colored cyan and gold in Figure
\ref{3dpl}\pant{i--j}.  Unlike the inner wall, the average profile is
concave, reaching its maximum radius roughly 1.4 ly along the axis.
This corresponds to a local maximum seen along the hourglass, and to
the expected positions of the ORs \citep{CH00}.  Noting the
conspicuous lack of echo signal between the hourglass and outer wall
at this position, we question whether this is indicative of episodic
mass loss, and if the OR formation mechanism is somehow related.

\subsection{The Geometry of the Outer Rings \label{sec-CS-rings}}

While great attention has been placed on the geometry of the ER
\citep{Jak91,Pla95,Bur95}, relatively little work exists on the ORs.
In \sect{sec-CS-geom-ER} the ER was used as a constraint to the
geometry of the CS nebula.  Now that a three-dimensional map of the
nebula exists, this problem can be inverted to ask instead how the
echoes can constrain the 3-D geometry of the ORs.

\citet{Bur95} reported the NOR is well fit by an ellipse with
$b/a=0.678$, and a semi-major axis of 1\farcs77 oriented at \pa
70\fdg7, while the SOR is fit by an ellipse with $b/a=0.513$, and
semi-major axis 1\farcs84 oriented at \pa 90.  Assuming both rings to
be circular in their reference frames, they are inclined roughly
43\degr and 31\degr to the line of sight, respectively.  Accordingly,
if the SOR has the same inclination as the ER and NOR, it cannot have
a circular shape.  They also note that the segment connecting
the centers of both ORs misses the SN and ER centroid by $\sim0\farcs4$.
\citet{CH00} use these orientations to fit a kinematic model to the
ring velocities as measured in STIS and ground-based
echelle spectra.  This model suggests the NOR and SOR are located 1.32
and 1.04--1.24 ly (depending on whether one forces the SOR to be
circular or not) from the SN along the inclination axis.  

\begin{figure}\centering
\includegraphics[height=2in,angle=-90]{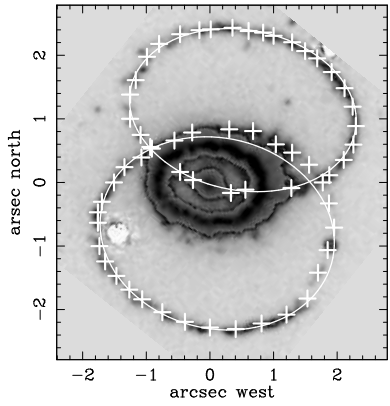}
\caption{The full WFPC2 image of the three rings, from which Fig.\
\ref{plrings_ER} is taken.  To increase the display range, brighter
pixels are displayed with a separate color stretch.
Heavy crosses show the positions of the rings as measured through
radial profiles.  The best-fit planar ellipses to these points are
traced by thin, solid lines.
\label{plrings_OR}} 
\end{figure}

Figure \ref{plrings_OR} shows the complete three-ring nebula from the
same WFPC2 image used to study the ER in \sect{sec-CS-geom-ER}.  We
measured the positions of each OR at various P.A.s by performing the
same radial-profile fitting as that used in measuring echoes.  With
the positions of the rings on the sky known to better than one pixel
(0\farcs023), the only unknown for each point is its line-of-sight
distance from the SN.  If one assumes the ring lies on or about the CS
hourglass, each $z$ position can be determined as the intersection of
the hourglass with the ``tube'' of all possible $z$ positions for a
given point on the ring.

\begin{figure}\centering
\includegraphics[width=3.in,angle=0]{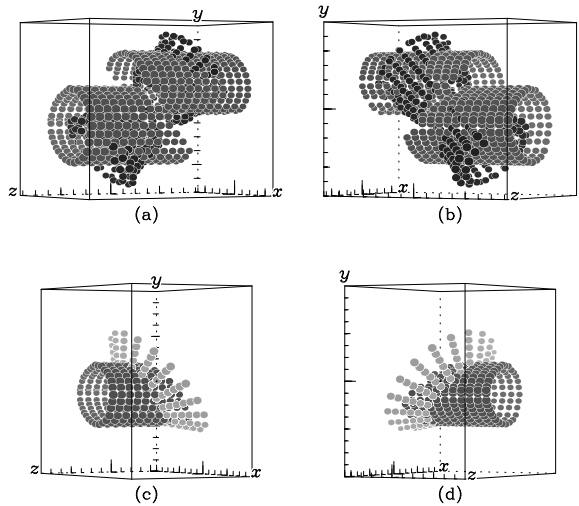}
\caption{Rendered points showing the determination of OR
positions. The NOR and SOR must lie along the upper (opening away from
the observer) and lower (opening toward the observer) medium-grey
tubes, respectively.  One possible location of the rings could be the
intersection of these tubes with the CS hourglass, shown in panels
\panp{a}--\panp{b}.  Panels \panp{c}--\panp{d} show the inner edge of
the belt (light grey), which is a good fit to the SOR.
 \label{3de_ORint}}
\end{figure}

Figures \ref{3de_ORint}a--b show the results of this exercise.  The
red hourglass from Figure \ref{3dpl} is shown in dark grey, while
lighter dots trace all possible $z$ values for each point along the
NOR (upper tube) and SOR (lower tube).  The hourglass intersects both
tubes at many points, but fails to do so along the western side of the
NOR and the entire northern half of the SOR.  If we instead consider
the intersection of each tube with the outer edge of the hourglass
(using the radial-profile error bars in Fig.\ \ref{plCS3}), the NOR is
well fit at all P.A.s by points located at $z'=1.37\pm0.05$, while the SOR
still fails.


We consider two possibilities for the SOR.  First, were the southern
lobe oriented with $i_y\sim0\degr$, it would cleanly intersect the SOR
tube.  It is only by construction that the hourglass has a single axis
of symmetry, unfortunately too little of the southern lobe was
illuminated by echoes to test for such a departure.  Second, rather
than simply lie on or around the hourglass, the SOR may result from
its intersection with another structure.  In Figure \ref{plCS3}, the
dashed line traces a sharp, inner edge to the belt, which nearly
intersects the southern lobe of the hourglass (traced with squares).
When the diagonal segment of this boundary is revolved around the
hourglass axis (light-grey in Fig.\ \ref{3de_ORint}c--d), it
intersects the SOR tube continuously at all P.A.s at a distance of
$z'=1.0\pm0.05$ ly from the SN.  We question whether the SOR is some
sort of density enhancement resulting from the interaction between the
hourglass flow and the inner edge of the belt.

This same belt material makes a poor fit to the NOR, and we can not
readily identify a similar intersection of observed structures.
However, echoes from material behind the SN could have been missed for a
variety of reasons, including source confusion and small scattering
efficiency.

\begin{figure}\centering
\includegraphics[height=2.5in,angle=0]{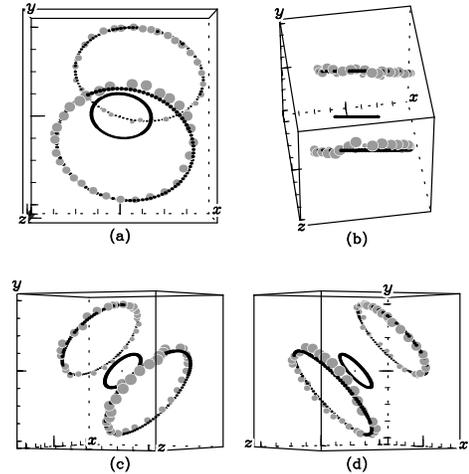}
\caption{Probable positions of the ORs (grey) and the planar,
  elliptical fits (black) to each ring, viewed \panp{a} face-on,
  \panp{b} 60\degr east, \panp{c} 60\degr west, and \panp{d}
  orthogonal to the inclination axis.
 \label{3de_ORs}}
\end{figure}

The 3-D positions of the ORs are rendered in Figure \ref{3de_ORs}, and
appear reasonably planar.  We approximate both rings as planar
ellipses (Appendix \ref{app-model-LM-ell}), fit to each
distibution of points viewed along the CS inclination axis.  The
resulting parameters are listed in Table \ref{tbl-CS-rings}, where
$(x'_0,y'_0)$ are the centroid offsets from the axis, $z'_0$ is the
distance to the SN along the axis, $(a,b)$ are the semi-major and
minor axes, and $\phi$ is the \pa of the major axis from north.  For
completeness, the parameters for the ER \secp{sec-CS-geom-ER} are also
listed.

\begin{deluxetable}{l c c c c c c}
\tablewidth{0pt}
\tablecaption{Best-fit ellipses to the CS rings \label{tbl-CS-rings}}
\tablehead{
\colhead{Ring} & \colhead{$x'_0$} & \colhead{$y'_0$}  & 
  \colhead{$z'_0$}  & \colhead{$a$}  & \colhead{$b$} & \colhead{$\phi$} \\
\colhead{} & \colhead{(ly west)} & \colhead{(ly north)}  & 
  \colhead{(ly)}  & \colhead{(ly)}  & \colhead{(ly)} & \colhead{(\degr)}
}
\startdata
ER   & 0.015    & 0.0    &  0.0   & 0.647 & 0.98   & 81.1  \\
NOR  & 0.26     & 0.04   &  -1.36 & 1.42  & 0.94   & 70.5  \\ 
SOR  & 0.19     & 0.06   &  1.00  & 1.59  & 0.92   & -1.1 
\enddata
\end{deluxetable}

These fits are indicated in black in Figure \ref{3de_ORs}, and by thin
white lines in Figure \ref{plrings_OR}.  The approximation to the NOR
is quite good, but the SOR ellipse fails to intersect the points that
pass just north and west of the ER.  This is not suprising, since no
single ellipse can fully reproduce the observed shape of the SOR
unless it is non-planar.    Since these ellipses are consistent with
both the CS structure and observed three ring nebula, we have used
them in the echo renderings throughout this work.

\begin{figure*}\centering
\includegraphics[width=\linewidth,angle=0]{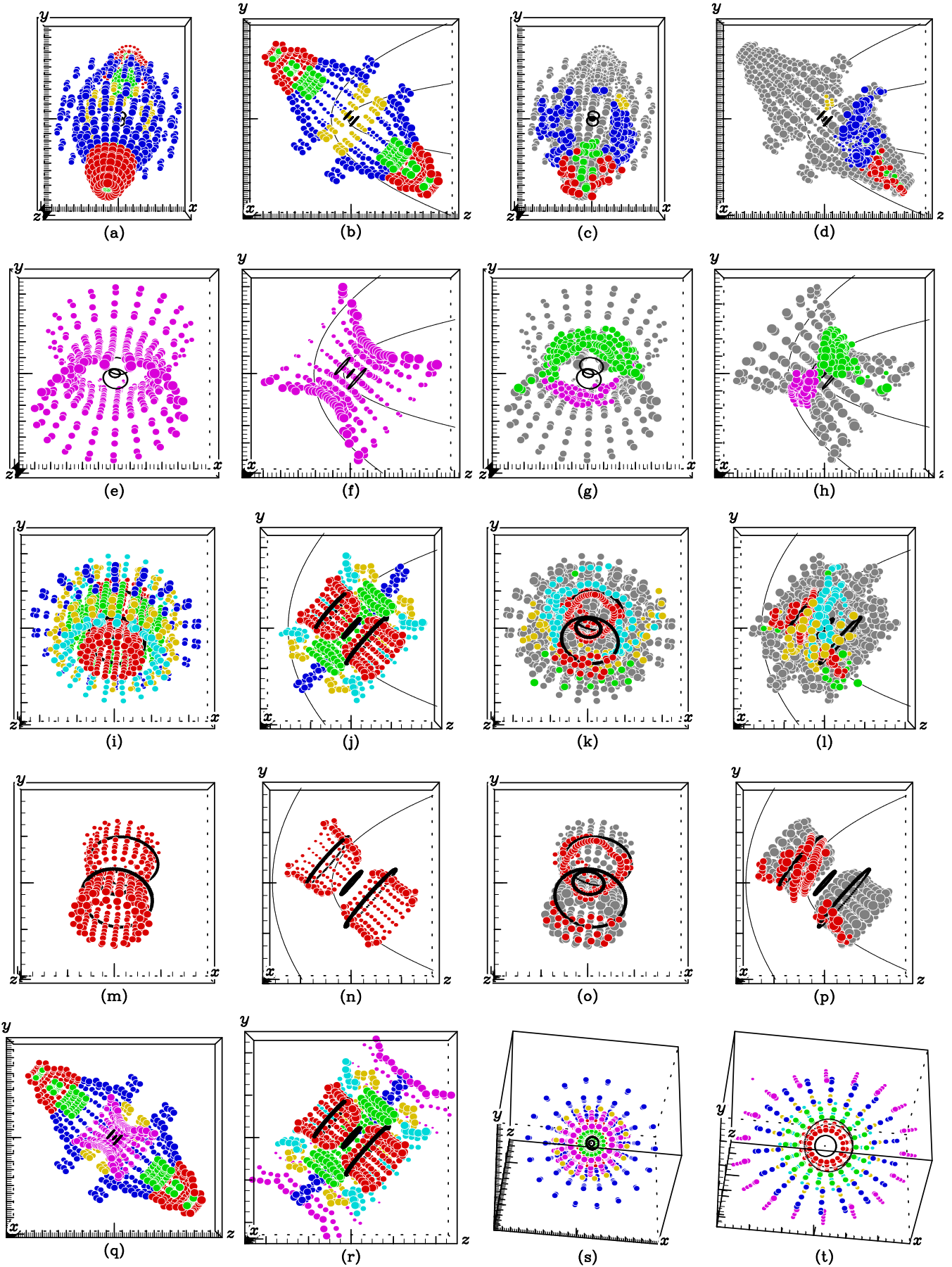}
\caption{Rendered views of the probable structures containing the
 observed echoes.  {\em Top Row}: The CD nebula.  {\em Second Row}:
 The NH nebula. {\em Third Row}: The CS nebula.  {\em Fourth Row}: The
 CS hourglass.  For these top four rows --- {\em Left column:} the
 face-on view of the complete structure; {\em Second column:} the
 western half, viewed from the east, showing a clear view of the
 interior; {\em Right columns:} face-on and side views of the complete
 structures in monotone grey, overlaid with actual echo points from
 Figure \ref{3dle}.  For the bottom row --- \panp{q} The western
 halves of the CD and NH, and \panp{r} NH and CS nebulae, viewed from
 the east.  \panp{s} The northern halves of the CD and NH, and
 \panp{t} the NH and CS nebulae, viewed along the inclination axes
 from the south.  See text for colors.
\label{3dpl}}
\end{figure*}

\section{Density and Mass of the CSE }\label{sec-density}

\subsection{A complete picture of the CSE of SN~1987A
  \label{sec-density-model}}

The probable geometry of the structures containing all echoes within
25\arcsec of SN~1987A are rendered in color in Figure \ref{3dpl}, and
summarized in Table \ref{tbl-density-geom}.  The final column gives
the approximate total volume of the structure containing each echo,
found by integrating the radial profiles in Figures \ref{plCD4},
\ref{plNH4}, and \ref{plCS3} along, and revolved $2\pi$ about, the
respective inclination axes.

\begin{deluxetable*}{l c c c  c c c  c c c c c c c }
\tablecaption{Summary of Echo Geometries \label{tbl-density-geom}}
\tablewidth{0pt}
\tablehead{
\colhead{Struc-} & 
 \multicolumn{3}{c}{Inclination\tablenotemark{a}} & 
 \multicolumn{3}{c}{Cross Section\tablenotemark{b}} &
 \multicolumn{6}{c}{Geometry\tablenotemark{c}} 
\\
\colhead{ture} & 
 \colhead{$i_x$} & \colhead{$i_y$} & \colhead{$i_z$} &
 \colhead{$b/a$} & \colhead{$\Delta x_0$} & \colhead{$\Delta y_0$} &
 \colhead{Shape} & \colhead{$r$} & \colhead{$z'$} & \colhead{$\pi'$} &
    \colhead{$\Delta\pi'$} & \colhead{$V_{tot}$} 
\\
\colhead{} & 
 \colhead{(\degr)} & \colhead{(\degr)} & \colhead{(\degr)} &
 \colhead{} & \colhead{(ly)} & \colhead{(ly)} &
 \colhead{} & \colhead{(ly)} & \colhead{(ly)} & \colhead{(ly)} &
    \colhead{(ly)} & \colhead{(ly$^3$)} 
}
\startdata
CS & 41\degr & -8\degr & -9\degr & 0.94 & $\lesssim 0.1$ & 0. &
   Hourglass & $1-2.9$ & $0.3-2.5$ & $0.8-1.6$ & $0.5-1$ & 20  \\
   &         &         &        &      &                &    &
   Belt      & $1.5-2.8$ & $0.-1.0 $ & $1.5-2.6$ & $1.5-2.0$ & 40  \\
   &         &         &        &      &                &    &
   Walls     & $1.8-4.2$ & $0.8-2.7$ & $1.8-3.6$ & $2.0-3.0$ & 120  \\
NH & 40\degr & -7\degr & 12\degr & 0.82 &  0. & 0. & 
   Hourglass & $3-11$  & $0.7-7.5$ & $2.8-7.5$ & $2-4$ & $1300$ & \\
CD & 40\degr & -8\degr & \nodata & $\gtrsim 0.95$ & $<-1$ &
   $\lesssim -0.1$ &
   Prolate   & $10-28$ & $4-28$ & $0-16$ & $4-5$ & $1.2\times10^4$  
\enddata
\tablenotetext{a}{PYR angles applied in order of
 roll-pitch-yaw \secp{sec-lea-analy-coo}.} 
\tablenotetext{b}{$\Delta x_0$ and $\Delta y_0$ are the western and
   northern offsets of the symmetry axis from the SN, which is at
   $x=0$, $y=0$.}
\tablenotetext{c}{See Fig.\ \ref{axis} for coordinate definitions.  
 $\Delta\pi'$ is the approximate width along $\pi'$. 
 $V_{tot}$ is the inferred total volume of the structure.}
\end{deluxetable*}

The east/west and north/south inclinations are quite consistent,
suggesting all circumstellar material shares a common inclination of
$i_x=40\degr$ south and $i_y=8\degr$ west.  The inner two structures
(CS and NH) both have elliptical cross sections (see Figure
\ref{3dpl}\pant{s--t}), although the ellipticities and orientations
differ.  However, it is unclear whether this particular result is
significant.  As noted in Appendix \ref{app-model}, $\chi^2$ minimization
does not readily allow for errors in multiple dimensions, yet the
measured echo positions have finite width and depth.  Rather than
attempt much more complicated (and error-prone) techniques, such as
iteratively reweighted least squares with robust estimators, we give
these results with the understanding that the CS and NH ellipticies
are most-likely bounded between the values reported for the two
structures.

The CS gas is nested neatly within the NH material.  The extended CS
equatorial plane connects to the waist of the NH hourglass, forming a
continuous structure from 1.5 to 5 ly from the SN.  Tracing NH out
from the waist, there are inward-pointing spurs above and below this
equatorial plane that connect to the outer CS wall, which itself
connects to the belt.  The CS belt, equatorial plane, outer wall, and
the waist of NH are therefore consistent with a single, uniform thick
waist, extending inward to the CS hourglass.  This is shown in Figure
\ref{3dpl}\pant{r}.  

The outer edges of the NH hourglass almost reach the inner spur of the
CD shell, and indeed the echoes as viewed on the sky do connect at
early times (Fig.\ \ref{dimages}).  As shown in Figure
\ref{3dpl}\pant{q}, no material was illuminated along the equatorial
plane exterior to NH.\ \ Unless this region was shadowed from the SN
by material at smaller radii \secp{sec-density-density-shadowing},
there is no higher density gas outside the waist of NH, suggesting the
NH hourglass is the pinched waist of the CD.

The one caveat to this argument is that just above the equatorial
plane and outside the NH hourglass are the northwestern WFPC2 echoes,
whose relationship to the rest of the CD material was unclear in
\sect{sec-CD}.  Shown as gold in Figure \ref{3dpl}\pant{m}, these
would have formed a cylindrical band outside the waist of the NH
hourglass.  We do not believe these are image artifacts, nor part of a
distinct, uniform, and rotationally-symmetric structure.  Rather, they
may indicate local inhomogeneities in the undisturbed CSM, or they may
simply trace the outermost edge of the contact discontinuity.  In the
following, we will exclude them from our analyses.

\begin{figure}\centering
\includegraphics[height=2in,angle=0]{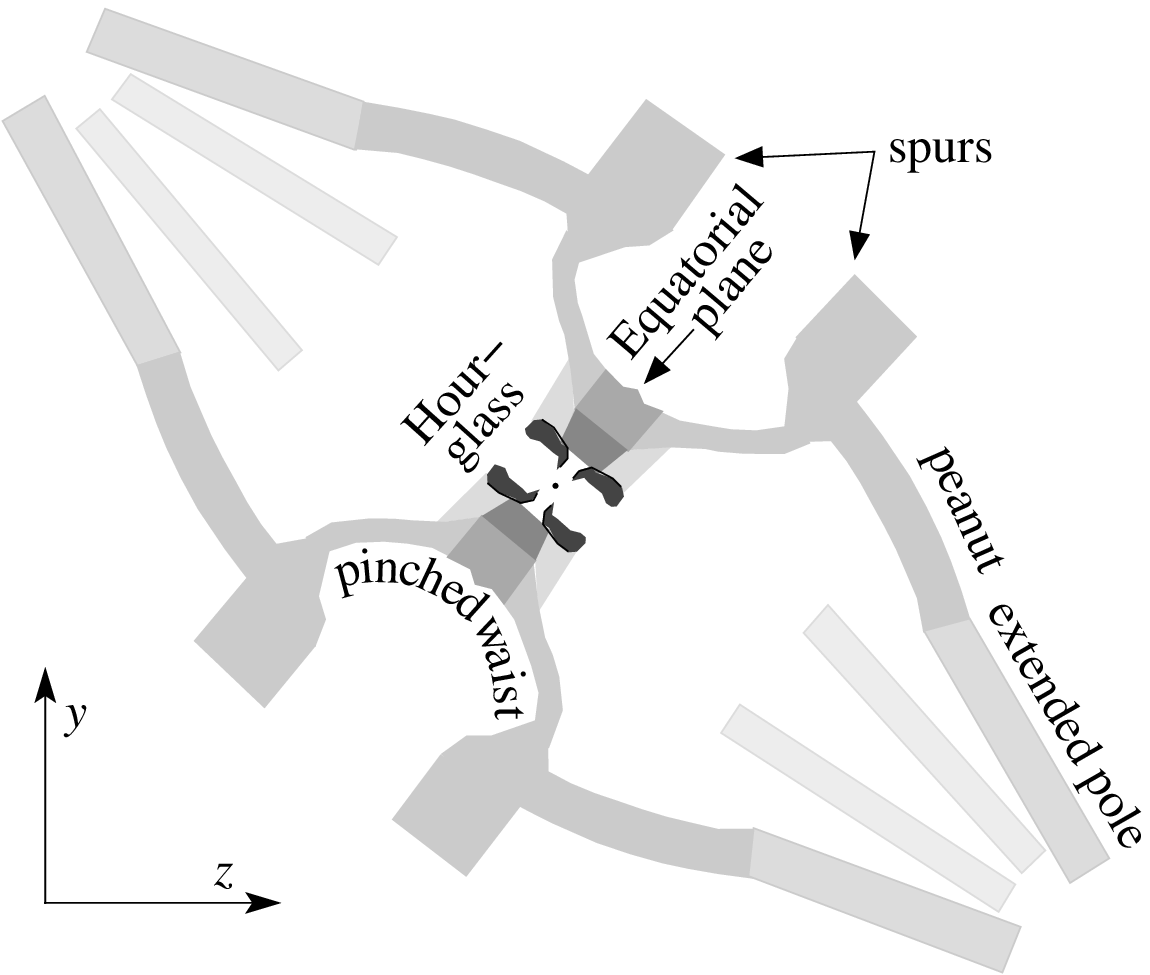}
\caption{Revised cartoon showing the simplified CSE suggested in
Figs.\ \ref{CD_toon}, \ref{NH_toon}, and \ref{CS_toon}.  Structures
are shaded to indicate density, which increases with greyscale.
Fig.\ is to scale, and orientation arrows indicate 10 ly.
 \label{all_toon2}}
\end{figure}

Figure \ref{all_toon2} shows all circumstellar material, simplified
according to the above arguments.  The outer structure is peanut-like,
extended along the poles to $r\lesssim28$ ly, and narrowly-pinched at
the waist at $r\sim5$ ly.  We will henceforth refer to this outer
shell (traced by the CD and outer NH echoes) as the ``Peanut.''
Roughly 35\degr from the equatorial plane are the spurs, extending
$3-4$ ly out from each lobe, perpendicular to the Peanut's axis.
Unlike the northwestern WFPC2 echoes, these spurs do appear at more
than one location, but they are limited enough in extent that it is
unclear whether they are uniform features that encircle the entire
structure, or are isolated clumps.

Inside the equatorial region of the Peanut is a thick ($\sim 3.5$ ly
in radial thickness, 2 ly in axial length) annular ring (or belt),
which extends inward until it terminates at the inner CS hourglass.
The extent of this waist in $z'$ is unclear.  Echoes were detected
outside the CS hourglass at most equatorial ($\pi'$) radii between 1.6
and 4 ly, with a vertical ($z'$) distribution extending at least as
far as the CS hourglass itself.  Still, the bulk of material is within
1 ly of the equator.

\subsection{Density of the Circumstellar Material
 \label{sec-density-density}}

To constrain the gas density and dust composition of the CSE, we apply
the dust-scattering model from \sect{sec-le-model} using the surface
brightnesses of each echo discussed in \sect{sec-CD}--\ref{sec-CS}.
Fluxes were only measured in the F470, F612, and F688 filters.  Our
integrated SN spectrum does not extend to 8090\AA, thus the model
cannot be applied to the F809 filter without making considerable
assumptions.  We chose not to include measurements from WFPC2 data for
a number of reasons.  First, echoes were very extended and faint,
resulting in large photometric errors. Second, intercomparing the
observations from wide-band WFPC2 filters with the narrow-band,
continuum-only ground-based data is challenging, due to e.g.\
potential line contamination, different filter wavelengths and widths,
etc.

\begin{figure}\centering
\includegraphics[height=3.25in,angle=-90]{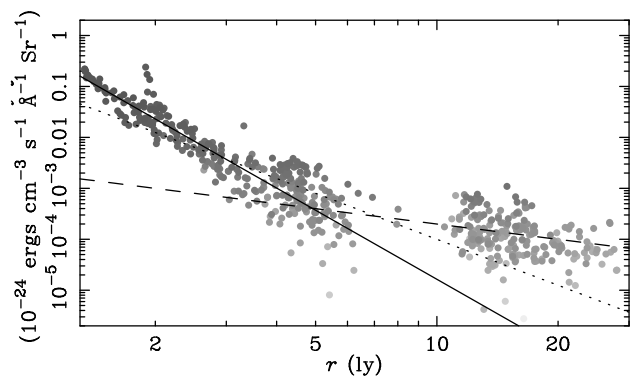} 
\caption{Flux density of scattered light per unit volume as a function
of 3-D radius from the SN, measured for each echo in the F612
filter.  Points are shaded in proportion to their signal-to-noise,
which increases with darker greyscale.  Lines are plotted as follows:
$j\propto r^{-1}$ (dashed), $j\propto r^{-3}$ (dotted), $j\propto
r^{-4.5}$ solid.
 \label{volemm}}
\end{figure}

Figure \ref{volemm} shows the volume emissivity $j$ (flux density per
unit volume) of all echoes in F612 as a function of radius.  Note that
2-D surface brightness was converted to 3-D flux density using the
line of sight echo depth $\Delta z$ given in Eq.\ \eqp{sec-le-width3}.
This can be compared with Figure 3 of \citet{CK91}.  If the density of
material is constant, $j$ will decrease as $r^{-1}$ (Eq.\ [12] of
Sugerman 2003), while a uniform wind blowing into a homogenous, low
density medium will produce an $r^{-3}$ emissivity profile.  These are
shown in the figure as dashed and dotted lines.  The outer-NH and CD
echoes are well-described as constant density, while some NH/CS
material may follow an $r^{-3}$ profile.  The best fit to the CS and
inner-NH points follows $r^{-4.5}$, which is suggestive of a disturbed
CSE, such as would occur via interacting stellar winds.

\begin{figure*}\centering
\includegraphics[width=0.95\linewidth,angle=0]{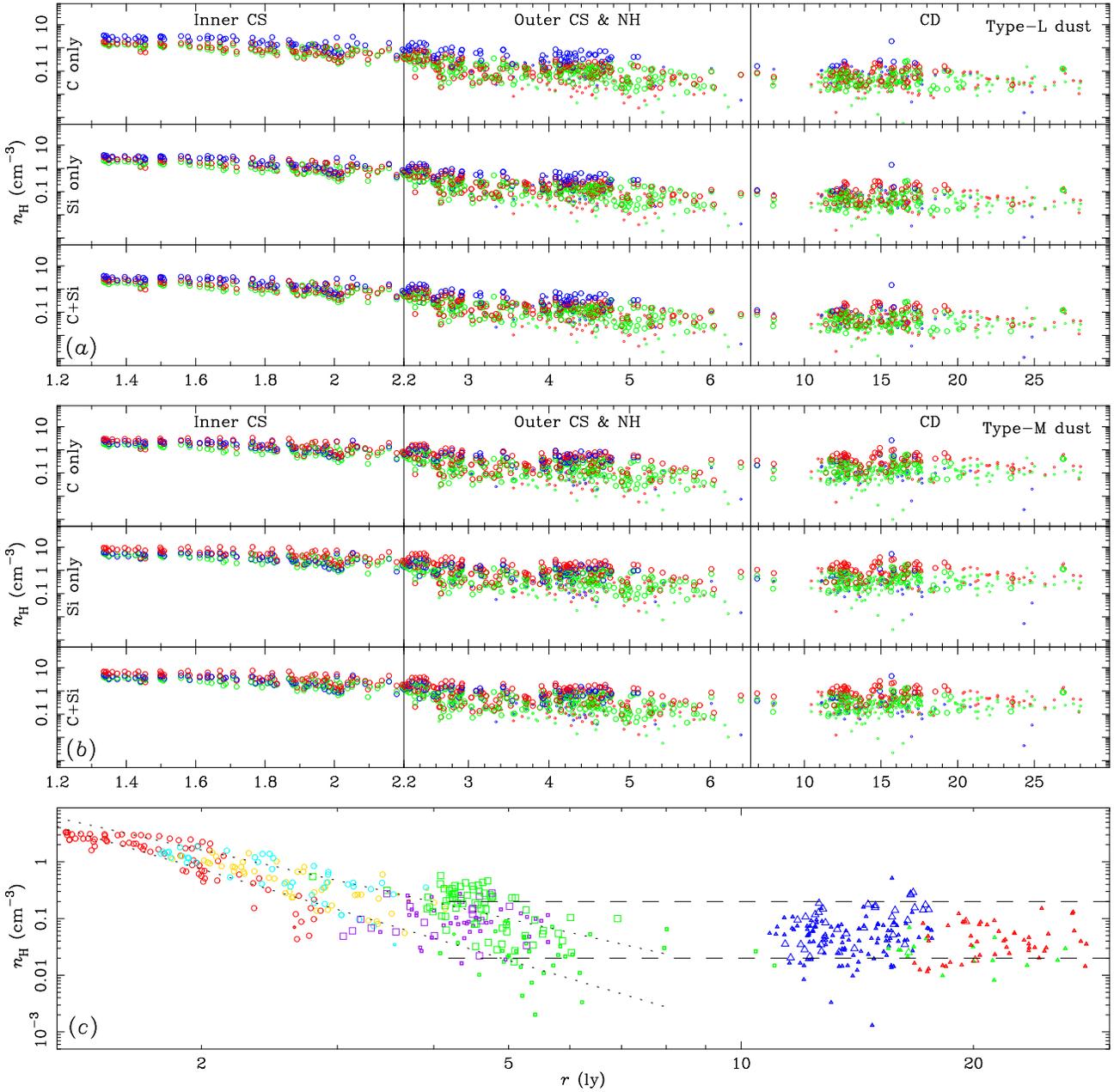} 
\caption{Gas density versus radius for the dust scattering model.
Model results are shown for \panp{a} type-L and \panp{b} type-M dust.
Each filter is shaded its (roughly) corresponding color: F470 in blue,
F612 in green, F688 in red.  Large circles denote signal-to-noise
above 2, and small ones show $S/N<2$.  The absissca changes between
columns, each of which contains the structure as noted in the top
subpanel.  Rows show the results for the dust compositions listed to
the far left.  \panp{c} Average gas densities that match observed echo
fluxes to the dust-scattering model.  Note that the abscissa is shown
in now logarithmic.  Point colors correspond to those in Fig.\
\ref{3dpl}.  CS material is marked by circles, NH by squares, CD by
triangles, and smaller symbols denote measurements with $S/N<2$.
Dotted lines are the best-fits through the upper and lower CS loci,
corresponding to $n\propto r^{-3.0}$ (upper) and $n\propto r^{-4.5}$
(lower).  Horizontal dashed lines delineate the rough density
boundaries of the CD gas.
\label{enh}}
\end{figure*}

Recall from \sect{sec-le-model} that we defined three dust-grain size
distributions: S, M, and L.  We begin with type-L dust.  The results
are shown for pure-carbonaceous (C-only), pure-silicate (Si-only), and
an LMC-abundance mixture (Ci+S) in Figure \ref{enh}\panp{a}.  For
clarity, densities are plotted logarithmically but positions are given
linearly, with the horizontal scale changing between each vertical
line, which roughly delineate CS, NH, and CD material.

The C-only dust yields a close fit between F612 and F688 for the CS
material, but the F470 is uniformly too dense.  An overdensity implies
a large scaling $n_{\rm H}$, which results from an underluminous
prediction.  Since this dust does not produce enough blue flux, the
model is too red.  The Si-only model reconciles the F470 and F688
points, but predicts too much flux in F612.  Adding carbon to
silicates reddens the spectrum, which again
pushes F470 densities to be slightly too high.  Overall, the Si-only
is the better fit.

For the most part, the NH densities vary in a similar fashion.  The
C-only model is too red, while the Si-only dust is a very good
fit. Adding C to Si reddens the model slightly, making the F470
densities slightly more discrepant with the red points.  The
Si-only model is again the better fit.  Finally, the CD echoes are
well fit by all three models, with the C+Si model having the lowest
scatter between filters.

Figure \ref{enh}\panp{b} shows the type-M model results.  For the CS
structures, the C-only model is a reasonable fit to the F470 and F688
data, yet the F612 flux is too high.  Si-only dust is overly red, and
adding carbon does little to improve the fit.  Since smaller particles
do not scatter enough red flux, the CS dust is better fit by a
distribution including some large grains.  The NH material is
reasonably-well fit by all three models, but the least
dispersion between filters occurs when silicates are present.  Type-M
dust is not a good fit for the CD, since this model is overluminous in
F612 and too blue.  Although not shown, the fits of the type-S model
are poor since the spectrum is too blue.

The CD colors require large grains, and are best-fit by the type-L
C+Si model.  Neither L nor M models are satisfactory fits to the NH
and CS echoes, which leads us to explore intermediate grain-size
ranges.  We tested a number of upper size limits between
$a=0.1-2.0$\micron, and find that smaller values fit better.  The CS
and inner NH data ($r<5$ ly) have the lowest dispersion between
filters for Si-only, $a_{max}=0.2$\micron dust.  We will denote this
new size range as ML, as it includes particles slightly larger than
the type-M dust.  Outer-NH data ($r>5$ ly) are more consistent with
Si-dominated dust with larger grains.  The general trend is that inner
echoes are better reproduced by smaller, Si-dominated dust, and with
increasing distance, the grain sizes and C-content increases.

\citet{Fis02} have modeled mid-IR emission from collisionally-excited
dust grains heated in the shocks between the \ion{H}{2} region
interior to the ER and the forward blast of the SN.  They find the
emission is best explained by small ($a\lesssim0.25\micron$) grains
with a Si-Fe or Si-C composition.  The dust abundance is quite low,
which they attribute to evaporation from the UV flash and sputtering
in the shocked gas.  This further constrains grain sizes to
$a\lesssim0.25\micron$ and excludes a pure-carbon composition.  The
dust in both this \ion{H}{2} region and the CS hourglass was formed
from material expelled very late in the progenitor's life, thus we
expect the dust properties to be similar.  Indeed, our CS-dust model
also favors silicate-dominated\footnote{``Astronomical silicate'' dust
parameters used in our model \citep{DL84,WD01} are derived from
olevine, a magnesium-iron-silicate.} dust with small grains
($a<0.2\micron$).

Our models suggest that dust which formed at later times (and is
consequently closer to the star) has a smaller carbonaceous-component.
This can result from a change in surface CNO abundances over the
star's late-stages of evolution, since carbon-rich envelopes create
carbon-rich dust, while oxygen-rich envelopes create silicate-rich
dust.  Such CNO processing is also inferred from early IUE spectra
\citep{Fra89} of the ER, which show nitrogen and oxygen to be
overabundant with respect to carbon.

The average dust density, taken from the best model results above is
plotted in Figure \ref{enh}\panp{c}.  The innermost CS hourglass
material has a relatively-constant gas density of $n_{\rm H}=2-3$
cm$^{-3}$ up to $r\sim1.6$ ly.  We note that this value is consistent
with the findings of \citet{Bur95}, who argued the CS hourglass must
have a density $\lesssim5$ cm$^{-3}$ to not be observed in
recombination in WFPC2 images.  

Beyond this position, the material splits into two distributions, for
which density drops either slowly ($n_{\rm H}\propto r^{-3.0\pm0.2}$)
or rapidly ($r^{-4.5\pm0.2}$) with radius.  Material outside the
hourglass tends to lie along one of these two distributions, up to a
radius of roughly 4 ly.  That the density varies more steeply than the
inverse-square expected for a freely-expanding wind suggests the
mass-loss mechanical luminosity ($\dot{M} v_{exp}^2$) increased with
time toward the end of the RSG.


The densest material traces the waist of the hourglass.  The hourglass
points along the $r^{-3}$ profile are those at large $\pi'$ for a
given $z'$, which lie outside the average hourglass position,
designated by the ``extended flux'' squares in Figure \ref{plCS3}.
Beyond $r=2$ ly, this profile traces the belt and equatorial plane
material.  The steeper, $r^{-4.5}$ profile includes most hourglass
material and extends to include the inner and outer ``walls.''
Allowing larger dust grains or increasing the carbonaceous content of
the dust shifts the densities uniformly lower by only 10\%, thus these
results are reasonably robust against the particular model chosen.

The outer CS gas makes a smooth transition to the inner NH points
around $r\gtrsim 3$ ly.  The two profiles described above also bound
the majority of the NH gas up to a radius of 6 ly.  Excluding a
density enhancement between $r=4-5$ ly, there is no evidence of a
structural distinction between the CS and NH gas, which justifies the
simplified model in Figure \ref{all_toon2}.  Returning to the
aforementioned enhancement, there is a thin shell or wall at $r\sim
4.5$ ly, of thickness $\sim1$ ly and density $n_{\rm H}=0.2-0.3$
cm$^{-3}$, roughly three times higher than the rest of the NH gas.
This high-density material is located primarily along the waist of the
NH hourglass around $\pi'=4-5$ ly, marking an outer edge to the
equatorial overdensity.

The CD density is constant with radius, and bounded by $0.02\le n_{\rm
H} \le 0.2$ cm$^{-3}$, which are delineated by dashed lines Figure
\ref{enh}\panp{c}.  The greater majority of the NH gas is also bounded
by these values, confirming the suggestion from Figure \ref{volemm}
that the outer NH and CD mark a constant-density boundary.  Although
not a significant trend, the inner CD gas (blue conical-shell
material), is at slightly higher density ($n_{\rm H}\sim0.06$) than
the red, radial material at larger radii ($n_{\rm H}\sim0.04$).  The
green (radial jet) points have the lowest density around $\sim0.02$
cm$^{-3}$.

The greyscale shading of structures in Figure \ref{all_toon2}
reflects the differences in density among the many echoing structures,
with darker grey indicating higher-density material.  The picture of
the CSE will be completed in the following subsections, in which we
reexamine the Peanut's pinched waist, measure the density of dust
between the bright echo structures, and estimate the total mass of all
CSE material.

\subsubsection{The Peanut: a Pinched Waist or Shadowing?
 \label{sec-density-density-shadowing}} 

According to Figure \ref{all_toon2}, the CD and NH form a Peanut with
a narrowly-pinched waist.  To interpret this as the contact
discontinuity between the RSG and MS winds, we must assume there is no
disturbed MS material beyond these structures.  The alternative is
that material is present outside the waist of the NH structure, but
that it was not illuminated by the SN pulse because of significant
extinction, or shadowing, from equatorial material interior to the
Peanut.  

Heuristically, this situation is unlikely.  That echoes were observed
within 0.5 ly of the equatorial plane (Fig.\ \ref{3dpl}\pant{h,l})
demonstrates SN light did illuminate this region, so it is difficult
to argue that the entire equatorial region is shadowed from, for
example, the ER.  An alternative is that all material inside NH has
shadowed any material outside it.  Let us assume the waist of the
Peanut (NH echoes) has a lower density than the CS equatorial disk.
Extinction scales with density, so the waist of the Peanut will cause
much less shadowing than the material inside it.  Otherwise stated,
the Peanut itself contributes very little to the total optical depth,
and therefore should not have been illuminated by echoes.  Since NH
echoes were observed at the equator, we conclude that material outside
NH should have been illuminated as well, which is not the case.

If, alternatively, the waist of the Peanut has very high density and
is optically thick, then less light will be scattered toward the
observer than expected in our optically-thin model, and we will
underpredict the actual dust densities by a large margin.  We find the
NH dust has about the same density as the CD shell of the Peanut
in the previous section, which means the {\em actual} NH density must be
just the right amount to {\em appear} to have the same properties as
the CD material.  This seems highly improbable, and leads us to doubt
that any strong shadowing has occurred.  

A conservative estimate of the column density of all equatorial dust,
by integrating $n_{\rm H}\times r$ to a radius of 6 ly in Figure
\ref{enh}\panp{c}, is $N_{\rm H}\sim6\times10^{18}$ cm$^{-2}$.  The ER
was not observed in echoes, however an average gas density of $10^4$
cm$^{-3}$ \citep{Son97} and observed width of 0.3 ly \citep{Bur95}
yields an additional column of $N_{\rm H}\lesssim3\times10^{21}$
cm$^{-2}$.  The total extinction \citep{Sug03} at 6120\AA\ for type-ML
dust will be $A_{6120}=5\times10^{-23}N_{\rm H}$ for C+Si particles,
and 30\% smaller for Si-only.  The total extinction from the SN to NH
will be less than 0.2 mags, which will only attentuate the SN pulse by
13\%.  Since the column density within the ER is over two orders of
magnitude larger than in the CSM, any shadowing from the ER would have
hidden the CS and NH echoes as well.

If the ER is only the ionized skin of a much more extended equatorial
disk \citep{Pla95} then the equatorial extinction will be
correspondingly higher.  Only if this disk has constant density and
extends out to 1.5 ly can the equatorial extinction approach an
optical depth of unity.  Again, this seems unlikely. The qualitative
and quantitative cases for shadowing are dubious, thus the narrow
waist appears to be a genuine boundary.

\subsubsection{An Unbiased Measurement of Density
 \label{sec-density-density-unbiased}} 

The previous results are biased, in the sense densities were measured
only in those regions known to be dense.  Any dust intersecting an
echo parabola will scatter light toward the observer, thus if any is
present between or external to the known structures, its signature
should be recorded in faint, background surface brightness.  To probe
these regions, we applied the dust-scattering model to unbiased
photometry, as follows.

\begin{figure}\centering
\includegraphics[height=3.25in,angle=-90]{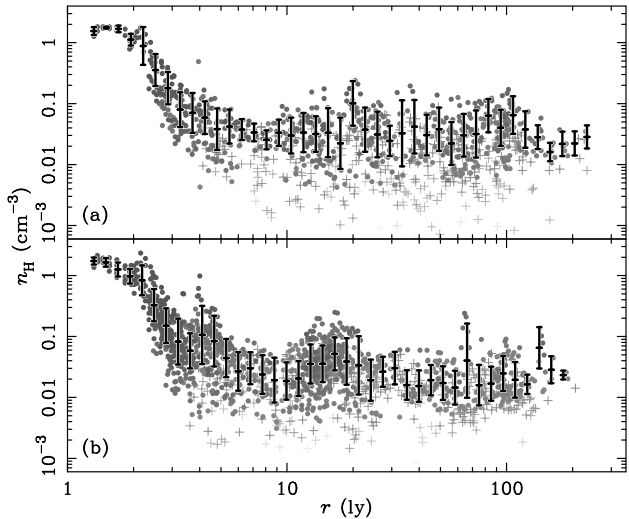} 
\caption{Type-L Si+C dust-scattering model densities for all flux
 within 50\arcsec of the SN, as measured through \panp{a}
 semi-circular annuli centered at \pa 0\degr and 180\degr, and
 \panp{b} 20\degr annular wedges centered at \pas 90\degr, 135\degr,
 180\degr, 225\degr, and 270\degr.  Signal-to-noise increases with
 greyscale, and points with $S/N<1$ are shown as crosses (included
 since our error estimates are very generous).  Weighted averages
 measured through bins of constant $\Delta(\log{r})$ are shown in black.
\label{plenh_ALL}}
\end{figure}

Surface brightness was remeasured in all difference images (Fig.\
\ref{dimages}) through circular-arc annuli from
$0\farcs3$ to 50\arcsec (or 45\arcsec until 1990, as explained below)
from the SN.  Photometry beyond $\rho\sim25\arcsec$ is intrinsically
unreliable, since at these radii we made no effort to keep the
reference images echo-free.  However, with the exception of echo R310
($\rho\sim45\arcsec$ until 1990), all interstellar echoes
\citep{Xu94,Xu95} are located beyond $\rho=50\arcsec$, and therefore
cause no contamination.  The results are shown in Figure
\ref{plenh_ALL}, using the type-L Si+C dust model. Note that this
figure extends to 300 ly in radius, a factor of 10 greater than the
spatial coverage in Figure \ref{enh}\panp{c}.

Panels \panp{a} and \panp{b} differ only in how surface brightnesses
were measured.  Semi-circular annuli, centered at \pa 0\degr and
180\degr, were used in panel \panp{a} to probe uniform, faint
variations in background flux while smoothing over discrete, non-echo
signal.  Panel \panp{b} shows the results from annuli of arclength
20\degr, centered at \pas 90\degr, 135\degr, 180\degr, 225\degr, and
270\degr, which better preserve small or non-circular echoes.  In both
cases, the average profile (black points and error bars) is determined
as the weighted average of densities binned in radius.

A number of known features are recovered, such as the CS material at
small radii, the denser waist of the Peanut around $r=4$ ly, and
the bright CD between $11-22$ ly.  The average density between
the CS and CD structures ($r=4-22$ ly) decreases as $r^{-1}$ in panel
\panp{a}, and $r^{-2.0\pm0.5}$ in panel \panp{b}.  The exact value
of the exponent is sensitive to the particular limits in radius.
Nonetheless, the interior of the CD is filled with diffuse material. 

As expected, the semi-circular apertures smoothed over most of the CD
structure, however, a drop in density around $r=30$ ly in panel
\panp{b} is consistent with its outer boundary.  External to the CD,
the gas density is very uniform with an average value of $0.01-0.02$
cm$^{-3}$, up to a radius of $80-90$ ly, at which there is a marginal
density peak.  These same densities correspond to that expected from a
background-noise-only model at $r>30$ ly, and are thus only upper
limits to the actual density in this region.

\subsection{Total Mass of the Circumstellar Environment
 \label{sec-density-mass}}

The total mass of gas and dust within the CSE is estimated from the
average density and volume of each structure.  The volumes of the
probable structures traced out by light echoes are measured by
integrating the binned radial profiles in Figures \ref{plCD4}, \ref{plNH4},
and \ref{plCS3}.  Each profile is assumed to represent a surface of
revolution, and the limits of integration in $\pi'$ are determined as
a convolution of the standard deviation of the average profile with
the the average spatial extent of each echo position, transformed into
the cylindrical coordinate system.  The resulting values are given in
the final column of Table \ref{tbl-density-geom}.  The volume interior
to the CD is estimated a few ways: as the volume of the ellipsoid
enclosed by the CD, as the volume of two cones enclosed by the
quasi-linear profile, and as the integrated volume under the same
profile (Fig.\ \ref{plCD4}).  All three methods give a consistent
result of about 6500 ly$^3$.

The estimated masses of each structure are listed in Table
\ref{tbl-density-mass}, yielding total mass in all structures of $\sim
1.7$\msun.  Since the gas-to-dust mass ratio is large (400--600), dust
contributes a negligible fraction to the total mass.  Number density
is converted to mass density by $\rho_g=\mu n_{\rm H}m_{\rm H}$ where
$\mu=1.4$ is the cosmic average value.  It is very difficult to
quantify all the sources of error that enter these values, but we
believe the tendency will be to underpredict the masses of the
structures, given conservative measurements of their volume and the
unknown extent to which dust was not observable, due to unfavorable
geometry, poor data quality, confusion, or possible shadowing.

\begin{deluxetable}{c c}
\tablewidth{0 pt}
\tablecaption{Mass of the CSM \label{tbl-density-mass}}
\tablehead{
\colhead{Structure} & \colhead{$M/M_\sun$}
}
\startdata
CS Hourglass & 0.04 \\
CS Belt  & 0.04 \\
CS Walls & 0.06 \\
NH Waist & 0.06 \\ 
NH Walls & 0.07 \\
Intra-CD & 0.2 \\
CD       & 1.2  \\ \hline
Total    & 1.7 
\enddata
\end{deluxetable}

\section{Conclusions}

In this work, we have presented the discovery and analysis of light
echoes illuminating material within 30 ly of SN~1987A.  These results
are also summarized in the first half of \citet{Sug05}, after which we
show that no existing formation scenarios can explain this
newly-revealed circumstellar nebula.

As noted in the introduction, the outflows from Sk $-69$\degr~202 are
one example of a much wider class of bipolar outflows seen in many
different classes of evolved stars.  A small sampling of these include
the supermassive LBV $\eta$ Car, the B1.5ab BSG star 25 Sher in NGC
3603, the symbiotic star He 2-104, the ``Etched Hourglass'' MyCn 18,
and the ``butterfly'' nebula M2-9.  The underlying physical processes
producing these objects are still far from understood.  If the shaping
of bipolar outflows occurs very early in the mass-loss history of a
star, then the seeds of bipolarity or asymmetry have a strong effect
on subsequent outflows, but are also erased by those same winds.
Reconstructing these objects' mass-loss histories is very difficult
since there are few observational constraints extant from their early
stages of evolution.  Yet such ``stellar paleontology'' is critical to
understanding the mechanisms of aspherical stellar-mass loss.

As one researcher recently noted at a conference on asymmetric
planetary nebulae \citep{APN}, ``most astronomers would gladly give a
month's post-doc salary to know the three-dimensional structure of a
bipolar nebula.''  The light echoes from SN~1987A have finally
provided just that, not only for the visible nebula, but additionally
for the products of stellar mass loss from the last
$\gtrsim3\times10^5$ years of the star's life.  It is our hope that
this work will provide both a motivation and a means to readdress first
the mechanisms of mass loss in Sk $-69\degr$~202, and subsequently the
phenomena of bipolar mass loss in many more systems.


\acknowledgements

B.E.K.S. wishes to thank Alex Bergier for assistance in ground-based
image reduction, Robert Uglesich for assistance with difference
imaging, and Eric Blackman, Roger Chevalier, Adam Frank, Peter
Lundqvist, and Geralt Mellema for useful discussions.  We gratefully
acknowledge our referee, Richard McCray, for his critical reading and
insightful feedback on all the manuscripts in this series.  This
research was based in part on observations made with the NASA/ESA
Hubble Space Telescope, obtained from the Data Archive at the Space
Telescope Science Institute, which is operated by the Association of
Universities for Research in Astronomy, Inc., under NASA contract NAS
5-26555.  This paper also made use of publically-available data from
the SUSPECT supernova spectrum archive.  3-D rendering was
accomplished with {\em pgplot}, maintained by Tim Pearson at the
California Institute of Technology.  {\em Synphot} is developed and
maintained by the Science Software Group at STScI.  This work was
generously supported by STScI grants GO 8806, 8872, 9111, 9328, 9428,
\& 9343; NASA NAG5-13081; NSF AST 02 06048; and by Margaret Meixner
and STScI DDRF grant 82301.


\appendix

\section{The {\em difimphot} Data Pipeline }\label{sec-dip-pipeline}

The difference imaging pipeline requires the following steps: (1)
geometric registration, (2) creating the reference image, (3) building
the PSF, and (4) PSF-matching and differencing.  Step (1) is discussed 
in \sect{sec-reduc-reg-reg}; the rest are described below.

\subsection{The Reference Image \label{sec-dip-pipeline-ref}}

A successful implementation of PSF-matched difference imaging will
remove all constant sources of flux, leaving the desired signal from
time-variable sources, and unavoidable contributions from statistical
noise, residuals from bright stars, and cosmetic defects.  Images
within a dataset may be differenced against each other, to test for
relative changes.  Preferably a ``reference'' (or master) image, which
represents the median flux of all stars, may be subtracted from all
input images.  In this implementation, we combine all images in a
given filter to create an ``echo-free'' reference image, as described
in \sect{sec-reduc-reg-stack}.

\subsection{PSF Building \label{sec-dip-pipeline-psf}}

PSF-matching two images requires high signal-to-noise empirical PSFs
with clean profiles well into the wings.  Building such PSFs can be
difficult in sparse and crowded fields.  In sparse fields, finding a
sufficient number of bright stars to build a PSF representing the
average stellar profile can be problematic.  In crowded fields,
isolating bright stars from many nearby, faint sources is the dominant
challenge.  We address both problems by using the PSF-building
software from {\em daophot} in an iterative technique, as follows.

A candidate list of PSF stars is generated from the reference-star
list \secp{sec-reduc-reg-stack}, using only those members that do not
have any bright neighbors within $\gtrsim3\times$FWHM pixels.
Saturated stars are rejected, but stars with non-linear profiles are
kept.  The test PSF is built from the $N$ brightest linear-profile
stars, with non-linear stars used only to supplement the wings.
The {\em psf} algorithm fits an analytic function to each
profile and computes the residuals of the star from that function.  A
weighted average of all profiles, normalized to that of the brightest
star, is the temporary PSF.  This is compared to each input star, and
those with residuals that deviate from the model are down-weighted in
successive iterations.

The output model PSF is fit to each candidate star in the original
list, and up to $N$ of the best fit stars are used as a second input
list to the PSF-building algorithm.  Applying a cut-off to the fit
quality is highly effective at removing stars with faint neighbors
that would otherwise contaminate the wings of the PSF, as well as
removing non-linear and blended stars.

\subsection{PSF Matching and Differencing \label{sec-dip-pipeline-dip}}

Both the input and reference images are evaluated for seeing,
and the better-seeing (i.e.\ lower FWHM) of the two is chosen as 
image $r$.  PSF-matching proceeds by first using equation
\eqt{sec-dip3} to build the convolution kernel from the input and
reference PSFs.

Although we take great care to build high-quality PSFs, the noise in
the wings can be quite high, since these are the lowest-flux
components of the radial profile and are therefore the most-poorly
determined.  As the PSF of a star is roughly gaussian, a fixed
percentage of the PSF wings is modeled and replaced by the FT of an
elliptical gaussian.  The particular percentage to replace depends on
the number and flux of stars used to build the PSF.  In most
applications, we find no more than 10\% of a PSF need be replaced,
with a well-sampled ($\sim20$ stars) PSF needing only a 3\%
replacement.

Image $r$ is convolved with the kernel $k$ to PSF-match it to the
poorer-seeing image $i$.  The photometric scaling $c$ between $i$ and
$r$ is evaluated from the reference stars \secp{sec-reduc-reg-stack}
present in both images.  The difference image is then given by
$i-c(r\ast k)$.  It has been found that images that are already close in
seeing often need to be PSF-matched twice, with the first set of
matched images as input for a second matching iteration.

\section{Numeric Modeling of Data }\label{app-model}

 The generalized problem of 
finding the parameters ${\bf a}=a_j,\:j=1,M$ that best-fit the function
$y(x)=y(x,{\bf a})$ to the $N$ data points $(y_i,x_i)$ is solved by
minimizing the chi-squared merit function 
\begin{equation}
 \chi^2\equiv\sum_{i=1}^{N}\left(\frac{y_i-y(x_i,{\bf a})}{\sigma_i} \right)^2,
\label{app-model1}
\end{equation}
where $\sigma_i$ is the weight of $y_i$.  In cases in which the
function is not linear in $x_i$,  chi-squared
minimization requires an iterative procedure that relaxes $\chi^2$
by variation of parameters ${\bf a}$.  Our
preferred technique is the use of Levenburg-Marquardt (LM)
minimization \citep{Press}.  Briefly, one varies a trial set of
parameters ${\bf a}$ using the method of steepest descent, by
computing the gradient and Hessian of $\chi^2({\bf a})$.

Although very powerful, LM minimization has numerous (yet unavoidable)
pitfalls.  First, the topography defined by $\nabla\chi^2$ can have
many local minima into which the fit can relax, missing the absolute
minimum.  The minimization algorithm is known to wander near a minimum
if the topography is flat. Complicated functions with many parameters
often require that the user hold certain ones fixed while
varying others, lest the fit reside permanently in local minima.
Occasionally, the function can realize a minimized $\chi^2$
through an entirely unreasonable set of parameters.  Despite these
difficulties, the technique is robust and the standard of nonlinear
fitting.

Since the datapoints may not have normally-distributed errors, and
many fits are to functions of more than one dimension, we adopt the
bootstrap technique \citep[see][]{Press} to estimate the uncertainties
in the fitted parameters.  Below, we describe explicitly a number of
non-trivial minimizations used in this work.

\subsection{Gaussians and Moffats \label{app-model-LM-gauss}}

A moffat is a modified Gaussian, and can be expressed in terms of its
peak $A$, center $x_0$, and FWHM $\sigma$ as
\begin{equation}
 y=A\left[1+\frac{4(2^{1/\beta}-1)}{\sigma^2}(x-x_0)^2\right]^{-\beta}
\label{app-model-LM-gauss2}
\end{equation}
where $\beta\rightarrow\infty$ yields a
Gaussian. Any number of moffats can be fit simultaneously, with four
parameters per function.

\subsection{Ellipses \label{app-model-LM-ell}}

\begin{figure}\centering
\includegraphics[width=0.6\linewidth,angle=0]{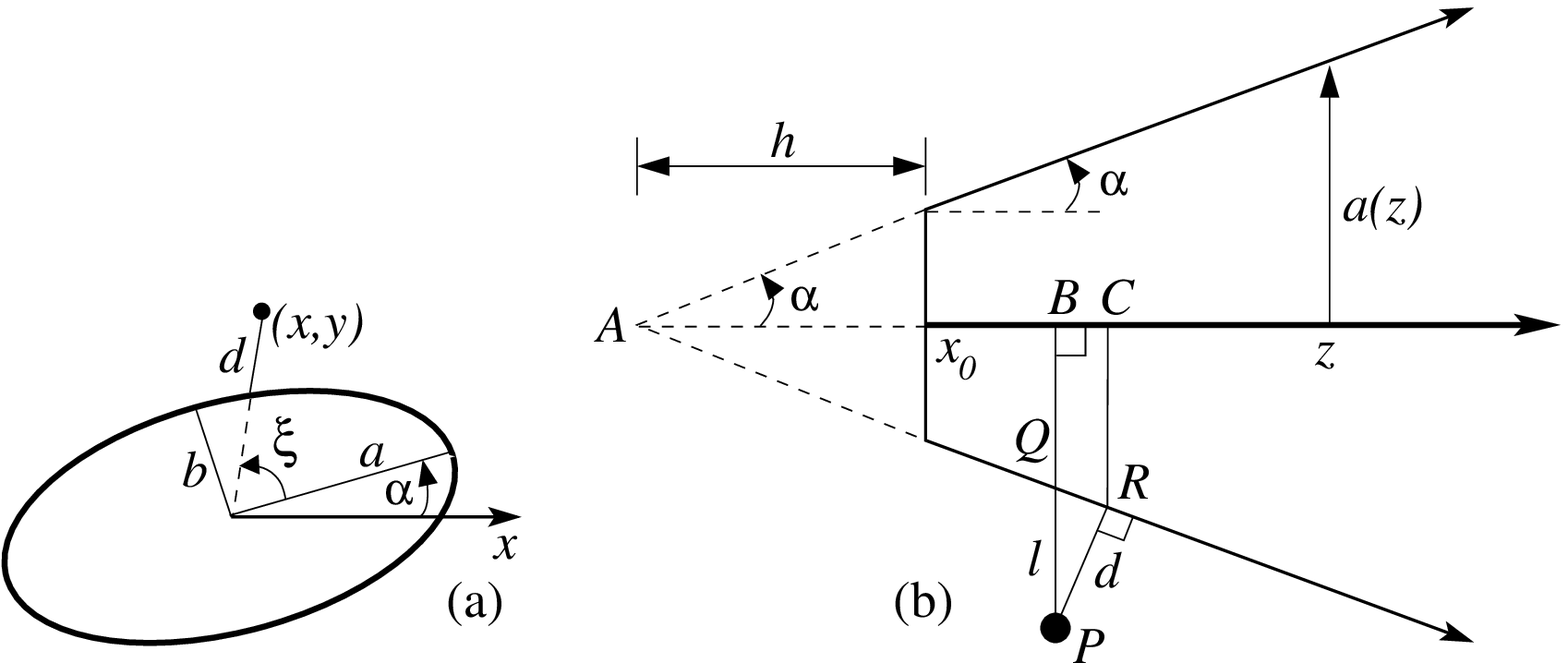}
\caption{Schematic geometries for Levenburg-Marquardt fitting, for
  \panp{a} and ellipse, and \panp{b} a cone or frustum. \label{LMfig}}
\end{figure}

The cartesian expression for an ellipse with center $(x_0,y_0)$ and
semi-major and minor axes $(a,b)$ is not a form condusive to LM
fitting for many reasons, not least of which is that it is
particularly complicated to account for rotations by some angle
$\alpha$ (see Figure \ref{LMfig}a).  Rather, the ellipse can be
expressed in parametric form as
 $(x  =  a\cos{\xi},\:
 y  =  b\sin{\xi} )$
with parameter $\xi$.  To fit an ellipse to a locus of points, we
minimize the geometric distance between the ellipse and the data as
follows.  Data are shifted by $(-x_0,-y_0)$ and rotated by $-\alpha$,
then for each datum, the value of $\xi$ is found that minimizes the
point's distance from the ellipse defined by $(a,b)$.  This specific
minimization can be accomplished by simple numeric bracketing, and as
$\xi$ is only used to find points on the ellipse, not to define it, it
is not fit by the LM algorithm.  The $\chi^2$ merit function is the
sum of the distance of all transformed data from the
ellipse, and is minimized by varying the five parameters
$\{(x_0,y_0),(a,b),\alpha\}$.  

Explicitly, this is a two-dimensional fit for the distance
function
$d=d({\bf x}_i,{\bf a})\equiv\sqrt{X_i^2+Y_i^2}$, where 
\begin{eqnarray*}
X_i & = & x_i-(x_0+a\cos{\alpha}\cos{\xi_i}+b\sin{\alpha}\sin{\xi_i}) \\
Y_i & = & y_i-(y_0-a\sin{\alpha}\cos{\xi_i}+b\cos{\alpha}\sin{\xi_i}). 
\end{eqnarray*}
The absolute minimum distance of all data to the ellipse is zero, thus
we wish to find the parameters such that $d({\bf x}_i,{\bf a})\approx
0$.  In the relevant analog to equation \ref{app-model1},
$\chi^2=\sum_i d({\bf x}_i,{\bf a})^2$.  N.B.\ Here, $\chi^2$ is a
measure of the total distance of all points from the ellipse, and does
not carry statistical interpretations such as absolute
goodness-of-fit.  Rather, $\sqrt{\chi^2/(N-1)}$, where $N$ is the
number of points to be fit, is the root-mean-squared (RMS) scatter of
the fit.

\subsection{Three-Dimensional Lines \label{app-model-LM-line}}

Finding the orientation and position of a 3-D line that
fits data is equivalent to finding the transformation that aligns the
data with a fixed line.  Define that transformation as the vector
offset $(x_0,y_0,z_0)$ and pitch-yaw-roll (PYR) rotation through
angles $(\theta,\phi,\psi)$.  As for the ellipse, we seek to find the
six parameters that minimize the total distance from all transformed
points to the $z$ axis. The same caveat discussed above, regarding the
statistical interpretation of $\chi^2$, applies.

\subsection{Cones and Frusta \label{app-model-LM-cone}}

As with the ellipse, the traditional cartesian definition of
a cone is not condusive to LM minimization.  Rather, we implement a novel
combination of the previous two subsections.  A cone can be considered
as a series of nested ellipses aligned along a given vector $\hat{n}$,
where the axes of each ellipse are functions of the distance of its center
from some point ${\bf x_0}$.  If the ellipses have non-zero major and
minor axes at ${\bf x_0}$, the figure is a truncated cone or
frustum.  The distance from the apex to the truncated top edge is
defined as $h$ (see Figure \ref{LMfig}b).

Proceeding as before, the data are rotated and translated to fit them
to a cone aligned along the $z$ axis.  The major and
minor axes are defined to be $a(z)=a_0(|z|+h)$, $b(z)=b_0(|z|+h)$,
where $h$ is defined above.  Defined in terms of $|z|$, an
hourglass (two cones reflected about the $x-y$ plane) can also be fit.

Consider the point $P$ as shown in Figure \ref{LMfig}\pant{b}.  In
truth, it is a distance $d$ from the cone, however calculation of the
point $R$, or its projection $C$ on the $z$ axis, is a complicated
function with extremely complicated partial derivates.  Rather, note
that the triangles $\triangle AQB$ and $\triangle PRQ$ are similar,
and if the cone has a fixed half-opening angle ($\angle BAQ=\alpha$),
$d/l=\cos{(\angle QPR)}$ is a fixed ratio, thus minmizing $l$ is
equivalent to minimizing $d$.  The application of this is simple,
since the distance $l$ can be measured using the technique from
\S\ref{app-model-LM-ell}.  Note however that the half-opening angle
$\alpha$ is fixed only for a cone with circular cross section
($a_0=b_0$).  For the general case of an elliptic cross section, $d/l$
is not constant with angle about the $z$ axis.  However, restricting
this technique to small eccentricities ($a\lesssim b$), $d/l$ is
approximately constant, and this technique is still valid.

Requiring that the axis of the cone pass through the origin, the
center ${\bf x_0}=k\hat{n}$, where $\hat{n}$ is the unit vector
$\hat{k}$ rotated by the PYR angles $(\psi_z,\theta_x,\phi_y)$.  This
cone (or frustum) can be fit by seven parameters: three rotation
angles, two ellipse axes, $h$ and $k$.  For a general center ${\bf
x_0}$, the fit requires nine parameters.  However, in the adopted PYR
rotation convention, $\psi_z$ has no effect for a circular cross
section, for which case the above fits are reduced to five and seven
parameters, respectively.


\section{Visualization Transformations }\label{app-3de}

\begin{figure}\centering
\includegraphics[width=0.6\linewidth,angle=0]{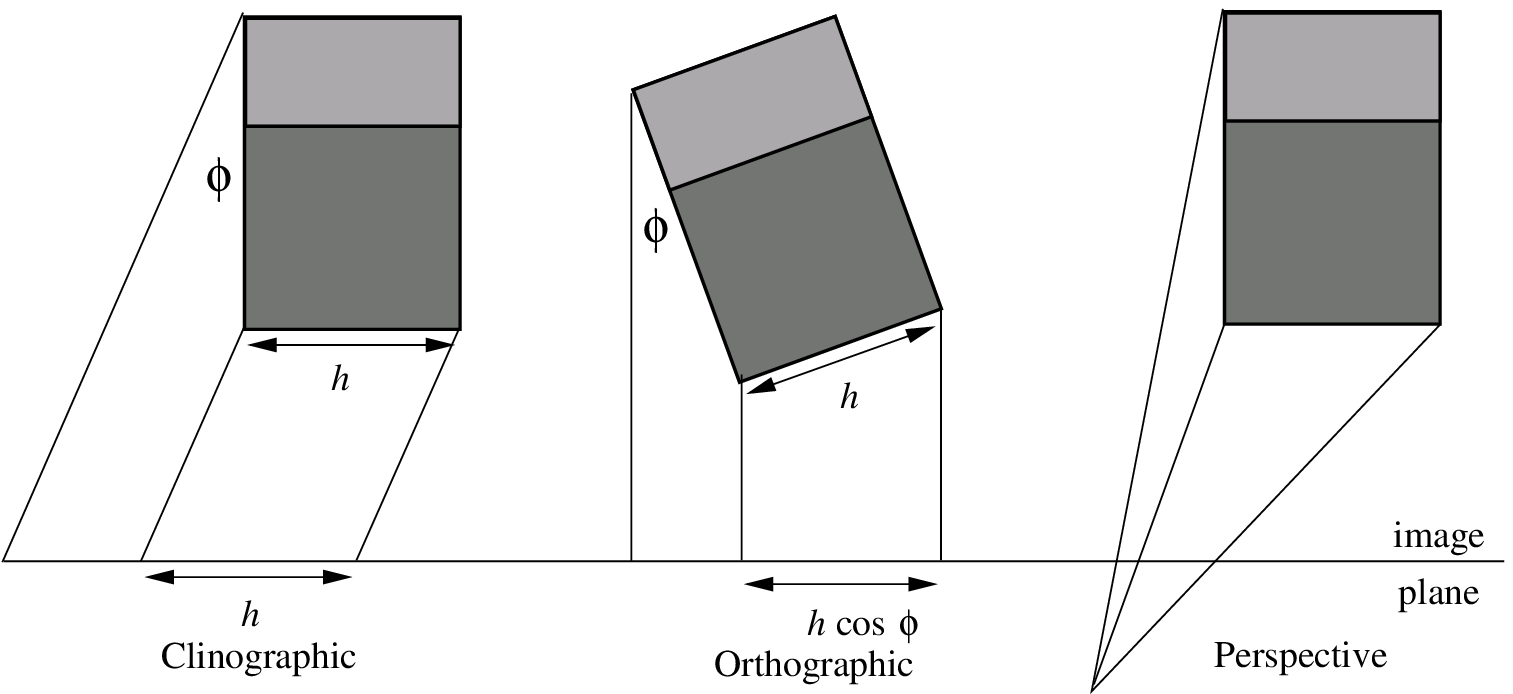}
\caption{Schematic showing the
principles of clinographic, orthographic, and perspective projections
onto an image plane.  In each case the figure is a cube, rotated
such that two opaque faces can be seen (dark and light grey) from
above.  After Fig.\ 1 of \citet{AM84}.
\label{proj}}
\end{figure}

A very large number of projections exist to represent a
3-D object on the two-dimensional image plane of a piece
of paper. Three of these are shown in Figure \ref{proj}.  A
clinographic projection views an object from an angle $\phi$ with each
point on the cube projected onto the image plane by a parallel line.
The cube is now seen including the top face, and the size $h$ of the
cube is preserved.  An orthographic projection first rotates the cube
by the angle $\phi$ and then projects the image onto the plane with
orthogonal lines, such that the side $h$ now has a length
$h\cos{\phi}$.  Such ``parallel projections'' are difficult to
interpret visually since we do not observe 3-D structures this way.
Rather, our eyes apply conical or perspective projections, by which
the relative sizes of points diminish with increasing distance from
the observer.  Since parallel lines project to an imaginary apex,
zonal relationships appear slightly distorted, but are also easier to
understand.

To view a 3-D object, one first applies orthogonal transformations
(positional shifts, rotations, scalings), then the resulting object is
projected onto the image plane using either a parallel or perspective
transformation.  The general case provides for: a linear shift in
three dimensions, rotations about three axes, a uniform scaling,
followed by a dimetric and perspective projection.  Consider the
image plane as a vertical pane of glass, with the object held in front
of the pane by a metal rod, connected orthogonally to the glass.  The
dimetric projection is an orthographic projection applied in two
dimensions, and establishes where along the glass the observer
is located with repsect to that rod.  The perspective projection
corrects for the distance of the observer behind the glass.   

All of these can be easily specified as $4\times4$ matrices.  A
rotation $R$, shift $S$, and translation $T$ are given by 
$$
R=
\begin{pmatrix}
a & b & c & 0 \\ 
d & e & f & 0 \\
g & h & i & 0 \\
0 & 0 & 0 & 1 \\
\end{pmatrix} 
\quad
S=
\begin{pmatrix}
s & 0 & 0 & 0 \\ 
0 & s & 0 & 0 \\
0 & 0 & s & 0 \\
0 & 0 & 0 & 1 \\
\end{pmatrix} 
\quad
T=
\begin{pmatrix}
1 & 0 & 0 & 0 \\
0 & 1 & 0 & 0 \\
0 & 0 & 1 & 0 \\
\Delta x & \Delta y& \Delta z & 1 \\
\end{pmatrix} 
$$ where for $R$ we have left the sin and cos coefficients out since
they vary with the particular convention used (i.e.\ Euler angles
versus pitch-yaw-roll).  A dimetric projection $P_d$ extrudes
projected surfaces in the direction of a viewing vector $\vec{v}$.  A
perspective projection $P_p$ is defined by the parameter $\alpha$,
where $\alpha\rightarrow\infty$ reduces to an orthographic projection
with $\phi=0$.  They are given by
$$
P_d=
\begin{pmatrix}
1 & 0 & 0 & 0 \\
0 & 1 & 0 & 0 \\
{-v_x}/{v_z} & {-v_y}/{v_z} & 0 & 0 \\
0 & 0 & 0 & 1 \\
\end{pmatrix}
\qquad
P_p=
\begin{pmatrix}
1 & 0 & 0 & 0 \\
0 & 1 & 0 & 0 \\
0 & 0 & {\alpha}^{-1} & {\alpha}^{-1} \\
0 & 0 & 0 & 1 \\
\end{pmatrix} 
$$
To apply a perspective projection, the position vector $(x,y,z,1)$ is
multiplied by $P_d$ and then normalized by the fourth component,
i.e. a perspective-transformed point is given by
$$
\left(\frac{\alpha x}{z+\alpha},\frac{\alpha
y}{z+\alpha},\frac{z}{z+\alpha}  \right). 
$$ Any combination of orthogonal and perspective transformations can
be applied by finding the matrix product of all relative
transformations, and then applying that new matrix to each 3-D datum,
loaded into a 4-vector as above.  The projected position onto the
image plane is given by the first two components of the final vector,
the third component fills the $z$-buffer to store depth information,
and the fourth component returns to 1.

In our implementation of the 3-D rendering described in
\sect{sec-lea-analy-3de}, we find the dimetric projection is never
needed.  Most renderings in this work have been generated with
$\alpha=10$, which creates a shallow perspective that aids the eye in
understanding the correct orientation of the coordinate box, but has a
small effect on zonal relationships.




\begin{thebibliography}{}

\bibitem[Alard(2000)]{Ala00} Alard, C.\ 2000, \aaps, 144, 363

\bibitem[Alcock et al.(1995)]{Alc95} Alcock, C.~et al.\ 1995,  
 \aj, 109, 1653

\bibitem[Angell \& Moore(1984)]{AM84}Angell, I.~O.~\& Moore, M., 1984,
 Teaching Pamphlets 12: Projections of Cubic C\ rystals, International
 Union of Crystallography, (Cardiff: University College C\ ardiff
 Press)

\bibitem[Arnett(1987)]{Arn87} Arnett, W.~D.~1987, \apj, 319, 136

\bibitem[Arnett(1988)]{Arn88} Arnett, W.~D.\ 1988, \apj, 331, 377

\bibitem[Arnett et al.(1989)]{Arn89}
 Arnett, W.D., Bahcall, J.N., Kirshner, R.P., Woosley, S.E.~1989,
 \araa,  27, 629


\bibitem[Balick, Preston \& Icke(1987)]{Bal87}
    Balick, B., Preston, H.L., Icke, V.~1987, \aj, 94, 164

\bibitem[Bjorkman \& Cassinelli(1993)]{BC93}Bjorkman, J.~E.,
   Cassinelli, J.~P.~1993, \apj, 409, 429

\bibitem[Blondin \& Lundqvist(1993)]{BL93}
   Blondin, J.~M., Lundqvist, P.~1993, \apj, 405, 337

\bibitem[Bond et al.(1989)]{Bon89} Bond, H.~E., Gilmozzi, R., Meakes,
 M.~G., Panagia, N.~1989 \iaucirc, 4733

\bibitem[Bond et al.(1990)]{Bon90} Bond, H.~E., Gilmozzi, R., Meakes,
 M.~G., \& Panagia, N.\ 1990, \apjl, 354, L49 

\bibitem[Bond et al.(2003)]{Bon03} Bond, H.~E., et al.\ 2003, \nat,
  422, 405

\bibitem[Burrows et al.(1995)]{Bur95}
    Burrows, C.J., et al.~1995, \apj, 452, 680

\bibitem[Cardelli et al.(1989)Cardelli, Clayton \& Mathis]{Car89}
 Cardelli, J.~A., Clayton, G.~C.~\& Mathis, J.~S.\ 1989, \apj, 345, 245

\bibitem[Castor et al.(1975)Castor, McCray, \& Weaver]{CMW75} Castor,
 J., McCray, R., \& Weaver, R.\ 1975, \apjl, 200, L107 

\bibitem[Chevalier(1986)]{Che86} Chevalier, R.~A.\ 1986,
 \apj, 308, 225

\bibitem[Chevalier \& Emmering(1989)]{CE89}
 Chevalier, R.~A.\ \& Emmering, R.~T.~1989, \apjl, 342, L75

\bibitem[Collins et al.(1999)]{Col99} Collins, T.~J.~B., Frank, A.,
Bjorkman, J.~E., Livio, M.~1999, \apj, 512, 322

\bibitem[Couch \& Malin(1989)]{CM89}Couch, W.~J., Malin, D.~F.~1989,
 \iaucirc 4739

\bibitem[Couderc(1939)]{Cou39}Couderc, P.~1939, Ann d'Ap, 2, 271

\bibitem[Crotts(1992)]{Cro92} Crotts, A.~P.~S.\ 1992, \apjl, 399, L43 

\bibitem[Crotts \& Heathcote(1991)]{CH91}
  Crotts, A.P.S., \& Heathcote, S.R.~1991, \nat, 350, 683

\bibitem[Crotts \& Heathcote(2000)]{CH00}Crotts, A.~P.~S.~\& Heathcote,
  S.~R.\ 2000, \apj, 528, 426 

\bibitem[Crotts \& Kunkel(1989)]{CK89} Crotts, A.~P.~S., Kunkel,
  W.~E.~1989, \iaucirc 4741 

\bibitem[Crotts \& Kunkel(1991)]{CK91}Crotts, A.~P.~S.\ \& Kunkel,
  W.~E.\ 1991, \apjl, 366, L73

\bibitem[Crotts et al.(1995)Crotts, Kunkel \& Heathcote]{CKH95}Crotts,
  A.~P.~S., Kunkel, W.~E.~\& Heathcote, S.~R.\ 1995, \apj, 438, 724 

\bibitem[Crotts et al.(1989)Crotts, Kunkel \& McCarthy]{CKM89}
    Crotts, A.P.S., Kunkel, W.E., \& McCarthy, P.J.~1989, \apj, 347,
    L61

\bibitem[Draine \& Lee(1984)]{DL84} Draine, B.~T.~\&  Lee, 
 H.~M.\ 1984, \apj, 285, 89 (Erratum: 318, 485)

\bibitem[Feast(1999)]{Fea99} 
    Feast, M.\ 1999, \pasp, 111, 775

\bibitem[Fischera et al.(2002)Fischera, Tuffs, \& V{\" o}lk]{Fis02} 
Fischera, J., Tuffs, R.~J., \& V{\" o}lk, H.~J.\ 2002, \aap, 395, 189 

\bibitem[Fransson et al.(1989)]{Fra89}
     Fransson, C., et al.~1989, \apj, 336, 429

\bibitem[Fruchter \& Hook(2002)]{FH02} Fruchter, A.~S.~\& Hook, R.~N.\
  2002, \pasp, 114, 144

\bibitem[Gould \& Uza(1998)]{Gou98}
    Gould, A., \& Uza, O.~1998, \apj, 494, 118

\bibitem[Groth(1986)]{Gro86} Groth, E.~J.\ 1986, \aj, 91, 1244

\bibitem[Hamuy et al.(1988)]{Ham88} 
Hamuy, M., Suntzeff, N.~B., Gonzalez, R., \& Martin, G.\ 1988, \aj,
  95, 63 

\bibitem[Henyey \& Greenstein(1941)]{HG41} Henyey, L.~C.~\&
  Greenstein, J.~L.\ 1941, \apj, 93, 70  

\bibitem[Jakobsen et al.(1991)]{Jak91}
    Jakobsen, P., et al.~1991, \apjl, 369, L63

\bibitem[Meixner et al.(2004)]{APN} Meixner, M., Kastner, J.~H.,
 Balick, B., \& Soker, N.\ 2004, Asymmetrical Planetary Nebulae III:
 Winds, Structure and the Thunderbird, ASP
 Conference Proceedings, Vol.\ 313 (San Francisco, PASP) 2004
 
\bibitem[Kwok(1982)]{Kwo82}
  Kwok, S.O~1982, \apj, 258, 280

\bibitem[Li \& Draine(2001)]{Li01} Li, A.~\& Draine, B.~T.\ 2001,
  \apj, 554, 778  

\bibitem[Livio \& Soker(1988)]{LS88}
  Livio, M., Soker, N.~1988, \apj, 329, 764

\bibitem[Lloyd, O'Brien, \& Kahn(1995)]{LOK95} Lloyd, H.~M., 
  O'Brien, T.~J., \& Kahn, F.~D.\ 1995, \mnras, 273, L19

\bibitem[Luo \& McCray(1991b)]{Luo91} Luo, D., \& McCray, R.~1991b,
 379, 659

\bibitem[Luo et al.(1994)Luo, McCray, \& Slavin]{Luo94} Luo, D.,
 McCray, R., \& Slavin, J.\ 1994, \apj, 430, 264

\bibitem[Martin \& Arnett(1995)]{MA95}
 Martin, C.~L., Arnett D.~1995, \apj, 447, 378

\bibitem[Mathis et al.(1977)Mathis, Rumpl, \& Nordsieck]{MRN77}
  Mathis, J.~S., Rumpl, W., \& Nordsieck, K.~H.\ 1977, \apj, 217, 425 (MRN)


\bibitem[Michael et al.(2002)]{Mic01} Michael, E., et al.\ 2002, \apj,
  574, 166  

\bibitem[Misselt et al.(1999)Misselt, Clayton, \& Gordon]{Mis99}
 Misselt, K.~A., Clayton, G.~C., \& Gordon, K.~D.\ 1999, \pasp, 111, 1398 

\bibitem[Panagia et al.(1996)]{Pan96}
   Panagia, N., et al.~1996, \apjl, 459, L17

\bibitem[Plait et al.(1995)]{Pla95}
   Plait, P., Lundqvist, P., Chevalier, \& R., Kirshner, R.~1995,
   \apj, 439, 730

\bibitem[Podsiadlowski, Fabian, \& Stevens(1991)]{Pod91}
Podsiadlowski, P., Fabian, A.~C., \& Stevens, I.~R.\ 1991, \nat, 354, 43

\bibitem[Press et al.(1992)]{Press}
   Press, W.H., et al.\ 1992, Numerical Recipes in Fortran, 2nd
   ed.\ (Cambridge: University Press)

\bibitem[Romaniello et al.(2000)]{Rom00}
   Romaniello, M., Salaris, M., Cassisi, S., \& Panagia, N.\ 2000,
   \apj, 530, 738

\bibitem[Sanduleak(1969)]{San69}Sanduleak, N.~1969, Contr. CTIO, 89, 1

\bibitem[Scuderi et al.(1996)]{Scu96} Scuderi, S., Panagia, N.,
  Gilmozzi, R., Challis, P.~M., \& Kirshner, R.~P.\ 1996, \apj, 465, 956 

\bibitem[Shigeyama et al.(1987)]{Shi87} Shigeyama, T., Nomoto, K.,
Hashimoto, M., \& Sugimoto, D.\ 1987, \nat, 328, 320 

\bibitem[Sonneborn et al.(1997)]{Son97} Sonneborn, G., 
Fransson, C., Lundqvist, P., Cassatella, A., Gilmozzi, R., Kirshner, 
R.~P., Panagia, N., \& Wamsteker, W.\ 1997, \apj, 477, 848

\bibitem[Sonneborn et al.(1998b)]{Son98b}
   Sonneborn, G., et al.~1998, \apjl, 492, L139

\bibitem[Stetson(1987)]{Ste87}
   Stetson, P.B.~1987, \pasp, 99, 191

\bibitem[Sugerman(2003)]{Sug03}Sugerman, B.~E.~K., 2003, \aj, 126, 1939

\bibitem[Sugerman et al.(2005)]{Sug05}Sugerman, B.~E.~K., Crotts,
  A.~P.~S., Kunkel, W.~E., Heathcote, S.~R., \& Lawrence, S.~S.~2005,
  \apj, in press; astro-ph/0502268

\bibitem[Sugerman et al.(2002)]{Sug02}Sugerman, B.~E.~K., Lawrence,
  S.~S., Crotts, A.~P.~S., Bouchet, P., Heathcote, S.~R., 2002, \apj, 572, 209 

\bibitem[Tomaney \& Crotts(1996)]{TC96}Tomaney, A., \& Crotts,
  A.~P.~S.\ 1996, \aj, 112, 2872 

\bibitem[Walborn et al.(1989)]{Wal89} Walborn, N.~R.,
Prevot, M.~L., Prevot, L., Wamsteker, W., Gonzalez, R., Gilmozzi, R.,
\& Fitzpatrick, E.~L.\ 1989, \aap, 219, 229 

\bibitem[Walker \& Suntzeff(1990)]{WS90} Walker, A.~R.~\& Suntzeff,
N.~B.\ 1990, \pasp, 102, 131  

\bibitem[Wampler et al.(1990a)]{Wam90a} Wampler, J., D'Odorico, 
S., Gouiffes, C., Tarenghi, M., \& Wang, L.-F.\ 1990, \iaucirc, 4943, 1

\bibitem[Wampler et al.(1990b)]{Wam90b} Wampler, E.~J., Wang, 
L., Baade, D., Banse, K., D'Odorico, S., Gouiffes, C., \& Tarenghi, M.\ 
1990, \apjl, 362, L13 


\bibitem[Wang, Dyson, \& Kahn(1993)]{WDK93} Wang, L., Dyson, 
J.~E., \& Kahn, F.~D.\ 1993, \mnras, 261, 391 

\bibitem[Wang \& Mazzali(1992)]{WM92}
  Wang, L., Mazzali, P.A.~1992, \nat, 355, 58

\bibitem[Wang \& Wampler(1992)]{WW92} Wang, L.~\& Wampler, 
E.~J.\ 1992, \aap, 262, L9 

\bibitem[Weingartner \& Draine(2001)]{WD01}
  Weingartner, J.~C.~\& Draine, B.~T.\ 2001, \apj, 548, 296 (WD01)

\bibitem[Woosley(1988)]{Woo88}
  Woosley, S.E.~1988, \apj, 330, 218

\bibitem[Woosley et al.(1987)]{Woo87} Woosley, S.~E., Pinto, P.~A.,
 Martin, P.~G., Weaver, T.~A.~1987, \apj, 318, 664

\bibitem[Woosley, Pinto \& Weaver(1988)]{WPW88} Woosley,
 S.~E., Pinto, P.~A., Weaver, T.~A.~1988, Pub.\ ASAu, 7, 355

\bibitem[Xu et al.(1994)Xu, Crotts \& Kunkel]{Xu94}Xu, J., Crotts,
  A.~P.~S.~\&  Kunkel, W.~E.~1994, \apj, 435, 274
\bibitem[Xu et al.(1995)Xu, Crotts \& Kunkel]{Xu95}Xu, J., Crotts,
  A.~P.~S.~\& Kunkel, W.~E.~1995, \apj, 451, 806 (Erratum: 463, 391) 
\bibitem[Xu \& Crotts(1999)]{Xu99}Xu, J.~\& Crotts, A.P.S.~1999, \apj,
 511, 262

\bibitem[Zacharias et al.(2004)]{Zac04} Zacharias, N., Urban, S.~E.,
 Zacharias, M.~I., Wycoff, G.~L., Hall, D.~M., Monet, D.~G., \&
 Rafferty, T.~J.\ 2004, \aj, 127, 3043

\bibitem[Zacharias et al.(2000)]{Zac00} Zacharias, N., et al.\ 2000,
  \aj, 120, 2131


\end{thebibliography}
\end{document}